\documentclass[twocolumn,superscriptaddress]{revtex4-2}
\usepackage{amsmath,amssymb,amsfonts}
\usepackage{xcolor,graphicx}
\usepackage[bookmarks=false,linkcolor=cyan,urlcolor=blue,colorlinks,citecolor=cyan]{hyperref}

\renewcommand{\d}{\mathrm d}
\newcommand{\unit}[1]{\,\mathrm{#1}}
\newcommand{\ket}[1]{|#1\rangle}
\newcommand{\bra}[1]{\langle #1|}
\newcommand{\ketbra}[2]{\ket{#1}\!\bra{#2}}
\DeclareMathOperator{\spn}{span}

\begin{document}

\title{Erasure qubits: Overcoming the $T_1$ limit in superconducting circuits}

\author{Aleksander Kubica}
\thanks{These authors contributed equally to this work.}
\affiliation{AWS Center for Quantum Computing, Pasadena, CA 91125, USA}
\affiliation{California Institute of Technology, Pasadena, CA 91125, USA}
\author{Arbel Haim}
\thanks{These authors contributed equally to this work.}
\affiliation{AWS Center for Quantum Computing, Pasadena, CA 91125, USA}
\affiliation{California Institute of Technology, Pasadena, CA 91125, USA}
\author{Yotam Vaknin}
\thanks{These authors contributed equally to this work.}
\affiliation{AWS Center for Quantum Computing, Pasadena, CA 91125, USA}
\affiliation{Racah Institute of Physics, The Hebrew University of Jerusalem, Jerusalem 91904, Givat Ram, Israel}
\author{Fernando Brand\~ao}
\affiliation{AWS Center for Quantum Computing, Pasadena, CA 91125, USA}
\affiliation{California Institute of Technology, Pasadena, CA 91125, USA}
\author{Alex Retzker}
\affiliation{AWS Center for Quantum Computing, Pasadena, CA 91125, USA}
\affiliation{California Institute of Technology, Pasadena, CA 91125, USA}
\affiliation{Racah Institute of Physics, The Hebrew University of Jerusalem, Jerusalem 91904, Givat Ram, Israel}
\date{\today}

\begin{abstract}
The amplitude damping time, $T_1$, has long stood as the major factor limiting quantum fidelity in superconducting circuits, prompting concerted efforts in the material science and design of qubits aimed at increasing $T_1$.
In contrast, the dephasing time, $T_{\phi}$, can usually be extended above $T_1$ (via, e.g., dynamical decoupling), to the point where it does not limit fidelity.
In this article we propose a scheme for overcoming the conventional $T_1$ limit on fidelity by designing qubits in a way that amplitude damping errors can be detected and converted into erasure errors.
Compared to standard qubit implementations our scheme improves the performance of fault-tolerant protocols, as numerically demonstrated by the circuit-noise simulations of the surface code.
We describe two simple qubit implementations with superconducting circuits and discuss procedures for detecting amplitude damping errors, performing entangling gates, and extending $T_\phi$.
Our results suggest that engineering efforts should focus on improving $T_\phi$ and the quality of quantum coherent control, as they effectively become the limiting factor on the performance of fault-tolerant protocols.
\end{abstract}
\maketitle

\section{Introduction}
\label{sec:intro}

Due to the fragile nature of quantum states, the path towards a fault-tolerant universal quantum computer will most likely go through implementing robust and efficient schemes for quantum error correction (QEC).
Protocols for QEC can largely benefit from the knowledge of the type of noise affecting the physical system~\cite{Aliferis2008, Aliferis2009,Sarvepalli2009,Brooks2013,Stephens2013,Webster2015,gefen2017enhancing}.
For example, it was shown that bias between the $X$, $Y$ and $Z$ components of Pauli noise can be exploited to dramatically increase the error-correction threshold and reduce the qubit overhead for various variants of the surface code~\cite{tuckett2019tailoring,tuckett2018ultrahigh,tuckett2020fault,bonilla2021xzzx,Dua2022,Higgott2022,Xu2022}.
More recently, such an error bias was effectively engineered in the context of bosonic cat qubits, which exhibit a bias between photon-loss errors and photon-dephasing errors~\cite{cochrane1999macroscopically,mirrahimi2014dynamically,ofek2016extending,puri2019stabilized, guillaud2019repetition}.

One type of noise bias that is fundamental to many quantum technologies is the bias between amplitude damping and dephasing errors.
The corresponding bias $T_\phi/T_1 > 1$ between the dephasing time, $T_{\phi}$, and the amplitude damping time, $T_1$, can be further magnified since standard techniques, such as dynamical decoupling, can prolong $T_{\phi}$, whereas extending $T_1$ usually necessitates QEC. 
Indeed, $T_\phi$ was recently measured to be more than an order of magnitude larger than $T_1$~\cite{PhysRevLett.120.260504,burnett2019decoherence}, and it is believed that with dynamical decoupling $T_\phi/T_1$ can be substantially extended. 
Moreover, in many quantum computing platforms the coherence time approaches the $T_1$ limit, which is a strong indication that $T_\phi$ can be made to surpass $T_1$ considerably.
These include NV centers in diamond \cite{cao2020protecting,balasubramanian2009ultralong,herbschleb2019ultra} (under the assumption that phonon-induced dephasing is negligible), and excited states of ions and atoms \cite{Sosnova_Thesis}.

\begin{figure}[ht!]
\centering
(a)\includegraphics[width=0.91\columnwidth]{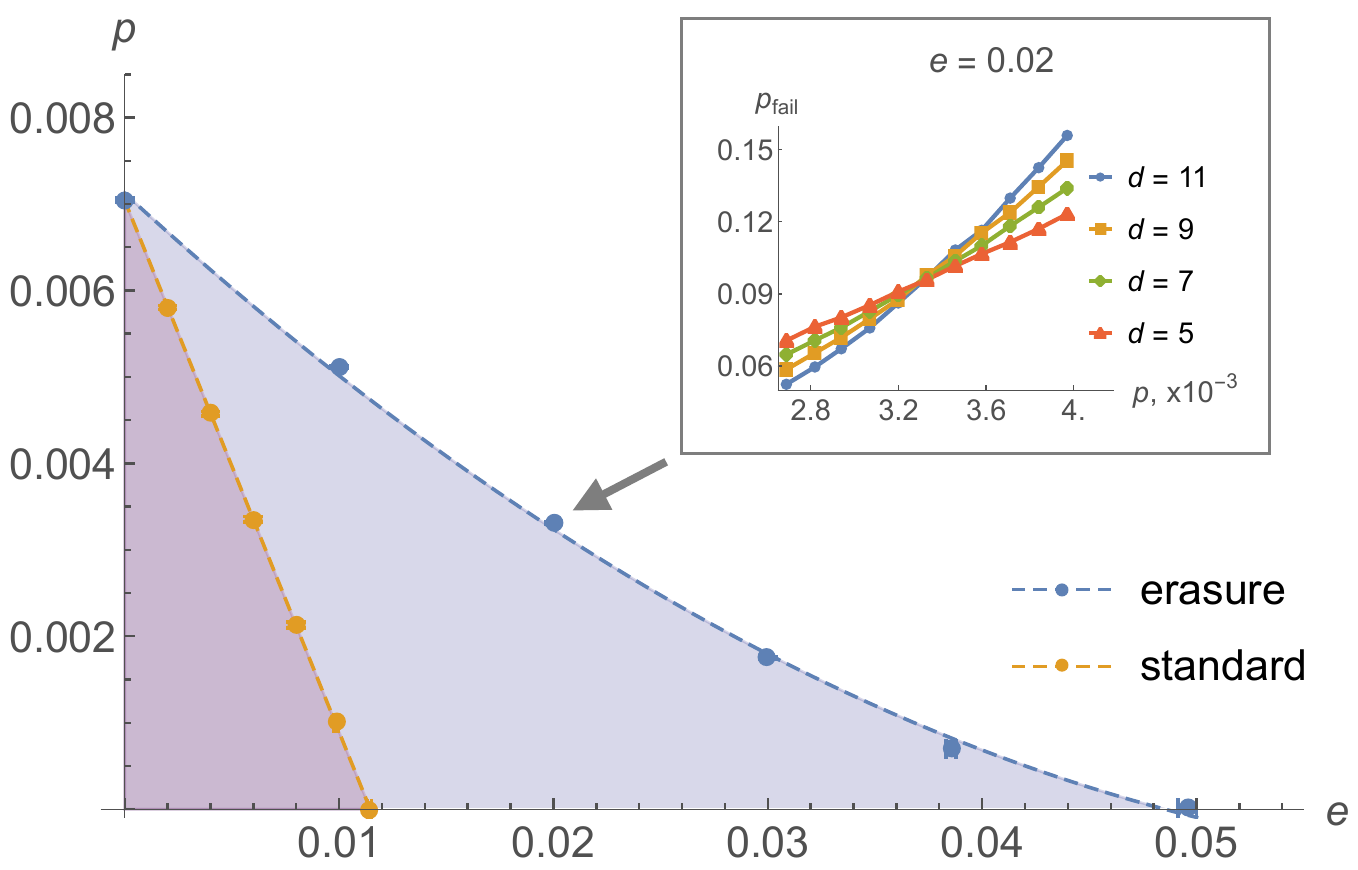}\\
(b)\includegraphics[width=0.91\columnwidth]{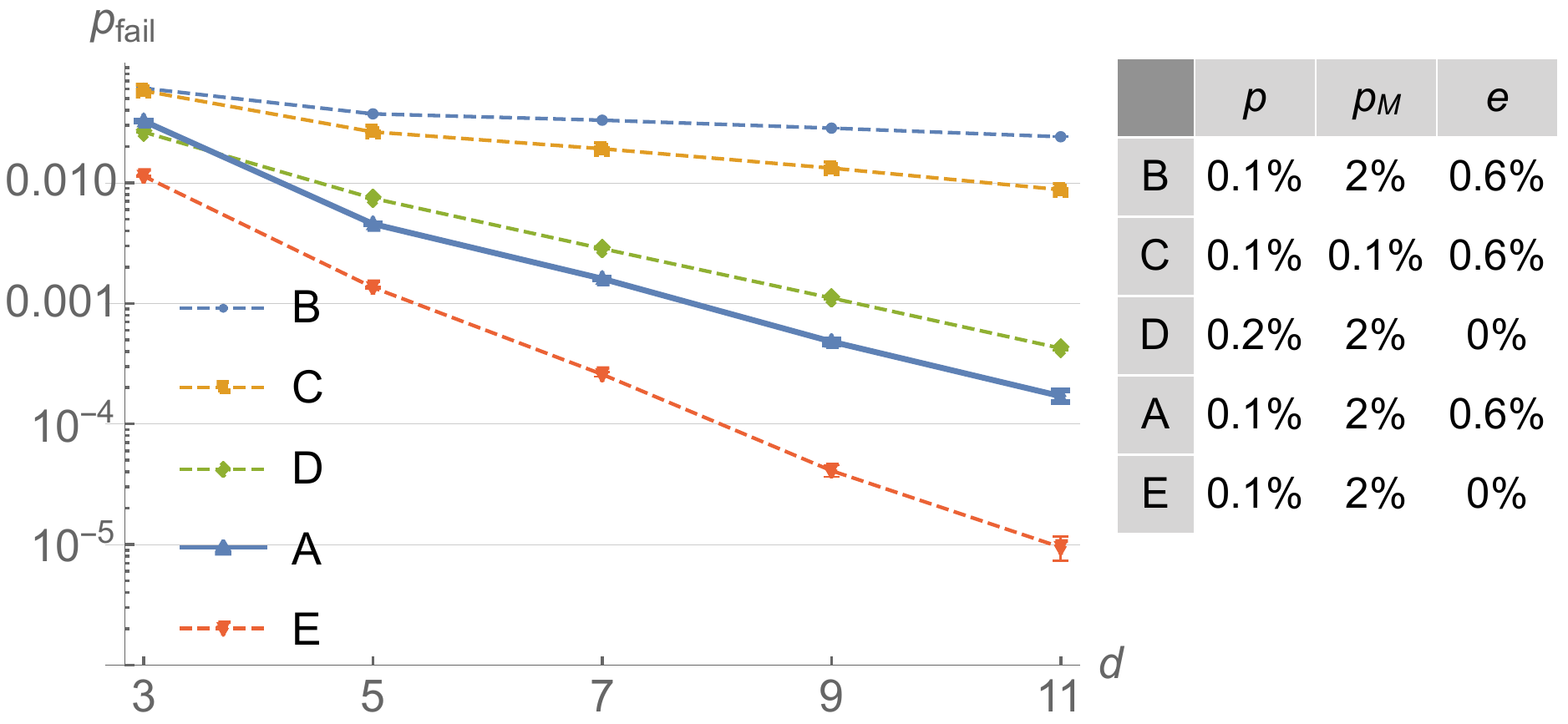}
\caption{
(a) The surface code threshold for the heralded erasure and Pauli errors with rate $e$ and $p$ (assuming the measurement error rate $p_{\rm M} = 2p/3$).
The light (dark) area indicates the correctable region for the erasure (standard) scheme.
(inset) The logical error rate $p_\text{fail}$ of the erasure scheme for $e=0.02$ and different surface code distance $d$ as a function of $p$.
(b) Comparison of the logical error rate $p_\text{fail}$ as a function of distance $d$ for the erasure (A) and standard (B--E) schemes for various values of noise parameters $p$, $p_{\rm M}$ and $e$.
We can directly compare A and B as they use the same parameters; we include C--E to test how changes in $p$, $p_{\rm M}$ and $e$ affect $p_\text{fail}$.
\vspace*{-5mm}
}
\label{fig_threshold}
\end{figure}

A compelling argument that the bias $T_\phi/T_1$ may simplify QEC protocols and subsequently reduce the required resources is the existence of the four-qubit code that approximately corrects any unknown single amplitude damping error~\cite{Leung1997}.
In contrast, the smallest QEC code that exactly corrects any unknown single-qubit Pauli error requires five qubits~\cite{Laflamme1996}.
Moreover, the thresholds for the amplitude damping noise can be substantially higher than for the depolarizing noise.
For instance, the code-capacity threshold of the surface code~\cite{kitaev2003fault,bravyi1998} for the amplitude damping noise is 0.39(2)~\cite{Darmawan2017}, compared to 0.189(3) for the depolarizing noise~\cite{Bombin2012}, which is yet another reason to expect the usefulness of the bias $T_\phi/T_1$.

In this article, we explore the bias $T_\phi / T_1$ and its impact on QEC protocols.
Given a multi-level system that suffers mostly from the amplitude damping errors we define a qubit in such a way that these errors can be detected.
Subsequently, the effective noise affecting the qubit becomes the heralded erasure noise~\cite{Grassl1997}, where the locations of erasures are known.
We call such a qubit an \emph{erasure qubit}.
Our scheme is a specific realization of a more general approach to simplifying QEC protocols---given the knowledge of the hardware noise we engineer qubits in such a way that the effective noise can be corrected efficiently by the QEC code (implemented using these engineered qubits).
Using erasure qubits significantly improves the performance of QEC protocols, as exemplified by our numerical estimates of the memory threshold and the logical error rate for the surface code (see Fig.~\ref{fig_threshold} and Sec.~\ref{sec:SurfCode}).
Lastly, we discuss in detail two simple realizations of erasure qubits with superconducting circuits and describe fault-tolerant implementations of single- and two-qubit gates that preserve the noise structure (see Sec.~\ref{sec:implement}).

\section{Surface code with erasure qubits}
\label{sec:SurfCode}

In this section we show the benefits of using erasure qubits for QEC protocols.
We illustrate our discussion with numerical simulations of the memory threshold of the surface code and demonstrate significant improvements over standard schemes.

\subsection{Why erasure errors?}

\begin{figure}
\includegraphics[width=0.5\columnwidth]{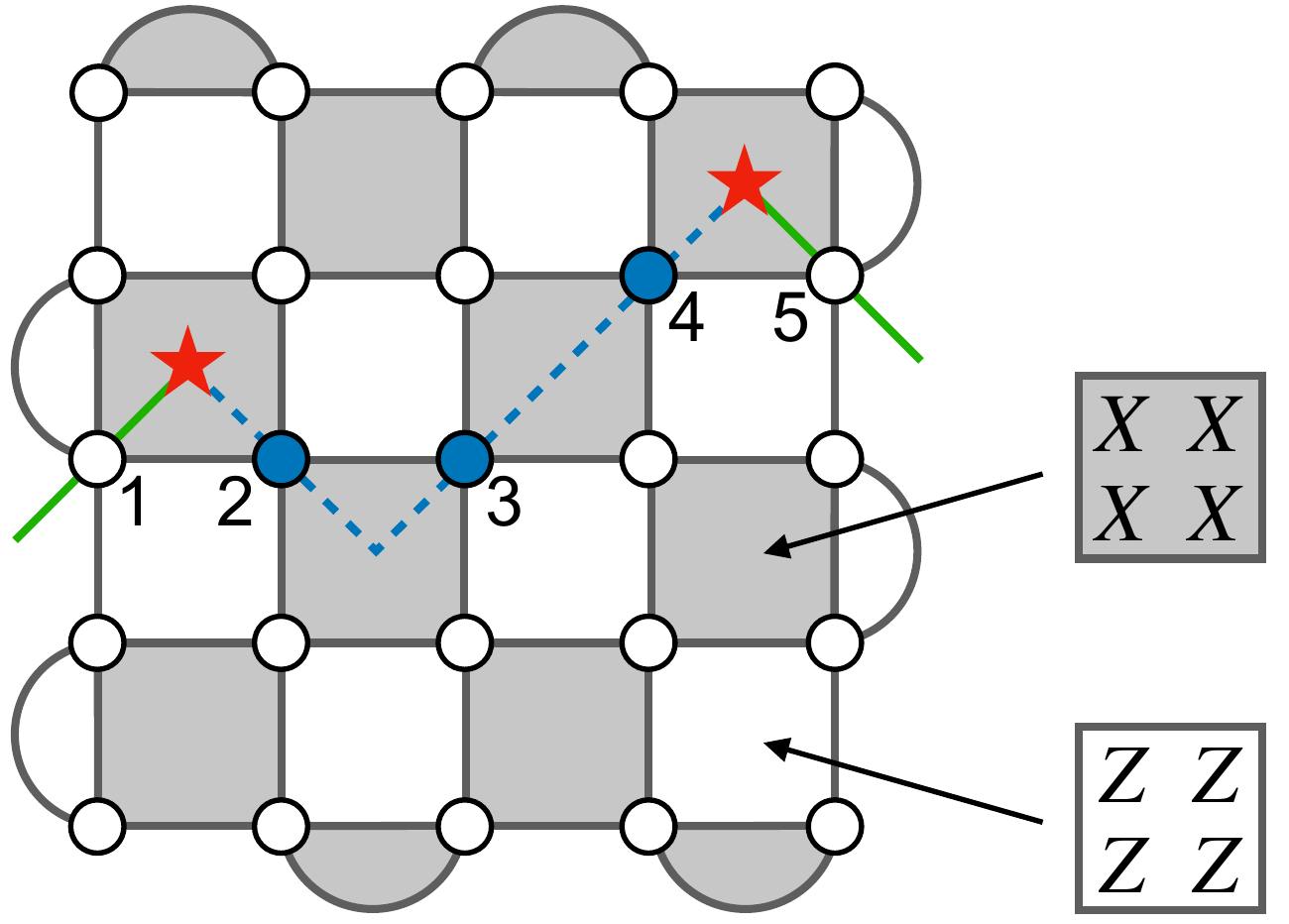}
\caption{
Using erasure qubits can significantly benefit QEC protocols.
We illustrate it with the surface code of distance $d=5$ and one of the possible error syndromes comprising ``point-like excitations'' (red stars) due to erasure errors on some qubits (blue dots).
Without the knowledge of error locations, one may attempt to fix the errors by moving the excitations to the boundary (along the solid green lines) by applying a Pauli operator $\sigma^\text{Z}_1\sigma^\text{Z}_5$, which would result in a logical error.
In contrast, with the knowledge of error locations one correctly chooses to fix the errors by pairing up the excitations (along the dashed blue lines) by applying $\sigma^\text{Z}_2\sigma^\text{Z}_3\sigma^\text{Z}_4$.
}
\label{fig_surface}
\end{figure}

We stress that at the core of our scheme lies a conversion of amplitude damping errors into erasure errors.
Such a conversion can be beneficial from the perspective of QEC.
Unlike the amplitude damping errors, we do know the locations of erasure errors and thus can correct them more effectively, as illustrated in Fig.~\ref{fig_surface}.
Also note that, by definition of code distance, the distance-$d$ code allows to successfully correct any $d-1$ (or fewer) erasure errors.
On the other hand, without the information about the error locations, we are guaranteed to successfully correct only up to $\lfloor(d-1)/2\rfloor$ arbitrary single-qubit errors.
Lastly, the thresholds for the heralded erasure noise can be extremely high. 
For instance, the code-capacity threshold of the square-lattice surface code is $0.5$, as it is equivalent to the bond-percolation threshold on the square lattice~\cite{Stace2009}.

\subsection{Surface code simulations}

To demonstrate the usefulness of erasure qubits for QEC protocols, we numerically estimate the performance of the surface code under the assumption that amplitude damping errors can be efficiently detected, which, in turn, converts them into heralded erasures; see Fig.~\ref{fig_threshold}.
Concretely, we use the minimum-weight perfect matching decoder~\cite{Dennis2002,Fowler2012} to find the memory threshold against the heralded erasure and Pauli (HEP) noise.
We define the HEP noise as follows:
(i) state preparation, idling and CNOT gates are modelled as ideal operations followed by a single- or two-qubit Pauli channel with error rate $p\in[0,1]$ (that is uniformly distributed over all non-trivial single- or two-qubit Pauli operators),
(ii) each Pauli measurement returns a wrong outcome with probability $p_{\rm M}\in[0,1]$,
(iii) additionally, with probability $e\in[0,1]$ each CNOT gate is followed by a fully-depolarizing two-qubit Pauli channel~\footnote{Note that a fully-depolarizing two-qubit Pauli channel has error rate of $15/16$.} and we know when this happens.
We expect the HEP noise to be a reasonable approximation of the effective noise in QEC protocols with erasure qubits.
Namely, erasure errors stem from the amplitude damping errors, whereas Pauli errors stem from other noise mechanisms, presumably dominated by dephasing.
For more simulation details, see Appendix~\ref{sec_numerics}.

We observe that the correctable region for the erasure protocol is roughly $3.5\times$ larger than the corresponding region for the standard protocol, for which we use the same HEP noise but assume no knowledge about the erasure locations; see Fig.~\ref{fig_threshold}(a).
Subsequently, the surface-code threshold of the erasure protocol exceeds the threshold of the standard protocol.
For example, for erasure rate $e=0.01$ the erasure protocol tolerates Pauli error rate $p$ up to 0.0051 (assuming $p_{\rm M} = 2p/3$), which constitutes $5.2\times$ improvement over the standard protocol value.
As exemplified by Fig.~\ref{fig_threshold}(b), the logical error rate for the erasure protocol (solid line) can be significantly lower than for the standard protocol (dashed lines) with the same or even better noise parameters $p$, $p_{\rm M}$ and $e$.
In turn, realizing the break-even point with erasure qubits may be possible with the state-of-the-art hardware.

\subsection{Imperfect detection of erasures}
\label{sub:ImperfecDetecion}

So far we assume that erasure errors can always be detected.
It is likely, however, that this assumption will not be satisfied.
For concreteness, let us consider a scenario of the HEP noise with parameters $p$, $p_{\rm M}$ and $e$, together with false positive and false negative errors for detection of erasures that happen with probability $q_+$ and $q_-$, respectively.
Note that false negative detection errors are equivalent to standard two-qubit Pauli errors (that follows certain CNOT gates), and thus can be crudely incorporated in our simulations by changing $p$ to $p+eq_-$.
On the other hand, false positive detection errors affect the edge weights that are used in the minimum-weight perfect matching decoder.
To roughly estimate their impact, one may replace $e$ with $e+q_+$ in our simulations.

\begin{figure}
\begin{tabular}{lr}
(a)\includegraphics[height=0.19\textheight]{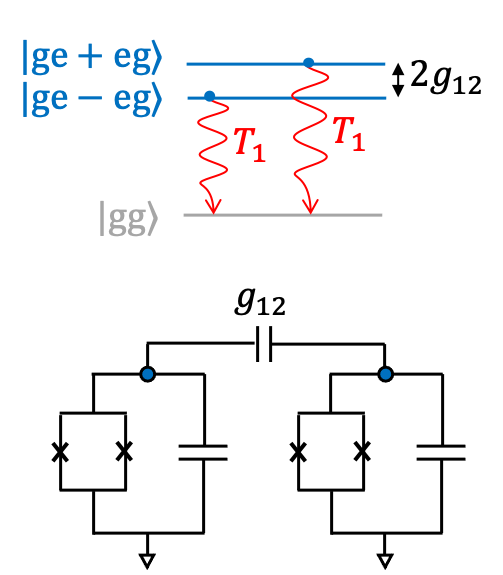}
& (b)\includegraphics[height=0.19\textheight]{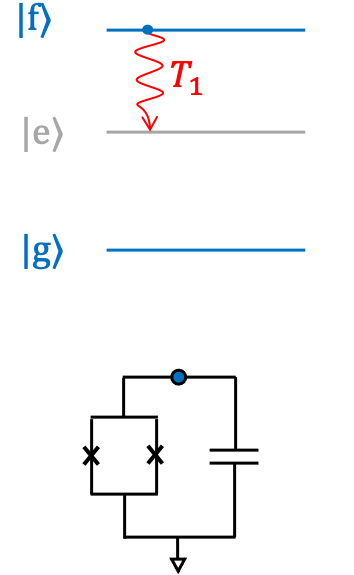}
\end{tabular}
\caption{
Two realizations of the erasure qubit.
(a) The dual-rail qubit is defined as a subspace $\spn\left\{(\ket{\rm ge}\pm\ket{\rm eg})/\sqrt{2}\right\}$ of two coupled transmons.
(b) The \rm{g-f} qubit is defined as a subspace $\spn\{\ket{\rm g},\ket{\rm f}\}$ of a single transmon.
The amplitude damping errors for the dual-rail and {\rm g-f} qubits are detected and converted into erasures by measuring the population in the $|{\rm gg}\rangle$ and $|{\rm e}\rangle$ states, respectively.
}
\label{fig:implementations}
\end{figure}

\section{Realizing erasure qubits}
\label{sec:implement}

Having demonstrated the benefits of erasure qubits, we now discuss possible realizations in the framework of superconducting circuits.
We focus on two different realizations: (i) the dual-rail qubit encoded in a composite system of two transmons, and (ii) the {\rm g-f} qubit encoded in the first three levels of a single transmon (see Fig.~\ref{fig:implementations}).
For each of these qubit realizations we explain how amplitude damping errors can be detected---a necessary ingredient needed in the surface-code scheme that we described.
We then analyze the process of pure dephasing as it directly affects the physical error rate $p$. In particular, we argue that in each of these realizations one can achieve a large $T_\phi/T_1$ bias, required to fully benefit from the QEC scheme of Sec.~\ref{sec:SurfCode}. Finally, we discuss implementation of single-qubit (1Q) and two-qubit (2Q) gates for each of the qubit realizations.

We stress that our 1Q and 2Q gates are, in a sense, fault-tolerant.
To be more concrete, an amplitude damping error that happens at any time during the gate results in a detectable error (the probability of an undetectable error is strongly suppressed as analyzed in the remainder of this section). 
In turn, the structure of the noise and the bias $T_\phi/T_1$ are preserved.

\subsection{The dual-rail qubit}
\label{sub:DR}

We begin with an erasure qubit based on the quantum dual-rail code~\cite{Duan2010}.
Within the rotating-wave approximation the composite
system of two transmons is modeled by the following Hamiltonian~\cite{Shim2016semiconductor,campbell2020universal}
\begin{equation}\label{eq:Hdr}
H_{\rm DR}=\sum_{i=1,2}\left(\omega_{i} a_{i}^{\dagger}a_{i}
+\frac{\eta}{2}a_{i}^{\dagger}a_{i}^{\dagger}a_{i}a_{i}\right)
+g_{12}(a_{1}^{\dagger}a_{2}+{\rm h.c.}),
\end{equation}
where $i=1,2$ labels transmons, $a_i, a^\dagger_i$ are bosonic ladder operators,
$\omega_{i}$ is the transmon frequency, $\eta$ is the anharmonicity (assumed, for simplicity, to be the same for both transmons), and $g_{12}$ is the coupling constant between two transmons.

\begin{figure*}[ht!]
\begin{tabular}{lccr}
\hskip 0.4cm
\includegraphics[clip,trim=0cm -5cm 0cm 0cm,height=0.18\textwidth]{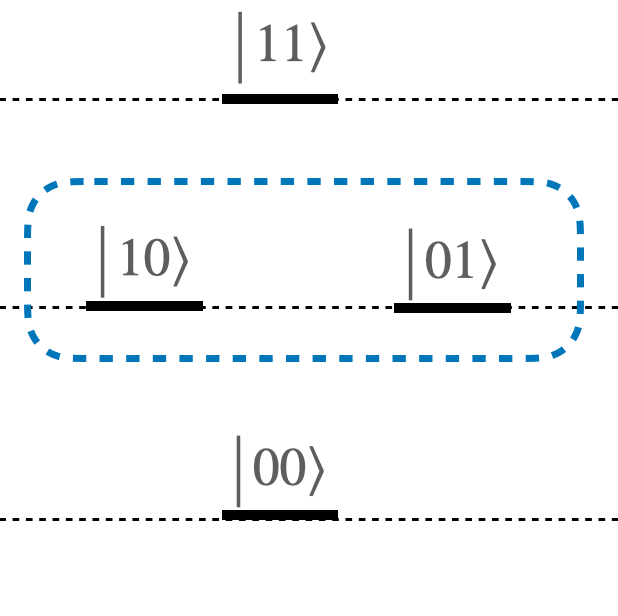}
\llap{\hskip 0mm \parbox[l]{5.9cm}{\vspace{-4.55cm}{\footnotesize $\Omega_0$}\Bigg\updownarrow }}
\llap{\hskip 0mm \parbox[l]{6.12cm}{\vspace{-2.75cm}{\footnotesize $\Omega_0$}\Bigg\updownarrow }}
\llap{\hskip 0mm \parbox[c]{6.5cm}{\vspace{-0.25cm}\small{(a)}}}
&
\hskip -0.35cm
\includegraphics[clip,height=4.0cm]{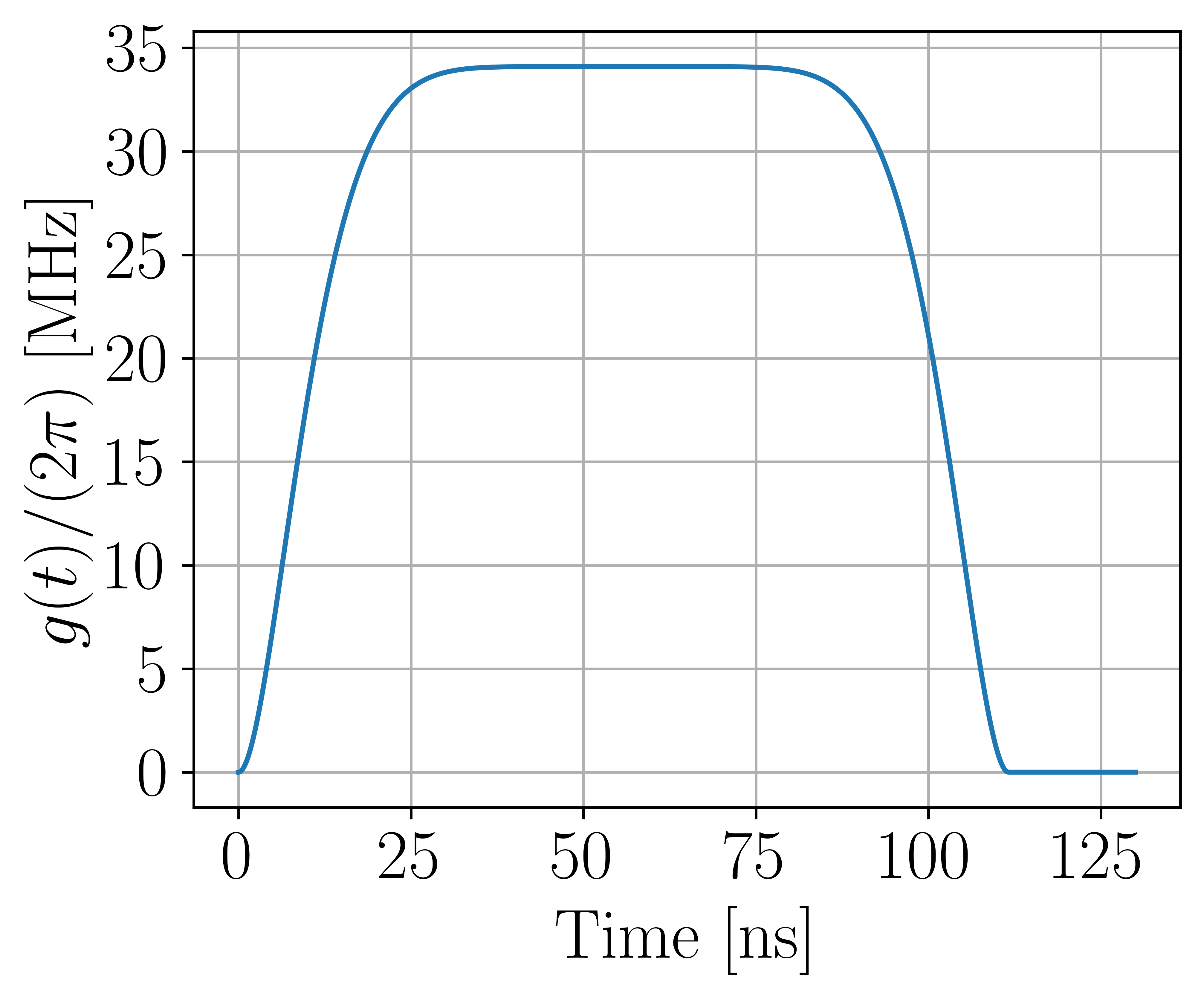}
\llap{\hskip 0mm \parbox[c]{9.5cm}{\vspace{-0.25cm}\small{(b)}}}
&
\hskip -0.3cm
\includegraphics[clip,height=4.0cm]{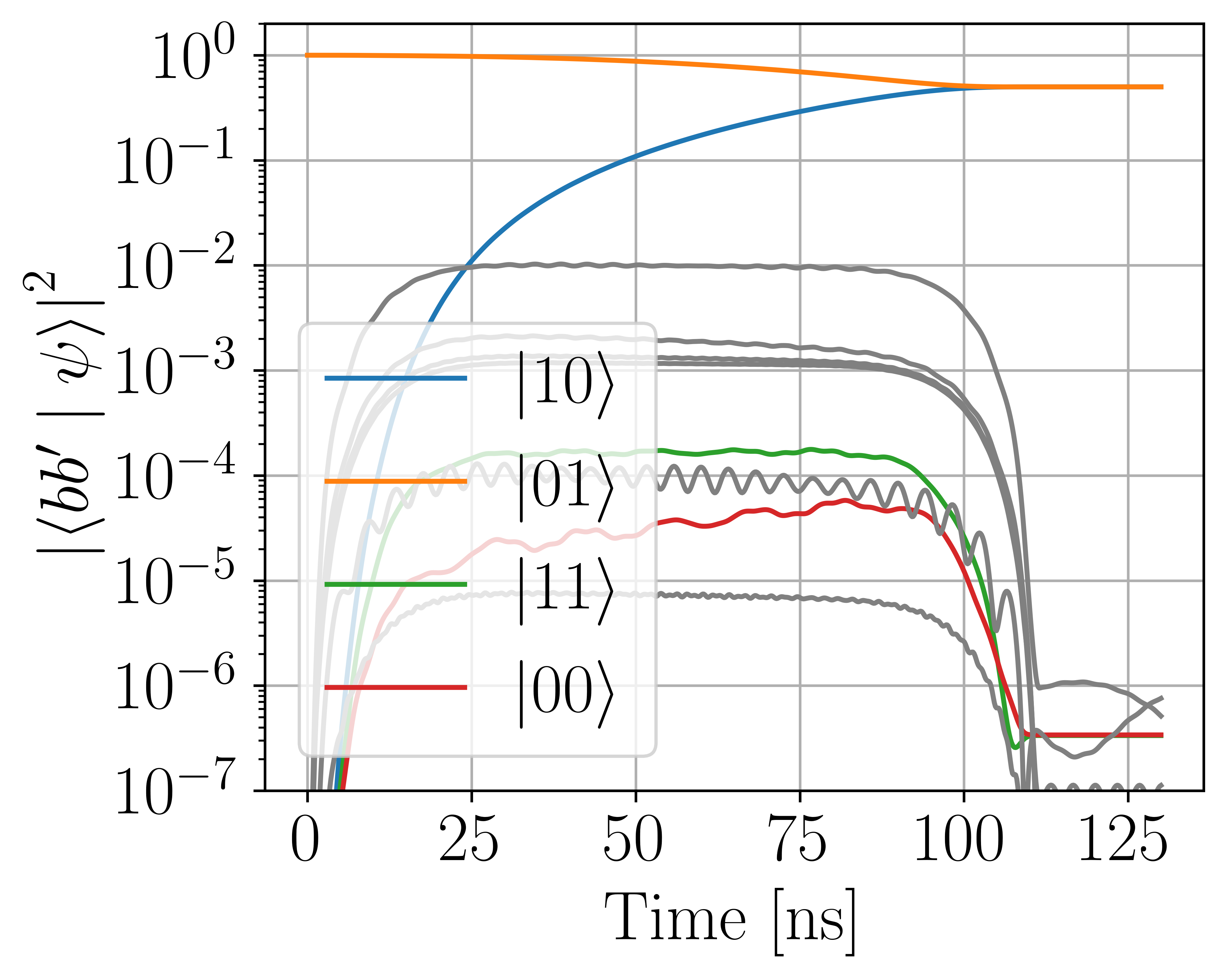}
\llap{\hskip 0mm \parbox[c]{9.5cm}{\vspace{-0.25cm}\small{(c)}}}
&
\hskip -0.3cm
\includegraphics[clip,height=4.0cm]{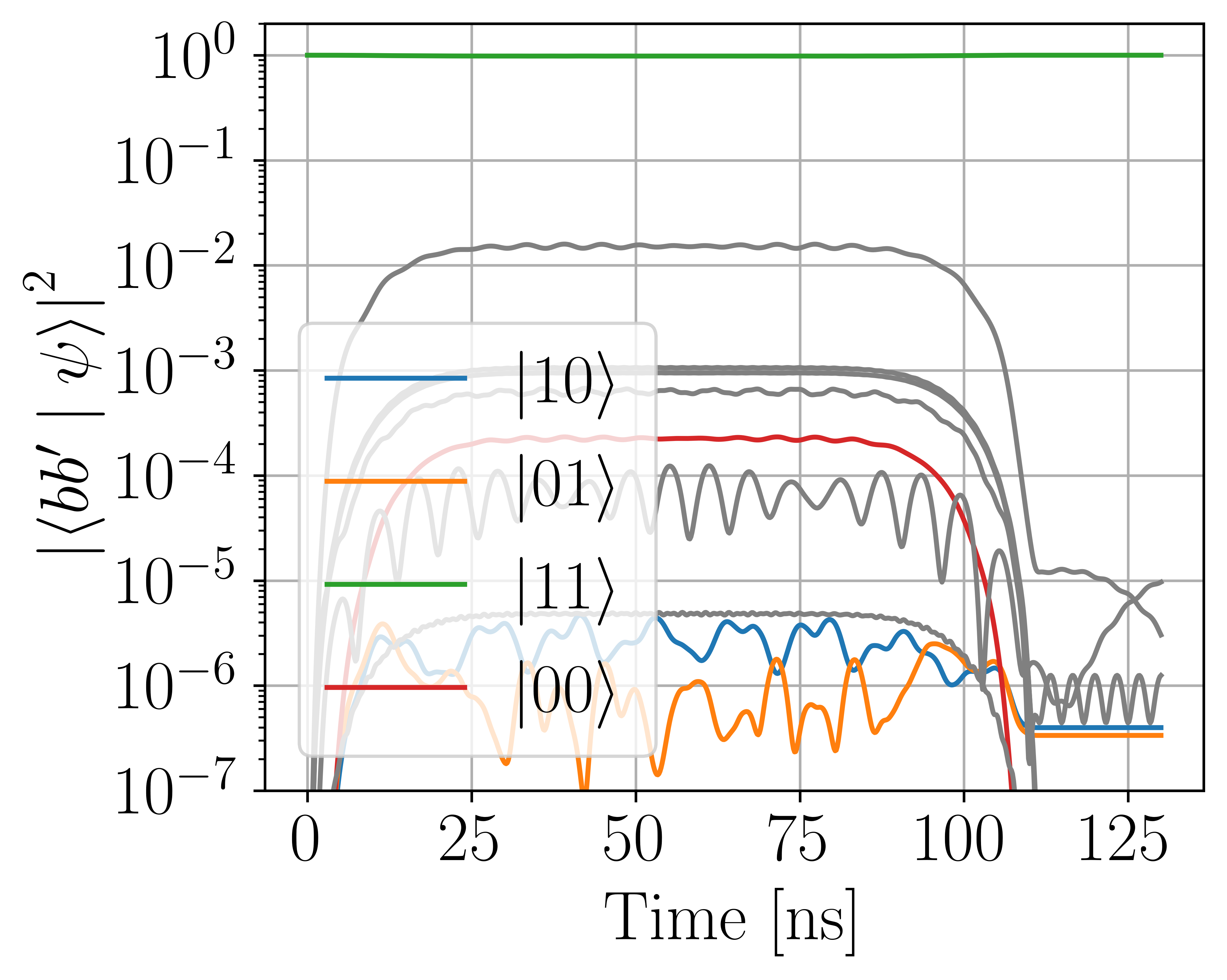}
\llap{\hskip 0mm \parbox[c]{9.5cm}{\vspace{-0.25cm}\small{(d)}}}
\end{tabular}
\caption{\label{fig:dispersive_composite}
Simulation of a $\sqrt{i{\rm SWAP}}$ gate between two dual-rail qubits with $\Omega_0/(2\pi)=80\unit{MHz}$, $\eta/(2\pi)=-250\unit{MHz}$, and $\Delta/(2\pi)=0.5\unit{GHz}$ (see Appendix~\ref{app:2qDR}).
(a) Energy levels of the computational basis states $\ket{bb'}$ for $b,b'\in\{0,1\}$.
(b) The coupling $g{\rm _c}(t) = g^{\rm max}_{\rm c} \left\{1-\left[1-\sin\left(\pi t/T_{\rm g}\right)\right]^{4}\right\}^{2}$ is turned on and off smoothly, reaching a maximum $g^{\rm max}_{\rm c}/(2\pi)=34\unit{MHz}$.
(c,d) Evolution of the system wavefunction $\ket{\psi(t)}$ during the gate, represented in an orthonormal basis with the computational basis states shown in color while other basis states are shown in gray.
In (c), the initial state of the system is $\ket{01}$, while in (d) it is $\ket{11}$ (similar behavior is obtained for $\ket{10}$ and $\ket{00}$). 
We can implement the ${\rm CX}$ gate by composing two $\sqrt{i{\rm SWAP}}$ gates with 1Q gates (see Appendix~\ref{app:2qDR}).
The infidelity of the $\sqrt{i{\rm SWAP}}$ gate (up to a spurious rotation about $\sigma_1^{\rm Z}\sigma_2^{\rm Z}$, which is cancelled in the implementation of the ${\rm CX}$ gate) is $5\times10^{-6}$.
The main contribution to the infidelity is leakage outside of the computational subspace, which in our scheme can be detected and converted to erasures.
}
\end{figure*}

The low-frequency part of the spectrum is shown in the upper panel of Fig.~\ref{fig:implementations}(a).
The computational basis states for our qubit are chosen to be the two single-excitation eigenstates, which for the resonant case, i.e., $\omega_1=\omega_2$, are given by
\begin{equation}\label{eq:drCompBase}
\ket b = \left[\ket{{\rm ge}}-(-1)^b \ket{{\rm eg}}\right]/\sqrt{2},\quad b=0,1,
\end{equation}
where $\ket{\rm g}$ and $\ket{\rm e}$ are, respectively, the ground and excited states of a single transmon.
Importantly, in this encoding amplitude damping errors will take the system outside of the qubit subspace and into the ground state of two coupled transmons, $\ket{\rm gg}$.
For any density matrix $\rho$ describing the dual-rail qubit we have
\begin{equation}
\label{eq_dualrail_amp}
\mathcal A^{\otimes 2}(\rho) = (1-\gamma)\rho + \gamma\ketbra{\rm gg}{\rm gg},
\end{equation}
where $\mathcal A$ is the amplitude damping channel that describes energy relaxation from $\ket{\rm e}$ to $\ket{\rm g}$ happening with probability $\gamma\in[0,1]$ and has Kraus operators
$K_0 = \ketbra{{\rm g}}{{\rm g}} + \sqrt{1-\gamma}\ketbra{{\rm e}}{{\rm e}}$ and $K_{1} = \sqrt\gamma \ketbra{{\rm g}}{{\rm e}}$,
assuming the same $T_1$ for both transmons~\footnote{$T_1$ may fluctuate and differ considerably for different transmons, leading to an extra term in Eq.~\eqref{eq_dualrail_amp} capturing dephasing at the rate $\Delta\Gamma = |\Gamma_1 - \Gamma_2|$, where $\Gamma_i$ is the amplitude damping rate for transmon $i = 1, 2$.
However, the strong coupling between the two transmons comprising the dual-rail qubit substantially reduces dephasing as long as $g_{\rm c} \gg \Gamma_1, \Gamma_2$.}.
Then, by measuring the population of the state $\ket{\rm gg}$ we can detect amplitude damping errors, and the effective noise affecting the dual-rail qubit is the heralded erasure noise~\cite{Duan2010}, as considered in Sec.~\ref{sec:SurfCode}.

Aside from the ability to detect amplitude damping errors,
the dual-rail qubit benefits from an inherent insensitivity to noise in the individual transmon frequencies.
In the presence of small fluctuations $\delta_1$ and $\delta_2$ in the transmon frequencies,
the splitting between the two qubit states becomes $\Omega = \sqrt{\Omega_0^2 + (\delta_1-\delta_2)^2}$, where $\Omega_0=2g_{12}$, and $\Omega$ is only second-order sensitive to the relative frequency fluctuation $\delta=\delta_1-\delta_2$~\cite{Shim2016semiconductor,campbell2020universal}.

We can go a step further and obtain the dephasing rate due to time-dependent fluctuations $\delta_1(t)$ and $\delta_2(t)$, considered here to be independent Gaussian processes. Then, $\delta(t)$ is a Gaussian process as well,
characterized by the power spectral density $S_{\delta}(\omega)=\int_{-\infty}^{\infty}\langle\delta(t)\delta(0)\rangle\exp(i\omega t) \d t$,
where $\langle\cdot\rangle$ stands for a statistical average.
We distinguish between the $t\ll\tau_{\rm c}$ and $t\gg\tau_{\rm c}$ regimes, where $\tau_{\rm c}$ is the characteristic correlation time over which $\langle\delta(t)\delta(0)\rangle$ decays.
The decoherence function, $W(t)=|\langle \exp[-i\phi(t)] \rangle|$, in each of these regimes obeys (see Appendix~\ref{app:dephas})
\begin{equation}\label{eq:dephas}
W(t) =
\begin{cases}
[1 + (\langle\delta^{2}\rangle t /\Omega_{0})^2]^{-1/4},  & t\ll\tau_{\rm c},\\
\exp[-S_{\delta}^{2}(0)t/(4\pi\Omega_{0}^{2}\tau_{c})], & t\gg\tau_{\rm c},
\end{cases}
\end{equation}
where $\phi(t)$ is the relative phase between the dual-rail qubit states, and $\left\langle \delta^{2}\right\rangle =\int_{-\infty}^{\infty}S_{\delta}(\omega)\d\omega/(2\pi)$ is the frequency mean squared deviation.
As can be seen, the coherence decays on a scale $T_\phi=2\Omega_0/\langle\delta^2\rangle$ for $t\ll\tau_{\rm c}$~\footnote{The numerical prefactor in the definition of $T_\phi$ was chosen such that the short-time expansion of the decoherence function is consistent with that of a single transmon, $\exp[-(t/T^{\rm trans}_\phi)^2] \simeq 1-(t/T^{\rm trans}_\phi)^2$.},
and $T_\phi=4\pi[\Omega_0/S_\delta(0)]^2\tau_c$ for $t\gg\tau_{\rm c}$.

These results can be compared with the dephasing time $T_{\phi}^{\rm trans}$ of the individual transmons composing the dual-rail qubit.
Since $\langle\delta^2\rangle=\langle\delta_1^2\rangle +\langle\delta_2^2\rangle$ and assuming, for simplicity, $\langle\delta_1^2\rangle = \langle\delta_2^2\rangle$,
we obtain (see Appendix~\ref{app:dephas}) $T_{\phi}/T_{\phi}^{\rm trans}=\Omega_{0}T_{\phi}^{\rm trans}/2$ for $t\ll\tau_{\rm c}$ and $T_{\phi}/T_{\phi}^{\rm trans}=2\pi\Omega^2_{0} T_\phi^\text{trans}\tau_{\rm c}$ for $t\gg\tau_{\rm c}$. 
As an example, consider a transmon with a dephasing time $T^{\rm trans}_\phi=10\unit{\mu s}$, which is due to a low-frequency noise ($\tau_{\rm c}\to\infty$).
For $\Omega_0=100\unit{MHz}$, the corresponding dephasing time of a dual-rail qubit comprising two such transmons will be $T_\phi=5\unit{ms}$.

To make use of erasure qubits one must be able to perform 1Q and 2Q gates in the encoded qubit subspace.
1Q gates can be implemented by controlling the transmon frequency difference $\delta(t)$, which we now consider as a tuning parameter. In the computational basis the dual-rail Hamiltonian reads $H_{\rm DR}=\frac{1}{2}\Omega_0\sigma^\text{Z}+\frac{1}{2}\delta(t)\sigma^\text{X}$. 
By making a fast excursion of $\delta(t)$ away from zero and back, one can realize rotations around $\sigma^\text{X}$~\cite{campbell2020universal}. Together with rotations around $\sigma^{\rm Z}$, which are realized by idling at $\delta(t)=0$, one can perform any rotation on the Bloch sphere.
Notice that during the time when $\delta(t)\neq0$ the immunity to frequency noise is lifted, however this time can be extremely short (potentially of the order of $2\unit{ns}$, limited only by control) and so dephasing during 1Q gates is not expected to be a limiting factor.

To perform 2Q gates between two dual-rail qubits, we couple together two transmons (e.g., capacitively)---one from each dual-rail qubit---with a coupling constant $g_{\rm c}$. Let the first pair be denoted by $j=1,2$ and the second pair by $j=3,4$.
Then, the coupling term reads 
\begin{equation}
\label{coupling_dual_rail}
H_{\rm c}=g_{\rm c} a_2^\dagger a_3+{\rm h.c.}  
\end{equation}
We consider the case where each pair is on resonance, but the pairs are detuned from each other, namely $\omega_1=\omega_2=\omega_3+\Delta=\omega_4+\Delta$. In the regime where $|\Delta|\gg|\eta|,g_{\rm c},\Omega_0$, the effective Hamiltonian is given by (see Appendix~\ref{app:2qDR})
\begin{equation}\label{eq:HeffDR}
H_{\rm eff} = \frac{\Omega}{2} (\sigma_1^\text{Z}+\sigma_2^\text{Z}) +g_{\rm XX}\sigma_1^\text{X}\sigma_2^\text{X} + \mathcal{O}(\Delta^{-3}),
\end{equation}
with $\Omega=\Omega_0(1+6g_{\rm c}^2 /\Delta^2)$ and $g_{\rm XX}=4g_{\rm c}^2\eta/(\Delta^2-\eta^2)$.

Turning the interaction $g_{\rm c}$ on and off adiabatically (e.g., using a tunable coupler) realizes a $\sqrt{i{\rm SWAP}}$ gate after an effective interaction time $T_{\rm g} = \pi/(4g_{\rm XX})$.
Together with 1Q gates, two $\sqrt{i{\rm SWAP}}$ gates can then be used to implement a ${\rm CX}$ gate (see Appendix~\ref{app:2qDR}), which allows for the surface-code syndrome extraction. 
The effective Hamiltonian of Eq.~\eqref{eq:HeffDR} does not account for leakage outside of the computational subspace, which can occur due to non-adiabatic transitions.
Considering an interaction $g_{\rm c}(t)$, which is ramped up over a time $T_{\rm ramp}$ to a value $g_{\rm c}^{\rm max}$, the leakage probability is roughly given by $P_{\rm leak}\simeq 2[g^{\rm max}_{\rm c}/(\Delta^2T_{\rm ramp})]^2$ (see Appendix~\ref{app:2qGateLeak}).
Importantly, however, leakage is to one of the states $\ket{\rm eegg}$, $\ket{\rm gfgg}$, $\ket{\rm ggee}$, and $\ket{\rm gggg}$, which can be detected by measuring the population of the $\ket{\rm gg}$ state of each dual-rail qubit. The probability of leakage therefore contributes to the erasure probability $e$, and is expected to be negligible compared with the amplitude-damping contribution.
Fig.~\ref{fig:dispersive_composite} presents a numerical simulation of a $\sqrt{i{\rm SWAP}}$ gate between two dual-rail qubits, realized in a system of four transmons with $T_{\rm g}=110\unit{ns}$, $\Omega_0/(2\pi)=80\unit{MHz}$, $\eta/(2\pi)=-250\unit{MHz}$, and $\Delta/(2\pi)=0.5\unit{GHz}$. The coupling $g_{\rm c}$ is turned on and off over a time scale $T_{\rm ramp}$ of the order of $20\unit{ns}$, reaching a value of $g^{\rm max}_{\rm c}/(2\pi)=34\unit{MHz}$.

As an alternative realization of 2Q gates, one can use a non-adiabatic transition to a state $\ket{\rm gg}\otimes(\ket{gf}-\ket{fg})/\sqrt{2}$, which is outside of the dual-rail qubit subspace, similar to the gate demonstrated by Campbell et al.~\cite{campbell2020universal}.
By bringing this state into resonance with $\ket{11}$, one realizes a ${\rm CZ}$ gate after an interaction time
$T_{\rm CZ}=100\unit{ns}$.
While the protection against dephasing is absent outside of the qubit subspace, such a gate could be advantageous if it can be performed fast compared to the unprotected dephasing time (which would be comparable with $T_\phi^\text{trans}$).
In this scheme, the need to avoid leakage into the $\ket{\rm gg}\otimes(\ket{gf}+\ket{fg})/\sqrt{2}$ state limits how fast the gate can be performed.

\begin{figure}[t!]
\includegraphics[trim=0cm 0cm 0cm 0cm, clip, scale=0.4]{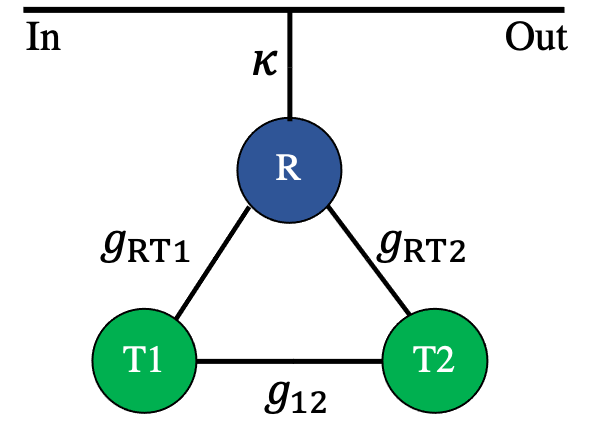}
\caption{
Detection of amplitude damping errors in the dual-rail qubit comprising two transmons (T1, T2).
The population of the $\ket{\rm gg}$ state is measured by a readout cavity (R) coupled to  T1 and T2.
The dephasing of the dual-rail qubit during the measurement is minimal when the couplings are the same, i.e., $g_{\rm RT1}=g_{\rm RT2}$ [see Eqs.~(\ref{eq:chiDRa},\ref{eq:chiDRb}) and the discussion following them].}
\label{fig:MeasGates} 
\end{figure}

We now analyze in detail the procedure for detecting amplitude damping into the $\ket{\rm gg}$ level.
We imagine doing that by performing a dispersive measurement that can distinguish between the $\ket{\rm gg}$ state and the computational subspace of the qubit.
It is important, however, that the measurement would not distinguish $\ket{0}$ from $\ket{1}$ as this would result in decoherence of the qubit.
To this end, we consider a readout cavity simultaneously coupled to both transmons of the dual-rail qubit, as shown in Fig.~\ref{fig:MeasGates}. By adiabatically eliminating the readout cavity (Appendix~\ref{app:DispersCoupling}), we find that it contributes the term
\begin{equation}\label{eq:Hchi}
    \begin{split}
        H_{\chi} &=\left[
        \chi_0\ketbra 0 0 +
        \chi_1\ketbra 1 1
        + \chi' (\ketbra 1 0 + \ketbra 0 1)
        \right]\tilde{c}^\dagger \tilde{c},
    \end{split}
\end{equation}
where
\begin{subequations}\label{eq:chisDR}
\begin{align}
\label{eq:chiDRa}
\begin{split}
\chi_{b} &= \frac{g_{\rm RT1}^2+g_{\rm RT2}^2}{\Delta^2}\eta
+\mathcal{O}(\Delta^{-3}),\quad b=0,1,
\end{split}
\\
\label{eq:chiDRb}
\chi' &= \frac{g_{\rm RT1}^2-g_{\rm RT2}^2}{\Delta^2}\eta + \mathcal{O}(\Delta^{-4}).
\end{align}
\end{subequations}
Here, $\tilde{c},\tilde{c}^\dagger$ are the ladder operators for the cavity mode dressed by its interaction with the transmons, $g_{\rm RT1}$, $g_{\rm RT2}$ are the cavity coupling constants to the two transmons, and $\Delta$ is its detuning relative to them.

The state-dependent shift to the cavity frequency allows us to distinguish the states in the qubit subspace from the state $\ket{\rm gg}$ by driving the cavity and measuring its quadratures in the usual way~\cite{Blais2004Cavity,Wallraff2004strong,Sete2015quantum}.
However, fluctuations in the cavity photon number, $n_{\rm c}=\tilde{c}^\dagger\tilde{c}$, can in principle decohere the qubit state by inducing fluctuations in the qubit frequency, $\Omega(n_{\rm c}) = \sqrt{ [\Omega_0+(\chi_1-\chi_0) n_{\rm c}]^2 + \chi'^2n_{\rm c}^2 }$~\footnote{We assumed here that the fluctuations in $n_{\rm c}=\tilde{c}^\dagger \tilde{c}$ happen on a long enough time scale such that we can diagonalize the qubit Hamiltonian in the adiabatic basis.}.
Since $\chi_0=\chi_1+\mathcal{O}(\Delta^{-3})$, the leading-order dependence of $\Omega(n_{\rm c})$ on $n_{\rm c}$ is determined by $\chi'$. 
For this reason, we propose designing the cavity coupling constants to obey the condition $g_{\rm RT1}=g_{\rm RT2}$, for which $\chi'=0$, rendering $\Omega(n_{\rm c})$ independent of $n_{\rm c}$ (see Appendix~\ref{app:DispersCouplingSymm} for the leading correction). This is a special case of creating a decoherence-free subspace~\cite{PhysRevLett.79.1953,PhysRevLett.81.2594}, in which both qubit states experience exactly the same noise.

We can now estimate the pure dephasing rate of a state 
$(\ket{0}+e^{i\phi}\ket{1})/\sqrt{2}$
during a dispersive measurement of the $\ket{\rm gg}$ population. In this process, the number of coherent photons in the cavity (induced by the readout tone) fluctuates, thereby leading to fluctuations in $\Omega(n_{\rm c})$ which decohere the qubit state. Given a cavity damping rate $\kappa$, the long-time behavior of the decoherence function for $t\gg \kappa^{-1}$ obeys $W_{\rm meas}(t) = \exp\left(-\Gamma^{\rm meas}_\phi t\right)$, with the decay rate given by~\cite{Blais2004Cavity}
\begin{equation}\label{eq:GammaMeas}
\Gamma^{\rm meas}_\phi = \frac{2}{\kappa}\left(\frac{\partial\Omega(\bar{n}_{\rm c})}{\partial n_{\rm c}}\right)^2 \bar{n}_{\rm c},
\end{equation}
accurate in the limit $|\partial\Omega(\bar{n}_{\rm c})/\partial n_{\rm c}|\ll\kappa$ and assuming, for concreteness, that the readout tone is applied at the resonance frequency of the cavity. 
As a realistic example, consider $g_{\rm RT1}$ and $g_{\rm RT2}$ that differ by $20\%$ and have an average of $g_{\rm RT}/(2\pi)=70\unit{MHz}$, $\Delta/(2\pi)=3\unit{GHz}$, $\eta/(2\pi)=-250\unit{MHz}$, $\Omega_0/(2\pi)=100\unit{MHz}$, and an average photon number $n_{\rm c}=20$. This results in $(\partial\Omega/\partial n_{\rm c})/(2\pi) \simeq 1.3\unit{kHz}$. For a cavity damping rate $\kappa=10\unit{MHz}$, the resulting measurement-induced dephasing rate $\Gamma_\phi^{\rm meas}$ is of the order of $0.25\unit{kHz}$.

We are now in position to estimate the noise parameters $p$ and $e$ used in the surface-code simulations (see Fig.~\ref{fig_surface}). Assuming an amplitude damping rate of $T_1=100\unit{\mu s}$, the main contributions to $e$ come from decay during the 2Q gate and during the $\ket{\rm gg}$ population measurement that follows it (manifesting as the false positive error $q_+$ discussed in Sec.~\ref{sub:ImperfecDetecion}). Taking the 2Q gate time to be $T_{\rm g}\simeq2\times110\unit{ns}$ (see Fig.~\ref{fig:dispersive_composite}) and the measurement time to be $T_{\rm meas}\simeq400\unit{ns}$, one has $e=T_{\rm g}/T_1+T_{\rm meas}/T_1\simeq6\times 10^{-3}$~\footnote{We assume here that false-positive errors are dominated by amplitude decay events during the measurement, as opposed to errors in resolving the cavity quantum state.}.
The contribution from leakage during the 2Q gate is of the order of $10^{-6}$ (see Fig.~\ref{fig:dispersive_composite}) and can be neglected.

Contributions to the Pauli error rate $p$ come from dephasing errors, control errors, and false-negative errors of the $\ket{\rm gg}$ population measurement. Dephasing due to low-frequency noise ($\tau_{\rm c}\to\infty$) is expected to be negligible, both because of the extension of $T_\phi$ offered by the dual-rail construction as well as due to the quadratic time-dependence of the decoherence function at short times [see Eq.~\eqref{eq:dephas}]. In contrast, photon shot noise in the readout cavity has a correlation time $\tau_{\rm c}\simeq\kappa^{-1}$ which can be short compared with both the 2Q gate time and the measurement time (of the $\ket{\rm gg}$ population), giving rise to a linear contribution to $p$. In particular, dephasing during measurement is expected to dominate as photon shot noise increases with the number of cavity photons. Using the dephasing rate estimated in Eq.~\eqref{eq:GammaMeas}, $T_\phi^{\rm meas}=1/\Gamma_\phi^{\rm meas}\simeq4\unit{ms}$, we have $T_{\rm meas}/T_\phi^{\rm meas}\simeq10^{-4}$.
Estimating a false-negative error rate of $q_-\simeq 10^{-3}$ (presumably dominated by finite resolving power of the measurement), its contribution to $p$ is $eq_-\simeq6\times 10^{-5}$. Assuming good enough quantum coherent control, we end up with $p\simeq1.6\times10^{-4}$.

\subsection{The {\rm g-f} qubit}
\label{sub:gf}

We now describe an alternative realization of an erasure qubit involving a single transmon, which we refer to as the g-f qubit.
The first three levels of the transmons are denoted $\ket{\rm g}$, $\ket{\rm e}$, and $\ket{\rm f}$ [see Fig.~\ref{fig:implementations}(b)], and it is described by the Hamiltonian
\begin{equation}
H_0 = \omega\ketbra{{\rm e}}{{\rm e}} + (2\omega+\eta)\ketbra{{\rm f}}{{\rm f}},
\end{equation}
where $\omega$ and $\eta$ are the transmon frequency and anharmonicity, respectively.
As the name suggests, the g-f qubit is encoded in the subspace spanned by the $\ket{\rm g}$ and $\ket{\rm f}$ levels.
In order to achieve a large $T_\phi/T_1$ bias, we propose using dynamical decoupling in the form of spin-locking~\cite{Yan2013rotating}, by driving oscillations between the $\ket {{\rm g}}$ and $\ket{{\rm f}}$ states, i.e.,
\begin{equation}\label{eq:gfDrive}
H_{\rm R} = \frac{\Omega_0}{2} \exp[i(\omega_{\rm gf}t-\varphi)] \ketbra {{\rm g}}{{\rm f}} + {\rm h.c.},
\end{equation}
where $\omega_{\rm gf}=2\omega + \eta$ is transition frequency between $\ket{\rm g}$ and $\ket{\rm f}$.
The interaction $H_{\rm R}$ can be engineered, for example, using a two-photon transition by introducing an XY drive at frequency $\omega_{\rm gf}/2$ (see Appendix~\ref{spinlock_g_f} for details). Alternatively, $H_{\rm R}$ can be realized in a tunable transmon by applying high-frequency flux modulation, giving rise to a non-linear interaction that resonantly couple the $\ket{\rm g}$ and $\ket{\rm f}$ states.

Under the influence of $H_{\rm R}$, the states
\begin{equation}
\label{eq:gfCompBase}
\ket b = \left[\ket{{\rm g}}-(-1)^b \ket{{\rm f}}\right]/\sqrt{2},\quad b=0,1,
\end{equation}
become eigenstates of the Hamiltonian for $\varphi=0$, when written in the interaction picture with respect to $H_0$.
We choose the states specified in Eq.~\eqref{eq:gfCompBase} to be the computational basis states of the {\rm g-f} qubit. 
The Hamiltonian in this basis (for $\varphi=0$) is then given by $H=\tfrac{1}{2}\Omega_0\sigma^{\rm Z}$. Thanks to the spin-locking drive, the g-f qubit is first-order insensitive to noise in the transmon frequency since the latter enters the Hamiltonian as $\tfrac{1}{2}\delta(t)\sigma^{\rm X}$, following the same analysis of Sec.~\ref{sub:DR} for the dual-rail qubit~\footnote{While the g-f qubit frequency is insensitive to noise in the transmon frequency, it is sensitive to noise in the spin-locking drive amplitude, Eq.~\eqref{eq:gfDrive}. The latter can nevertheless be dealt with by applying concatenated decoupling schemes~\cite{cai2012robust,Genov2019mixed}, which involves pulsing the phase of the drive.}.

The benefit of using the $\ket{\rm g}$ and $\ket{\rm f}$ levels to encode the qubit comes from the fact that, similar to the dual-rail qubit, amplitude damping errors take the system out of the computational subspace. Indeed, one can verify that for any density matrix $\rho$ describing the {\rm g-f} qubit the amplitude-damping channel is effectively described by (Appendix~\ref{t1_decay_g_f})
\begin{equation}
\label{eq_gfnoise}
\mathcal N(\rho) = (1-\gamma)\rho + \gamma\ketbra{{\rm e}}{{\rm e}},
\end{equation}
where $\gamma\in[0,1]$ is the probability for an amplitude damping error.
Then, by measuring the population of the state $\ket{\rm e}$ we can detect amplitude damping errors and effectively convert them into erasure errors. 
Such a measurement can be implemented, for example, by mapping the $\ket{\rm e}$ state to an extra transmon (without affecting the $\ket{\rm g}$ and $\ket{\rm f}$ states) and then measuring the latter.

In order to detect any amplitude decay error, it is important to avoid coherent leakage from the computational subspace to other states. Otherwise, an amplitude decay event could bring the system into the computational subspace of the {\rm g-f} qubit and result in an undetectable error.
For example, in the case when $H_{\rm R}$ is generated by a two-photon-drive, the $\ket{\rm e}$ state is virtually populated with probability  $P_{\rm e}= 3\Omega_0/(\sqrt{8}\eta)$, where $\Omega_0\ll|\eta|$ (Appendix~\ref{spinlock_g_f}). The probability of having an undetectable amplitude decay error over a time $T$ is then $P_{\rm e}T/T_1$.

1Q gates can be implemented in a similar fashion to the dual-rail qubit scheme.
By considering the transmon frequency to be a controlled parameter, $\omega\to\omega+\delta(t)$, the Hamiltonian becomes $H=\tfrac{1}{2}\sigma^{\rm Z}+\tfrac{1}{2}\delta(t)\sigma^{\rm X}$.
As in the dual-rail case, by controlling $\delta(t)$ one can realize arbitrary 1Q rotations.
The same effect is achieved by introducing, instead, a shift to the drive frequency, $\omega_{\rm R}\to\omega_{\rm R}-\delta(t)$.

The g-f qubit, however, allows for another route towards implementing 1Q gates. By introducing a non-zero phase $\varphi$ for the drive amplitude in Eq.~\eqref{eq:gfDrive}, the Hamiltonian becomes $H =\tfrac{1}{2}\Omega_0(\cos\varphi\sigma^{\rm Z}-\sin\varphi\sigma^{\rm Y})$.
Then, changing $\varphi$ in time allows for realizing arbitrary rotations on the Bloch sphere. This approach has the advantage of keeping the transmon frequency on resonance with the drive, thereby benefiting from the immunity to frequency noise offered by the spin-locking effect.

To perform 2Q gates between a pair of g-f qubits, we consider two transmons, each driven by a Hamiltonian of the form of Eq.~(\ref{eq:gfDrive}). The transmons are coupled to each other (e.g., capacitively), $H_{\text{c}}=g_{\rm c}(a_1^{\dagger}a_2+{\rm h.c.})$, where $a_1$ and $a_2$ are their respective annihilation operators.
Let the transmons frequencies be denoted by $\omega_1$ and $\omega_2$.
Then, the effective interaction in the two-qubit computational basis is given in the regime of $g_c\ll\left|\omega_{1}-\omega_{2}\right|$ by (see Appendix \ref{ZZ_interaction})
\begin{equation}
\label{eq:coupling_gf_qubits}
H_{\rm gf}=\frac{\Omega_0}{2}\sigma^{\rm Z}_{1} +\frac{\Omega_0}{2}\sigma^{\rm Z}_{2}+g_{\rm XX}\sigma^{\rm X}_{1}\sigma^{\rm X}_{2}+\mathcal{O}(\Delta^{-3}),
\end{equation}
similar to the effecive 2Q dual-rail Hamiltonian, but where now $g_{\rm XX}=4g^{2}\eta/(\Delta^{2}-\eta^{2})$ and $\Delta=\omega_1-\omega_2$. As in the dual-rail case, this effective interaction can be used to realize a $\sqrt{i{\rm SWAP}}$ gate, two of which can be composed to implement either a ${\rm CX}$ or a ${\rm CZ}$ gate (see also Appendix~\ref{app:2qDR}).

There is an alternative scheme for 2Q gates that does not involve the spin-locking drive of Eq.~\eqref{eq:gfDrive}. 
By bringing the $\ket{{\rm gf}}$ state into resonance with $\ket{{\rm ee}}$ and waiting for a time $\pi/(\sqrt{2}g_{\rm c})$, one obtains a $\pi$ phase, realizing a ${\rm CX}$ gate (up to 1Q rotations).
Lacking the spin-locking effect, this gate scheme does not benefit from an extended dephasing time, however, it could potentially be faster as it relies on a resonant interaction.
The speed of the gate is limited by the need to suppress the off-resonant transitions from $\ket{{\rm ff}}$ to the $\ket{{ \rm he}}$ state (where $\ket{\rm h}$ denotes the fourth level of the transmon), which is detuned by $\left|\omega_{\rm ff}-\omega_{\rm he}\right|=\left|\eta_{1}-\eta_{2}\right|$, where $\eta_i$ is the transmon nonlinearity and $\omega_{xy}$ is the transition frequency from level $\ket{x}$ to level $\ket{y}$. 
Finally, even though this gate scheme takes the system outside of the computational subspace, a single amplitude decay error is still detectable as the latter takes the state $\ket{{\rm ee}}$ into either $\ket{{\rm ge}}$ or $\ket{{\rm eg}}$, both of which are outside the computational subspace and can be detected.

\section{Discussion}
\label{sec:discussion}

In our work, we provide a simple scheme to detect and convert amplitude damping errors into erasure errors.
With our approach, we can overcome the conventional $T_1$ limit on fidelity and significantly enhance the QEC protocols, as demonstrated by our numerical simulations of the surface code. 
Our analysis indicates that it may be sufficient to make $T_1$ two orders of magnitude longer than the gate times, as with erasure qubits we can tolerate erasure rate up to nearly 5\% [see Fig.~\ref{fig_threshold}(a)].

We suspect that the QEC performance can be further improved by using adaptive schemes, which, for instance, abort any syndrome extraction circuit that suffers from erasure errors.
We defer the analysis of adaptive schemes to future work.

We emphasize that our approach can be extended to settings with different types of noise bias and hardware. 
In the context of atom or ions, for example, our results encourage the use of excited states as qubits.
More specifically, the $D$ excited state manifold could be used as a qubit in the ion trap settings while the decay to the $S$ state could be efficiently detected in ions such as $^{40}\mathrm{Ca}^+,$ $^{171}\mathrm{Yb}^+,$ $^{138}\mathrm{Ba}^+$ and $^{88}\mathrm{Sr}^+$ or for Rydberg atoms as suggested by Wu et al.~\cite{Wu2022erasure}, where an idea of converting the dominant physical noise into heralded erasures was explored.
This, in turn, implies that the restriction of long lifetime of excited state should not be a major limitation when choosing a qubit construction.

The approach we take in this work can be viewed as an example of a more general design principle---\emph{given the knowledge of the hardware and its dominant noise, engineer qudits in such a way that the effective noise is easier to correct by QEC protocols}.
In other words, we seek two QEC codes and concatenate them so that:
(i) the inner code defining qudits is a simple code operating on the hardware level that is tailored to the physical noise,
(ii) the outer code efficiently corrects the effective noise on the engineered qudits and can be implemented with quantum operations that preserve the effective noise structure.
This design principle is exemplified by QEC protocols with bosonic qubits~\cite{guillaud2019repetition,Chamberland2022building} or by Wu et al.~\cite{Wu2022erasure}.

For the amplitude damping noise, one may hope to improve on the dual-rail and g-f qubits by using the four-qubit code as the inner code, which, in principle, allows to correct a single amplitude damping error~\footnote{Concatenation of the four-qubit code and the surface code has already been considered in the context of the standard circuit noise with no amplitude damping errors in Ref.~\cite{Criger2016}.}.
Unfortunately, the need to incorporate dynamical decoupling would most likely only allow for error detection but not correction, and a large $T_\phi/T_1$ bias would be difficult to achieve (see Appendix~\ref{sec_dynamical}).
Thus, the qubit based on the four-qubit code would not offer any advantage over the substantially simpler dual-rail and g-f qubits that we propose (see Sec.~\ref{sec:implement})

Lastly, we point out that our scheme is well suited to deal with leakage errors, as they can be detected and converted to erasures in a similar way as amplitude damping errors.
In the dual-rail qubit, specifically, we observed that leakage during 2Q gates is, in fact, detected by the measurement that monitors for amplitude decay.
However, different qubit encodings and different gate realizations might require more judicious measurement protocols for detection of leakage errors.

\section*{Acknowledgements}

We acknowledge fruitful discussions with N. Alidoust, C. Chamberland, S. T. Flammia, A. L. Grimsmo, H. Levine and G. Refael.

\appendix
\onecolumngrid

\section{Details of surface code simulations}
\label{sec_numerics}

\begin{figure*}[ht]
\begin{tabular}{lcr}
(a)\includegraphics[height=0.16\textheight]{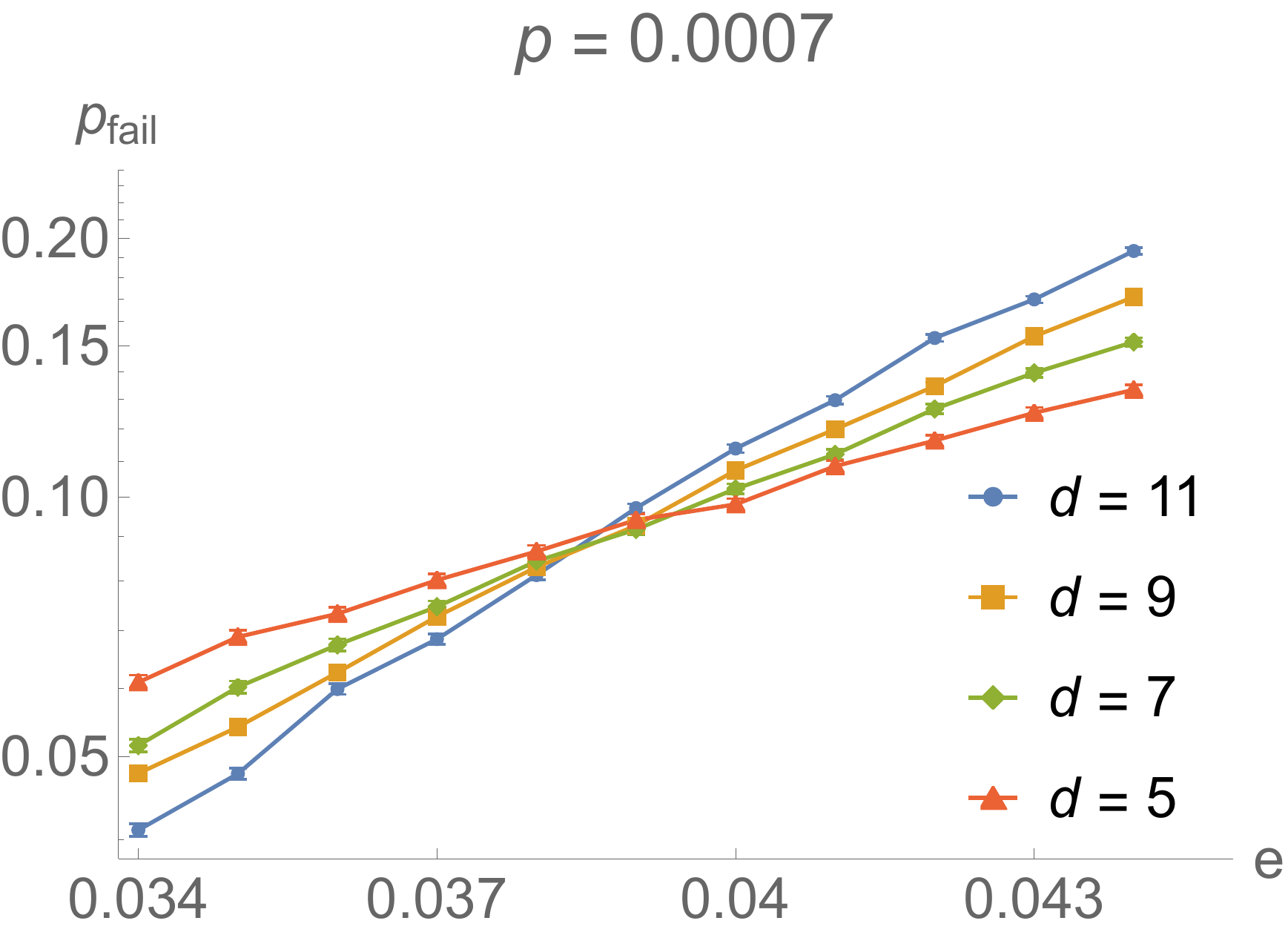}&
(b)\ \includegraphics[height=0.16\textheight]{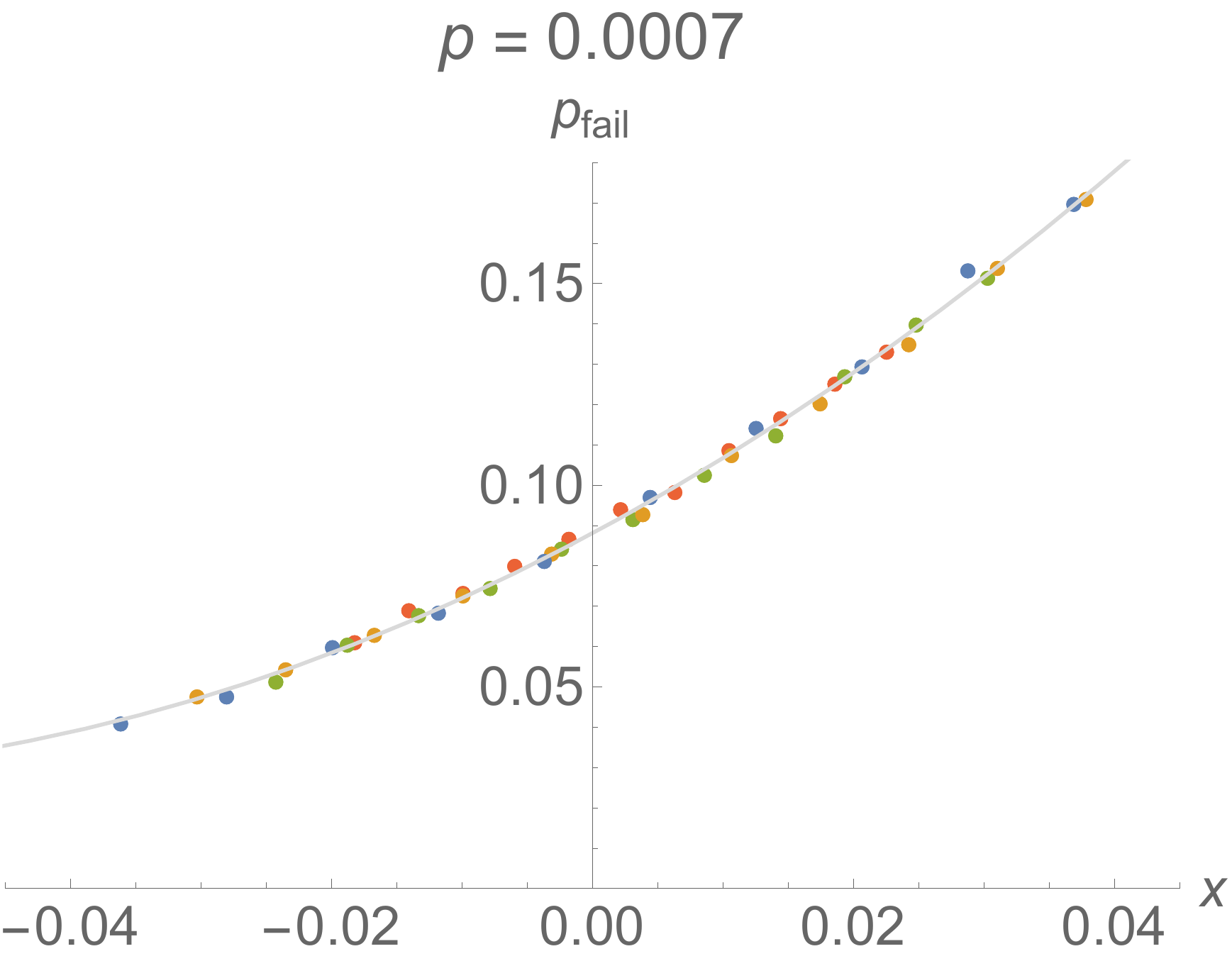}&
(c)\includegraphics[height=0.16\textheight]{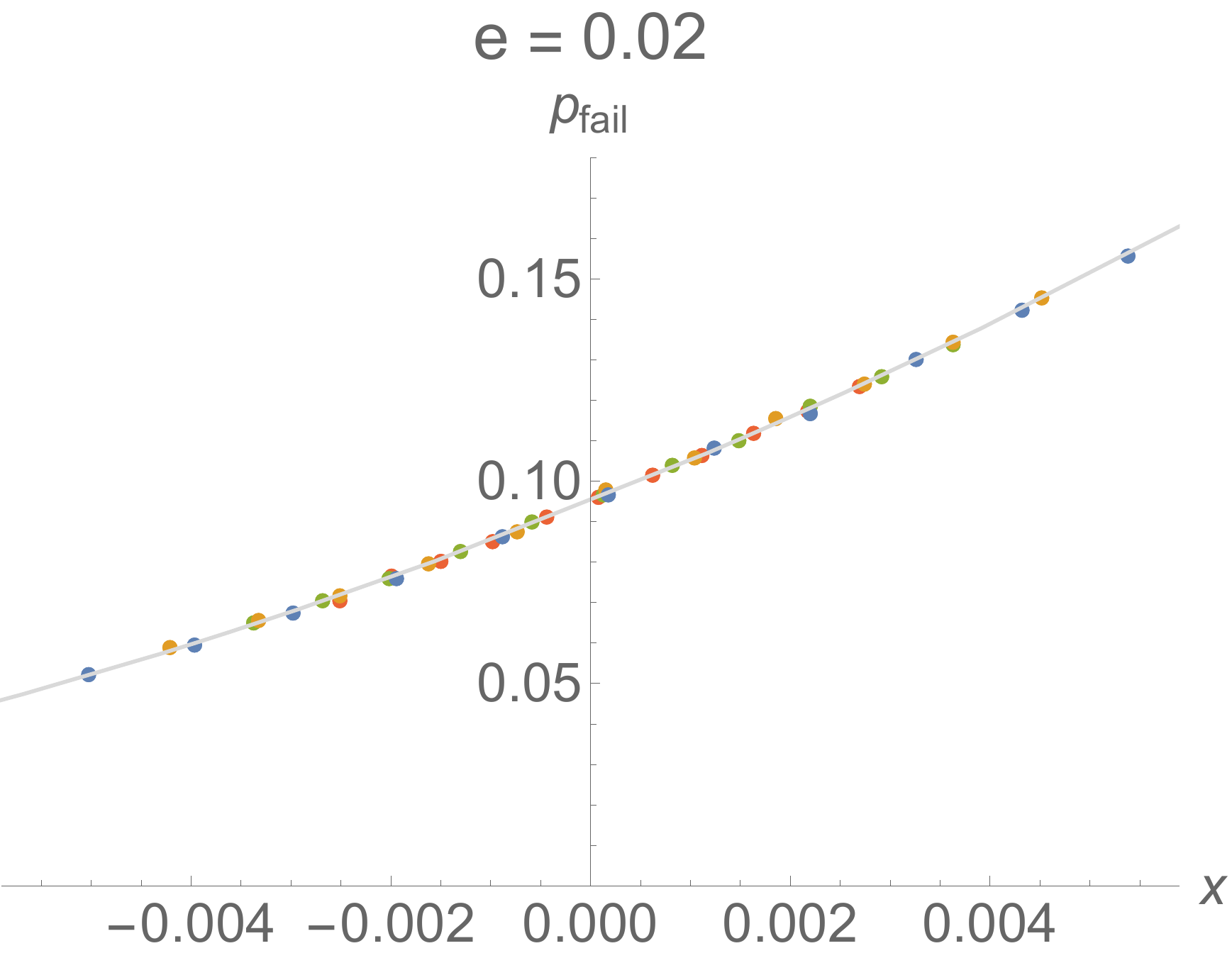}
\end{tabular}
\caption{
(a) The logical error rate $p_\text{fail}$ of the erasure scheme for $p = 0.0007$ and different code distance $d$ as a function of $e$.
In (b), we show the same data using the rescaled variable $x = [e-e_\text{th}(p)]d^\mu$, where the fitting parameters are $e_\text{th}(p) = 0.0386(2)$ and $\mu = 0.98(13)$.
(c) The rescaled data for the inset in Fig.~\ref{fig_threshold}(a) using the rescaled variable $x = [p-p_\text{th}(e)]d^\mu$, where $e=0.02$ and the fitting parameters are $p_\text{th}(e) = 0.00331(2)$ and $\mu = 0.87(3)$.
}
\label{fig_extranum}
\end{figure*}

To estimate the logical error rate of the surface code of distance $d$, where $d$ is an odd integer, for the heralded erasure and Pauli noise with parameters $p$, $p_{\rm M}$ and $e$, we simulate $d$ rounds of noisy syndrome extraction, followed by one noiseless round.
In the erasure scheme, for every code distance $d$ and erasure rate $e$ we first sample $10^4$ realizations of space-time locations where CNOTs are followed by a fully-depolarizing two-qubit Pauli channel.
Then, for every Pauli and measurement error rates $p$ and $p_{\rm M}$ we use each of these realizations $n_\text{rep} = 100,100,100,50,25$ times for $d=3,5,7,9,11$, respectively, to sample errors, run the MWPM decoder and estimate the logical error rate for the given realization.
Finally, the logical error rate $p_\text{fail}(p,p_{\rm M},e,d)$ is calculated as the average over the logical error rate for different realizations.
In the standard scheme, we perform standard circuit-noise simulations, where
(i) state preparation and idling are followed by a single-qubit Pauli channel with error rate $p$,
(ii) each CNOT gate is followed by a two-qubit Pauli channel with error rate $p + 15e/16 - e p$,
(iii) each Pauli measurement returns a wrong outcome with probability $p_{\rm M}$.
We then sample errors $2\times 10^6$ times, run the MWPM decoder and estimate the logical error rate $p_\text{fail}(p,p_{\rm M},e,d)$.

To find the boundary of the correctable region for the erasure and standard protocols in Fig.~\ref{fig_threshold}(a), we perform either vertical or horizontal sweeps in the $(e,p)$ phase space to find the logical error rate $p_\text{fail}$ as a function of, respectively, either $p$ or $e$ for fixed $e$ or $p$; see Fig.~\ref{fig_extranum}(a) and the inset of Fig.~\ref{fig_threshold}(a).
Note that we have only two noise parameters $e$ and $p$ as we set $p_{\rm M} = 2p/3$.
We estimate the threshold $p_\text{th}(e)$ from the data by fitting the following ansatz
\begin{equation}
p_\text{fail}(x) = a x^2 + b x + c,\quad\quad x = [p-p_\text{th}(e)]d^\mu,
\end{equation} 
where $a$, $b$, $c$, $p_\text{th}(e)$ and $\mu$ are the fitting parameters; we similarly estimate the threshold $e_\text{th}(p)$.
We illustrate this procedure in Fig.~\ref{fig_extranum}(b)(c).

\section{Dephasing in the dual-rail qubit}
\label{app:dephas}

In this appendix we derive the decoherence function for the dual-rail qubit given in Eq.~\eqref{eq:dephas}, and provide details on the comparison with the case of a single transmon.
We consider the Hamiltonian $H_{\rm DR}=\frac{1}{2}\Omega_0\sigma^{\rm Z}+\frac{1}{2}\delta(t)\sigma^{\rm X}$, where $\delta(t)=\delta_1(t)-\delta_2(t)$, and $\delta_{j}(t)$ describes the noise in the frequency of transmon $j=1,2$.
The transition rate between the qubit states due to $\delta(t)$ is given by the noise spectral density at the bare qubit frequency, $|S_{\delta}(\omega=\Omega_0)|$. We assume this rate to be small compared with all relevant energy scales such that we can diagonalize the Hamiltonian in the adiabatic basis, obtaining the instantaneous qubit frequency, $\Omega(t)=\sqrt{\Omega_{0}{}^{2}+\delta(t)^{2}}\simeq\Omega_{0}+\delta^{2}(t)/(2\Omega_{0})$.
The dephasing of the qubit is then described by the decoherence function,
\begin{equation}
W(t) = \left|\left\langle \exp[-i\phi(t)]\right\rangle\right|,\quad
\phi(t)=\int_0^t {\rm d}t' \Omega(t').
\end{equation}

We analyze $W(t)$ separately in the regimes $t\ll\tau_{\rm c}$ and $t\gg\tau_{\rm c}$, where $\tau_{\rm c}$ is the correlation time for the noise process $\delta(t)$, over which $\langle\delta(0)\delta(t)\rangle$ decays to zero.
In the regime $t\ll\tau_{c}$, $\delta(t)$ can be treated as a constant, i.e., $\delta(t)=\delta$.
The decoherence function then reads
$W(t) = |\langle \exp\left[-i\delta^{2}t/(2\Omega_{0})\right]\rangle|$.
Since $\delta$ is by assumption normally distributed, $\delta^{2}$ is described by the chi-squared distribution with one degree of freedom.
We can then obtain $W(t)$ directly from the characteristic function of the chi-squared distribution, i.e.,
\begin{equation}
W(t)=\left|(1+i\langle\delta^{2}\rangle\Omega^{-1}_{0}t)^{-1/2}\right|=\left[1+\left(\langle\delta^{2}\rangle\Omega^{-1}_{0}t\right)^{2}\right]^{-1/4},
\end{equation}
as given in Eq.~\eqref{eq:dephas} of the main text. We define the dephasing time from the short-time behavior of the decoherence function via $W(t)\simeq 1 - (t/T_\phi)^2$, yielding $T_\phi=2\Omega_0/\langle\delta^2\rangle$.

For comparison, we apply the same treatment to the case of a single transmon, whose decoherence function is analogously given by $W_{\rm trans}(t)=\left|\left\langle\exp\left[-i\int_0^t{\rm d}t'\delta_j(t')\right]\right\rangle\right|$.
In the regime of $t\ll\tau_{c}$, $\delta_j(t)$ can be treated as a constant, i.e., $\delta_j(t) = \delta_j$ and we have
$W_{\rm trans}(t)=|\langle \exp[-i\delta_jt] \rangle|=\exp(-\langle\delta_j^2\rangle t^2/2)$,
where we use the assumption that $\delta_j$ is normally distributed.
As in the dual-rail case, the dephasing time is extracted from $W_{\rm trans}(t) \simeq 1 - (t/T^{\rm trans}_\phi)^2$, yielding $T_\phi^{\rm trans}=\sqrt{2/\langle\delta_j^2\rangle}$.

We now move on to analyze the dephasing of the dual-rail qubit in the other regime, $t\gg\tau_{c}$.
In this regime, $\phi(t)=\int_{0}^{t}\d t'\Omega(t')$ can be considered as a sum of many independent random contributions, and is therefore normally
distributed as a result of the central limit theorem~\cite{Bergli2009decoherence}.
We can then adopt the result of Refs.~\cite{Bergli2009decoherence,Omalley2016superconducting} $W(t)=\exp\left[-t/T_\phi\right]$, where $T_\phi=2/S_{\Omega}(0)$, except that $S_{\Omega}(\omega)$ is now the power spectral density of the noise in $\Omega(t)$, i.e.,
\begin{equation}
S_{\Omega}(\omega)=\int_{-\infty}^{\infty}\d t\left(\left\langle \Omega(0)\Omega(t)\right\rangle -\left\langle \Omega\right\rangle ^{2}\right)\exp(i\omega t).
\end{equation}
Importantly, since $\delta(t)$ is normally distributed, $S_{\Omega}(\omega)$ can be related to $S_{\delta}(\omega)$ using Wick's probability theorem,
\begin{equation}
\left\langle \Omega(0)\Omega(t)\right\rangle -\left\langle \Omega\right\rangle ^{2}
=\frac{1}{2\Omega_{0}^{2}}\text{\ensuremath{\left\langle \delta(0)\delta(t)\right\rangle ^{2}}},
\end{equation}
resulting in $S_{\Omega}(0) = (2\Omega_{0}^{2})^{-1}\int |S_{\delta}(\omega)|^{2} \d\omega/(2\pi)$.
This expression can be roughly estimated if we consider that $S_{\delta}(\omega)$ decays over a time scale of $\tau^{-1}_{c}$, and therefore $S_{\Omega}(0)$ is of the order of $S_{\delta}^{2}(0)/\left(4\pi\Omega_{0}^{2}\tau_{c}\right)$, as given in Eq.~\eqref{eq:dephas}.
As before, we can compare the dual-rail result with that of a single transmon, which in this regime is given by $W_{\rm trans}(t)=\exp(-t/T_\phi^{\rm trans})$ with $T_\phi^{\rm trans}=2/S_{\delta_j}(0)$~\cite{Bergli2009decoherence,Omalley2016superconducting}.

\section{2Q gates for the dual-rail qubit}
\label{app:2qDR}

In this appendix we derive the effective Hamiltonian in Eq.~\eqref{eq:HeffDR} that describes the interaction between two dual-rail qubits, and explain how it can be used to implement an entangling 2Q gate. 
We begin by considering the Hamiltonian of the two dual-rail qubits, $H=H_0+H_{\rm c}$, where
\begin{subequations}\label{eq:SimulatedH}
\begin{align}
    &H_0=\sum_{i=1}^4\left(\omega_{i}a_{i}^{\dagger}a_{i}+\frac{\eta}{2}a_{i}^{\dagger}a_{i}^{\dagger}a_{i}a_{i}\right)+\left(g_{12} a_{1}^{\dagger}a_{2}+g_{34}a_{3}^{\dagger}a_{4}+\text{h.c.}\right),
    \\
    &H_{\rm c} = g_{\rm c} a_2 a_3^\dagger +\text{h.c.},
\end{align}
\end{subequations}
the transmon frequencies $\omega_i$ are chosen to obey $\omega_1=\omega_2=\omega_3+\Delta=\omega_4+\Delta$, $\eta$ is the transmon nonlinearity, and $g_{12}=g_{34}=\Omega_0/2$ are the internal couplings within each dual-rail qubit (which, for simplicity, are taken to be the same).
The four computational states defined as follows
\begin{equation}
\ket {bc} = \frac{1}{2}
\left[\ket{{\rm ge}}-(-1)^b \ket{{\rm eg}}\right]
\otimes
\left[\ket{{\rm ge}}-(-1)^c \ket{{\rm eg}}\right],\quad
b,c\in\{0,1\},
\end{equation}
are eigenstates of $H_0$.

Assuming $g_{\rm c}\ll\Delta$, we can treat $H_{\rm c}$ as a perturbation, and obtain an effective Hamiltonian by using the Magnus expansion to second order in $H_{\rm c}$, i.e.,
\begin{equation}
H_{\rm eff} = \frac{1}{2it}\int_0^t {\rm d}t_1 \int_0^{t_1} {\rm d}t_2 [H_{\rm I}(t_1),H_{\rm I}(t_2)],
\end{equation}
where $[A,B] = AB - BA$ denotes the commutator, $H_{\rm I}(t)$ is the Hamiltonian in the interaction picture with respect to $H_0$, and then keeping only terms which do not decay with time. Upon projecting onto the computational subspace and going back to the lab frame, we obtain 
\begin{equation}
\label{eq:HeffFull}
H_{\rm eff} = \frac{1}{2}\Omega\left(\sigma_{1}^{\rm Z}+\sigma_{1}^{\rm Z}\right)+g_{\rm XX}\sigma_{1}^{\rm X}\sigma_{2}^{\rm X} + h^{(1)}_{{\rm X}}\sigma_{1}^{\rm X}+h^{(2)}_{{\rm X}}\sigma_{2}^{\rm X}
+\mathcal{O}(\Delta^{-3}),
\end{equation}
where $\Omega=\Omega_0(1+6g_{\rm c}^2 /\Delta^2)$, $g_{\rm XX} = {\eta g_{\rm c}^2}/{\Delta^2}$, and $h^{(j)}_{\rm X} = - g_c^2/(2\Delta)[1+(-1)^j (\Omega_0/2-\eta)/\Delta]$. 
Finally, notice that one can cancel out the $h_{\rm X}^{(1)}$ and $h_{\rm X}^{(2)}$ terms in Eq.~\eqref{eq:HeffFull} by applying small corrective shifts
$\omega_2\to\omega_2+h_{\rm X}^{(1)}$ and $\omega_3\to\omega_3-h_{\rm X}^{(2)}$ (or alternatively $\omega_1\to\omega_1-h_{\rm X}^{(1)}$ and $\omega_4\to\omega_4+h_{\rm X}^{(2)}$). Doing so results in the effective Hamiltonian in Eq.~\eqref{eq:HeffDR}.

We now describe the implementation of the $\sqrt{i{\rm SWAP}}$ gate, which we can use to realize 2Q gates needed for the syndrome extraction.
We start by considering the effective Hamiltonian of Eq.~\eqref{eq:HeffDR} with a time-dependent coupling $g_{\rm XX}(t)$ [as a result, for example, of a tunable coupler that controls $g_{\rm c}(t)$] that is smoothly turned on and off at $t=0$ and $t=T_{\rm g}$, respectively.
The evolution of the states $\ket{01}$ and $\ket{10}$ from $t=0$ to $t=T_{\rm g}$ can be solved exactly, yielding
\begin{equation}\label{eq:iSWAPodd}
\ket{01} \mapsto \cos\theta \ket{01} -i\sin\theta\ket{10},\quad
\ket{10} \mapsto \cos\theta \ket{10} -i\sin\theta\ket{01},
\end{equation}
where $\theta=\int_0^{T_{\rm g}}g_{\rm XX}(t)\d t$. Furthermore, if $g_{\rm XX}(t)$ is varied slowly enough, the adiabatic theorem dictates the evolution of the states $\ket{00}$ and $\ket{11}$,
\begin{equation}
\label{eq:iSWAPeven}
\ket{00} \mapsto e^{i\varphi} \ket{00},\quad
\ket{11} \mapsto e^{-i\varphi} \ket{11},
\end{equation}
where $\varphi=\int_0^{T_{\rm g}}\sqrt{\Omega^2+g_{\rm XX}(t)}\d t$. If we choose $T_{\rm g}$ such that $\theta=\pi/4$, then the transformation of Eqs.~(\ref{eq:iSWAPodd},\ref{eq:iSWAPeven}) describes by the gate 
$\sqrt{i{\rm SWAP}}\exp[-i\varphi(\sigma_1^{\rm Z}+\sigma_2^{\rm Z})]$.
Note that we can easily cancel a spurious rotation about $\sigma_1^{\rm Z}+\sigma_2^{\rm Z}$ with 1Q gates by, e.g., idling for an appropriate amount of time.

By composing two $\sqrt{i{\rm SWAP}}$ gates and a single 1Q gate we can realize the 2Q gate $\exp(-i\pi\sigma_{1}^{\rm X}\sigma_{2}^{\rm X}/4)$ (that, together with 1Q gates, is sufficient to implement the syndrome extraction circuits), namely
\begin{equation}
\exp(-i\pi\sigma_{1}^{\rm X}\sigma_{2}^{\rm X}/4)
= \sqrt{i{\rm SWAP}}\sigma_1^{\rm X}\sqrt{i{\rm SWAP}}.
\end{equation}
Note that this decomposition is invariant under $\sqrt{i{\rm SWAP}}\to\sqrt{i{\rm SWAP}}\exp(-i\delta_{\rm ZZ}\sigma_{1}^{\rm Z}\sigma_{2}^{\rm Z})$, namely spurious rotations about $\sigma_{1}^{\rm Z}\sigma_{2}^{\rm Z}$ do not affect the infidelity of the $\exp(-i\pi\sigma_{1}^{\rm X}\sigma_{2}^{\rm X}/4)$ gate.
Such a small undesired rotation in the realization of the $\sqrt{i{\rm SWAP}}$ gate can arise as a result of higher-ordered terms not specified in Eqs.~\eqref{eq:H2Qgf} and \eqref{eq:HeffFull}.

In Fig.~\ref{fig:dispersive_composite} of the main text, we present the result of simulating the evolution of the Hamiltonian in Eq.~\eqref{eq:SimulatedH}, where we take $g{\rm _c}(t) = g^{\rm max}_{\rm c} \left\{1-\left[1-\sin\left(\pi t/T_{\rm g}\right)\right]^{4}\right\}^{2}$.
Furthermore, to cancel the $h_{\rm X}^{(j)}$ terms for all times $t$, we apply time-dependent shifts to $\omega_2$ and $\omega_3$ as explained above.
To evaluate the infidelity of the simulated gate, we define $P$ to be a projector onto the computational subspace and $U$ to be the simulated unitary evolution under the Hamiltonian in Eq.~\eqref{eq:SimulatedH}.
Since we are interested in evaluating the fidelity of the $\sqrt{i{\rm SWAP}}$ gate only up to spurious rotations about $\sigma_j^{\rm Z}$ and $\sigma_1^{\rm Z}\sigma_2^{\rm Z}$, we choose to define the infidelity of our gate as follows
\begin{equation}
    {\rm IF}=\min_{\left\{ \boldsymbol{\delta},\boldsymbol{\delta}',\delta_{{\rm ZZ}}\right\} }\left\{ 1-\frac{1}{16}\left|{\rm Tr}\left[PU^\dagger PU_{{\rm Z}}\left(\boldsymbol{\delta}\right)\sqrt{i{\rm SWAP}}U_{{\rm Z}}\left(\boldsymbol{\delta}'\right)\exp\left(-i\delta_{{\rm ZZ}}\sigma_{1}^{{\rm Z}}\sigma_{2}^{{\rm Z}}\right)\right]\right|^{2}\right\},
\end{equation}
where $U_{{\rm Z}}\left(\boldsymbol{\delta}\right)=\exp\left[-i\left(\delta_{1}\sigma_{1}^{{\rm Z}}+\delta_{2}\sigma_{2}^{{\rm Z}}\right)\right]$.
For the parameters used in our simulations this yields ${\rm IF}=5 \times 10^{-6}$.

\section{Leakage in the dual-rail two-qubit gate}
\label{app:2qGateLeak}

In this appendix we estimate the probability of leakage during a gate between two dual-rail qubits (see Sec.~\ref{sub:DR}). This leakage occurs during ramping up and down the interaction term $H_{\rm c}=g_{\rm c}(t)(a_2^\dagger a_3+{\rm h.c.})$ as a result of non-adiabatic transitions from the computational subspace into states outside of it. 
Applying $H_{{\rm c}}$ to the computational basis states, $\ket{00}$, $\ket{01}$, $\ket{10}$, and $\ket{11}$, we see that each of them is coupled equally to the four states $\ket{\rm ggee}$, $\ket{\rm eegg}$, $\ket{\rm gfgg}$, $\ket{\rm ggfg}$, with coupling strength $g_{\rm c}/2$.
These states are detuned from the computational states by $\Delta+\mathcal{O}(1)$, where the expansion parameter is $\Delta^{-1}$.

For simplicity, let us first consider a non-adiabatic transitions from a single computational basis state $\ket{00}$ into a single state outside $\ket{\rm ggee}$, ignoring the rest of the system.
We model this two-level subsystem by
\begin{equation}
H(t)=\frac{1}{2}\Delta\tau^{\rm Z}+\frac{1}{2}g_{\rm c}(t)\tau^{\rm X},
\end{equation}
where the Pauli matrices $\tau^{\rm X}$, $\tau^{\rm Y}$ and $\tau^{\rm Z}$ are defined in the standard way in the basis $\ket{00}$ and $\ket{\rm ggee}$ (where $\tau^{\rm Z}\ket{00}=\ket{00}$ and $\tau^{\rm Z}\ket{\rm ggee}=-\ket{\rm ggee}$).
Below, we focus on the (relevant) case of $g(t)\ll|\Delta|$.

Assuming the system starts at $t=0$ in the state $\ket{00}$, and $g_{\rm c}(0)=0$, we are after the probability of making a non-adiabatic transition into the state evolving adiabatically from $\ket{\rm ggee}$, as a function of time. To begin, we perform a transformation that diagonalizes the instantaneous Hamiltonian, i.e.,
\begin{subequations}
\begin{align}
    &U^{\dagger}(t)H(t)U(t)=\frac{1}{2}\varepsilon(t)\tau^{\rm Z},
    \\
    &U(t)=\exp[-i\theta(t)\tau^{\rm Y}/2],
\end{align}
\end{subequations}
where $\varepsilon(t)=\sqrt{\Delta^{2}+g^{2}(t)}$, and $\tan\theta(t)=g(t)/\Delta.$
The transformed wavefunction $|\tilde{\psi}(t)\rangle=U^{\dagger}(t)|\psi(t)\rangle$ obeys the Schr\"odinger equation with the following Hamiltonian
\begin{equation}
\tilde{H}(t)=U^{\dagger}(t)H(t)U(t)-iU^{\dagger}(t)\dot{U}(t)=\frac{\varepsilon(t)}{2}\tau^{\rm Z}-\frac{\dot{\theta}(t)}{2}\tau^{\rm Y},
\end{equation}
where $\dot{f}(t)$ denotes the derivative of a time-dependent function $f(t)$ with respect to time $t$.

If $g(t)$ changes slowly enough one can neglect the second term, resulting in the adiabatic theorem. In order to get a correction to the adiabatic theorem we, instead, perform another diagonalization of the instantaneous Hamiltonian $\tilde{H}(t)$. The new transformed wavefunction is $|\bar{\psi}(t)\rangle=\bar{U}^{\dagger}(t)|\tilde{\psi}(t)\rangle$, where $\bar{U}(t)=\exp[i\phi(t)\tau^{\rm X}/2]$, and its dynamics is governed by the Hamiltonian
\begin{equation}\label{eq:DiabHbar}
    \bar{H}(t)=\bar{U}^{\dagger}(t)\tilde{H}\bar{U}(t)-i\bar{U}^{\dagger}(t)\dot{\bar{U}}(t)=\frac{\nu(t)}{2}\tau^{\rm Z}+\frac{\dot{\phi}(t)}{2}\tau^{\rm X},
\end{equation}
where $\nu(t)=\sqrt{\varepsilon^{2}(t)+\dot{\theta}^{2}(t)}$ and $\tan\phi(t)=\dot{\theta}(t)/\varepsilon(t)$.
Finally, if $|\dot{\phi}|\ll|\nu(t)|$, then we can neglect the second term in Eq.~(\ref{eq:DiabHbar}) and obtain
\begin{equation}
|\psi(t)\rangle=
\exp\left(\frac{-i\theta(t)}{2}\tau^{\rm Y}\right)
\exp\left(\frac{i\phi(t)}{2}\tau^{\rm X}\right)
\exp\left(\frac{-i}{2}\int_{0}^{t}{\rm d}t'\nu(t')\tau^{\rm Z}\right)
\exp\left(\frac{-i\phi(0)}{2}\tau^{\rm X}\right)
\exp\left(\frac{i\theta(0)}{2}\tau^{\rm Y}\right)|\psi(0)\rangle.
\end{equation}
The probability of making a non-adiabatic transition is then given by 
\begin{equation}\label{eq:Pd}
P_{{\rm D}}(t)=
\left|\bra{{\rm ggee}}\exp(i\theta(t)\tau^{\rm Y})\ket{\psi(t)}\right|^{2}
=\cos^{2}\left[\frac{\xi(t)}{2}\right]\sin^{2}\left[\frac{\phi(t)-\phi(0)}{2}\right]+\sin^{2}\left[\frac{\xi(t)}{2}\right]\sin^{2}\left[\frac{\phi(t)+\phi(0)}{2}\right],
\end{equation}
where $\xi(t)=\int_0^t{\rm d}t'\nu(t')$.

Consider, for example, a linear ramp-up of the coupling, $g_{\rm c}(t)=g^{\rm max}_{\rm c}t/T_{\rm ramp}$, where $g^{\rm max}_{\rm c}\ll|\Delta|$.
Plugging it in the definition of $\theta$ one arrives at $\dot\theta=g^{\rm max}_{\rm c}/(\Delta T_{\rm ramp})$ and consequently $\phi(t)\simeq g^{\rm max}_{\rm c}/(\Delta^2T_{\rm ramp})$, $\nu(t)\simeq \Delta$. Plugging in Eq.~(\ref{eq:Pd}), one obtains
\begin{equation}
P_{{\rm D}}(t)\simeq\frac{(g^{\rm max}_{\rm c})^{2}}{\Delta^{4}T_{\rm ramp}^{2}}\sin^{2}\left[\frac{\Delta t}{2}\right].
\end{equation}
Going back to the full system, where each of the computational basis states is coupled to the four states $\ket{\rm ggee}$, $\ket{\rm eegg}$, $\ket{\rm gfgg}$, $\ket{\rm ggfg}$, we can obtain a crude estimate for the leakage by ignoring interference effects and  time-dependent oscillations, and simply multiply $P_{\rm D}(t)$ by 4 and average over time, to obtain $P_{\rm leak}\simeq 2(g^{\rm max}_{\rm c})^{2}/(\Delta^{4}T_{\rm ramp}^{2})$.

We can compare this estimate with the results of the simulation presented in Fig.~\ref{fig:dispersive_composite}.
In the simulation we consider varying the coupling according to $g_{\rm c}(t)=g^{\rm max}_{\rm c}\{1-[1-\sin(t/T_{\rm g})]^4\}$ with $g_{\rm c}^{\rm max}/(2\pi)=34\unit{MHz}$ and $T_{\rm g}=110\unit{ns}$ (see Fig.~\ref{fig:dispersive_composite}b). We can approximate the ramp-up period by $g^{\rm max}_{\rm c}t/T_{\rm ramp}$ with $T_{\rm ramp}\simeq20\unit{ns}$, and then use the above result for the leakage probability, yielding $P_{\rm leak}\simeq2\cdot 10^{-6}$, which reasonably agrees with the results of the simulation [see grey lines in Fig.~\ref{fig:dispersive_composite}(c,d) at the end of the evolution]. As noted in the main text, this leakage can be detected and converted to erasure, thereby contributing to the erasure error rate $e$ (see Sec.~\ref{sec:SurfCode}).

\section{Readout dispersive coupling of the dual-rail qubit}
\label{app:DispersCoupling}

In this appendix we derive the dispersive coupling, presented in Eqs.~(\ref{eq:Hchi},\ref{eq:chiDRa},\ref{eq:chiDRb}), between the dual-rail qubit and a readout cavity. We separate the Hamiltonian, $H=H_{\rm quad}+H_{\rm quart}$, of the combined system into two parts
\begin{eqnarray}
\label{eq:Hreadout}
H_{\rm quad}&=&\omega_{c}c^{\dagger}c + \sum_{j=1,2}\omega a_{j}^{\dagger}a_{j}
+ g_{12}(a_{1}^{\dagger}a_{2}+{\rm h.c.})
+ g_{{\rm RTj}}(c^{\dagger}a_{j}+{\rm h.c.}),\\
H_{\rm quart}&=&\sum_{j=1,2}\frac{\eta}{2}a_{j}^{\dagger}a_{j}^{\dagger}a_{j}a_{j},
\end{eqnarray}
where $c,c^\dagger$ are the ladder operators for the readout cavity mode, $\omega_{\rm c}$ is its frequency, and $g_{\rm RT1}, g_{\rm RT2}$ are its coupling to the two transmons.

To analyze the system, we first diagonalize the part of the Hamiltonian which is quadratic in the ladder operators,
\begin{equation}\label{eq:Hquad}
H_{{\rm quad}}=
\begin{pmatrix}a_{1}^{\dagger}, & a_{2}^{\dagger}, & c^{\dagger}\end{pmatrix}
\begin{pmatrix}
\omega & g_{12} & g_{{\rm RT}}\\
g_{12} & \omega & g_{{\rm RT}}\\
g_{{\rm RT}} & g_{{\rm RT}} & \omega_{c}
\end{pmatrix}
\begin{pmatrix}a_{1}\\a_{2}\\c^{\dagger}\end{pmatrix}
=\tilde\omega_{a}\tilde{a}^{\dagger}\tilde{a}+\tilde\omega_{b}\tilde{b}^{\dagger}\tilde{b}+\tilde\omega_{c}\tilde{c}^{\dagger}\tilde{c},
\end{equation}
where the transformation to the eigenmodes can formally be written as
\begin{subequations}\label{eq:Uquad}
\begin{align}
\label{eq:atilde}
    \tilde{a} &= \alpha_1 a_1 + \alpha_2 a_2 +\alpha_c c,
    \\
    \label{eq:btilde}
    \tilde{b} &= \beta_1 a_1 + \beta_2 a_2 +\beta_c c,
    \\
    \label{eq:ctilde}
    \tilde{c} &= \gamma_1 a_1 + \gamma_2 a_2 +\gamma_c c.
\end{align}
\end{subequations}
Here, $\tilde{a},\tilde{b}$ are the eigenmodes of the dual-rail qubit dressed by the interaction with the cavity. In the absence of this interaction, the eigenmodes of the dual-rail qubit would be $a=(a_1+a_2)/\sqrt{2}$ and $b=(a_1-a_2)/\sqrt{2}$.
Similarly, $\tilde{c}$ is the mode of the cavity dressed by its interaction with the transmons. Below, we shall invoke a series expression in $\Delta^{-1}$ for the coefficients in Eqs.~(\ref{eq:atilde}-\ref{eq:ctilde}) as well as for the eigenfrequencies $\tilde\omega_a$, $\tilde\omega_b$, and $\tilde\omega_c$.

We can now write the part of the Hamiltonian which is quartic in the ladder operators, $H_{\rm quart}$,  using the dressed eigenmodes,
\begin{equation}\label{eq:Hquart}
\begin{split}
H_{\rm quart}=&\sum_{j=1,2}\frac{\eta}{2}(\alpha_j \tilde{a}^{\dagger}+\beta_j \tilde{b}^{\dagger}+\gamma_j \tilde{c}^{\dagger})^{2}
(\alpha_j \tilde{a} + \beta_j \tilde{b} + \gamma_j \tilde{c})^{2}.
\end{split}
\end{equation}
To obtain the dispersive coupling to leading order we focus on the terms in $H_{\rm quart}$ that are proportional to $\tilde{c}^\dagger c$, yielding
\begin{equation}
    H_\chi = 2\eta \sum_{j=1,2} \gamma_j^2\left[
        \alpha_j^2 \tilde{a}^\dagger \tilde{a}  +\beta_j^2 \tilde{b}^\dagger \tilde{b} + 
        \alpha_j \beta_j \left(\tilde{a}^\dagger \tilde{b} + \tilde{b}^\dagger \tilde{a} \right)
    \right]\tilde{c}^\dagger \tilde{c},
\end{equation}
As noted above, the dual-rail computational basis states, dressed by the interaction with cavity, are given by $|1\rangle=\ket{1_a,0_b,n_c}$ and $|0\rangle=\ket{0_a,1_b,n_c}$, where the three integers represent the number of excitations in the $a$,$b$ and $c$ modes, respectively.
Upon projecting $H_\chi$ onto the computational subspace, one obtains the Hamiltonian of Eq.~(\ref{eq:Hchi}), with $\chi_1=2\eta\sum_j\alpha_j^2\gamma_j^2$, $\chi_0=2\eta\sum_j\beta_j^2\gamma_j^2$, and $\chi'=2\eta\sum_j\alpha_j\beta_j\gamma_j^2$. Finally, we apply second-order perturbation theory to Eqs.~(\ref{eq:Hquad},\ref{eq:Uquad}) to obtain $\alpha_1,\alpha_2,\beta_1=1/\sqrt{2}+\mathcal{O}(\Delta^{-1})$, 
$\beta_2=-1/\sqrt{2}+\mathcal{O}(\Delta^{-1})$, and $\gamma_j^2=g_{{\rm RT}j}^2/\Delta_j^2+\mathcal{O}(\Delta^{-3})$. Substituting these expressions in $\chi_0$, $\chi_1$, and $\chi'$ results in Eqs.~(\ref{eq:chiDRa}, \ref{eq:chiDRb}) of the main text.

\subsection{The symmetric case, $g_{\rm RT1}=g_{\rm RT2}$}
\label{app:DispersCouplingSymm}
 
In the symmetric case of $g_{\rm RT1}=g_{\rm RT2}=g_{\rm RT}$ one obtains $\chi'=0$ and $\chi_0=\chi_1$ to leading order in $\Delta^{-1}$, resulting in $\partial\Omega/\partial n_{\rm c}=0$ [see Eq.~(\ref{eq:chiDRa},\ref{eq:chiDRb})]. It then becomes important to obtain the next-order terms in the expansion of $\Omega(n_{\rm c})$. To this end we first reexamine the diagonalization of $H_{\rm quad}$.
Due to the symmetry, an analytical expression for the frequencies $\tilde\omega_a$, $\tilde\omega_b$ and $\tilde\omega_c$ can be derived, as well as for the coefficients in Eq.~(\ref{eq:Uquad}).
In particular, we have $\alpha_1=\alpha_2=\alpha$, $\beta_1=-\beta_2=\beta$, and $\gamma_1=\gamma_2=\gamma$. Substituting in $H_{\rm quart}$ allows one to rewrite the full Hamiltonian as $H=H_{0}+H^{(1)}+H^{(2)}+H^{(3)}$, where we have
\begin{subequations}
\begin{align}
\begin{split}
H_{0}&=\tilde{\omega_{a}}\tilde{a}^{\dagger}\tilde{a}+\tilde{\omega_{b}}\tilde{b}^{\dagger}\tilde{b}+\tilde{\omega_{c}}\tilde{c}^{\dagger}\tilde{c}
+4\eta\gamma^{2}(\alpha^{2}\tilde{a}^{\dagger}\tilde{a}+\beta^{2}\tilde{b}^{\dagger}\tilde{b})\tilde{c}^{\dagger}\tilde{c}\\
&+\eta\left[\gamma^{4}\tilde{c}^{\dagger}\tilde{c}^{\dagger}\tilde{c}\tilde{c} + \alpha^{4}\tilde{a}^{\dagger}\tilde{a}^{\dagger}\tilde{a}\tilde{a} + \beta^{4}\tilde{b}^{\dagger}\tilde{b}^{\dagger}\tilde{b}\tilde{b} + \alpha^{2}\beta^{2}\left(4\tilde{a}^{\dagger}\tilde{a}\tilde{b}^{\dagger}\tilde{b} + \tilde{a}^{\dagger}\tilde{a}^{\dagger}\tilde{b}\tilde{b} + \tilde{b}^{\dagger}\tilde{b}^{\dagger}\tilde{a}\tilde{a}\right)\right],\\
\end{split}\\
&H^{(1)}=2\eta\gamma\left[\alpha \tilde{a}^{\dagger}\left(\alpha^{2}\tilde{a}^{\dagger}\tilde{a}+2\beta^{2}\tilde{b}^{\dagger}\tilde{b}\right)
+\alpha\beta^{2}\tilde{b}^{\dagger}\tilde{b}^{\dagger}\tilde{a}\right]\tilde{c}+{\rm h.c.},\\
&H^{(2)}=\eta\gamma^{2}\left(\alpha^{2}\tilde{a}^{\dagger}\tilde{a}^{\dagger}+\beta^{2}\tilde{b}^{\dagger}\tilde{b}^{\dagger}\right)\tilde{c}\tilde{c}+{\rm h.c.},\\
&H^{(3)}=2\eta\gamma^{3}\alpha \tilde{a}^{\dagger}\tilde{c}^{\dagger}\tilde{c}\tilde{c} + {\rm h.c}
\end{align}
\end{subequations}
The first term, $H_0$, does not exchange excitations between the dual-rail qubit and the cavity; we shall treat it as the unperturbed Hamiltonian. The other terms in the Hamiltonian, $H^{(1)}$, $H^{(2)}$, and $H^{(3)}$, are perturbations ordered according to powers of $\gamma \simeq g_{\rm RT}/\Delta$ that they contain.

In analyzing $H_0$, we notice that $n_{c}=c^{\dagger}c$ is a good quantum number.
The zeroth-excitation state and its energy are given by
\begin{equation}
|0_{a},0_{b},n_{c}\rangle,\hspace{1em}E^{(0)}_{00n_{\rm c}}
= (\tilde\omega_{c}-\eta\gamma^{4})n_{c}+\eta\gamma^{4}n_{c}^{2}.
\end{equation}
In the single-excitation sector there are two eigenstates,
\begin{subequations}
\begin{align}
|1_{a},0_{b},n_{c}\rangle,\hspace{1em}E^{(0)}_{10n_{c}} & =\tilde\omega_{a}+E^{(0)}_{00n_{\rm c}}+4\eta\alpha^{2}\gamma^{2}n_{c},
\\
|0_{a},1_{b},n_{c}\rangle,\hspace{1em}E^{(0)}_{01n_{c}} & =\tilde\omega_{b}+E^{(0)}_{00n_{\rm c}}+4\eta\beta^{2}\gamma^{2}n_{c}.
\end{align}
\end{subequations}
As noted above, in the absence of the perturbations, these correspond to the computational basis states $\ket{1}$ and $\ket{0}$, respectively.

In the two-excitation sector, one eigenstate is given by
\begin{equation}
|1_{a},1_{b},n_{c}\rangle,\hspace{1em}E^{(0)}_{11n_{c}}=\tilde\omega_{a}+\tilde\omega_{b}+E^{(0)}_{00n_{\rm c}}+4\eta\alpha^{2}\beta^{2}+4\eta(\alpha^{2}+\beta^{2})\gamma^{2}n_{c},
\end{equation}
and the other two eigenstates are obtained by 
diagonalizing the Hamiltonian that is obtained by projecting $H_0$ onto the subspace spanned by \{$|0_{a},2_{b},n_{c}\rangle$ and $|0_{a},2_{b},n_{c}\rangle\}$, i.e.,
\begin{equation}
H_{2{\rm exc}}=\begin{pmatrix}
E^{(0)}_{00n_{\rm c}} + 2(\tilde\omega_{a}+4\eta\alpha^{2}\gamma^{2}n_{c})+2\alpha^{4}\eta & 2\eta\alpha^{2}\beta^{2}\\
2\eta\alpha^{2}\beta^{2} & E^{(0)}_{00n_{\rm c}} + 2(\tilde\omega_{b}+4\eta\beta^{2}\gamma^{2}n_{c})+2\beta^{4}\eta
\end{pmatrix}.
\end{equation}
The above $2\times2$ matrix is diagonalized by
\begin{subequations}
\begin{align}
|\phi_{\rm I},n_{\rm c}\rangle & =\cos[\lambda(n_{\rm c})]|2_{a},0_{b},n_{\rm c}\rangle+\sin[\lambda(n_{\rm c})]|0_{a},2_{b},n_{c}\rangle,\\
|\phi_{\rm II},n_{\rm c}\rangle & =\sin[\lambda(n_{\rm c})]|2_{a},0_{b},n_{\rm c}\rangle-\cos[\lambda(n_{\rm c})]|0_{a},2_{b},n_{\rm c}\rangle,
\end{align}
\end{subequations}
where
\begin{equation}
    \lambda(n_{\rm c})=\tan^{-1}\left(
    \frac{2\eta\alpha^{2}\beta^{2}}
    {\tilde\omega_{a}-\tilde\omega_b+4\eta(\alpha^{2}-\beta^2)\gamma^{2}n_{c}+\eta(\alpha^{4}-\beta^4)}
    \right)
\end{equation}
and the corresponding eigenenergies are given by
\begin{subequations}
\begin{align}
\begin{split}
E^{(0)}_{{\rm I},n_{c}}&=
E^{(0)}_{00n_{\rm c}}
+
\tilde\omega_{a}+\tilde\omega_b+4\eta(\alpha^{2}+\beta^2)\gamma^{2}n_{c}+\eta(\alpha^{4}+\beta^4)\\
&+ 
\sqrt{[ \tilde\omega_{a}-\tilde\omega_b+4\eta(\alpha^{2}-\beta^2)\gamma^{2}n_{c}+\eta(\alpha^{4}-\beta^4) ]^2 + 4\eta^2\alpha^{4}\beta^{4} }
\end{split}
\\
\begin{split}
E^{(0)}_{{\rm II},n_{c}}&=
E^{(0)}_{00n_{\rm c}}
+
\tilde\omega_{a}+\tilde\omega_b+4\eta(\alpha^{2}+\beta^2)\gamma^{2}n_{c}+\eta(\alpha^{4}+\beta^4)\\
&- 
\sqrt{[ \tilde\omega_{a}-\tilde\omega_b+4\eta(\alpha^{2}-\beta^2)\gamma^{2}n_{c}+\eta(\alpha^{4}-\beta^4) ]^2 + 4\eta^2\alpha^{4}\beta^{4} }
\end{split}
\end{align}
\end{subequations}

We are interested in the leading corrections to $E^{(0)}_{10n_{\rm c}}$ and $E^{(0)}_{01n_{\rm c}}$, which arise from second-order perturbation theory. We neglect $H^{(2)}$ and $H^{(3)}$ as these would only contribute a correction of order $g_{\rm RT}^4\Delta^{-5}$. We obtain
\begin{subequations}
\begin{align}
\begin{split}
E_{10n_{c}}^{(2)}&=E^{(0)}_{10n_{\rm c}}+\frac{8n_{c}\eta^{2}\alpha^{2}\gamma^{2}\left(\alpha^{2}\cos[\lambda(n_{\rm c-1})]+\beta^{2}\sin[\lambda(n_{\rm c}-1)]\right)^{2}}{E^{(0)}_{10n_{\rm c}}-E^{(0)}_{{\rm I},n_{c}-1}}\\
&+\frac{8n_{c}\eta^{2}\alpha^{2}\gamma^{2}\left(\alpha^{2}\sin[\lambda(n_{\rm c}-1)]-\beta^{2}\cos[\lambda(n_{\rm c}-1)]\right)^{2}}{E^{(0)}_{10n_{\rm c}}-E^{(0)}_{{\rm II},n_{c}-1}},
\end{split}\\
E_{01n_{c}}^{(2)}&=E^{(0)}_{01n_{c}}+\frac{16\eta^{2}\gamma^{2}\alpha^{2}\beta^{4}n_{c}}{E^{(0)}_{01n_{c}}-E_{11n_{c}-1}}.
\end{align}
\end{subequations}

Substituting the expressions for $\alpha$, $\beta$, $\gamma$, and expanding up to fourth order in $\Delta^{-1}$, we have
\begin{equation}
\Omega(n_{\rm c})=E^{(2)}_{10n_{\rm c}}-E^{(2)}_{01n_{\rm c}} = -4 \frac{\eta g_{\rm RT}^4}{\Delta^4}n_{\rm c}
-4\frac{\eta^2g_{\rm RT}^2g_{12}}{\Delta^4}n_{\rm c}+\mathcal{O}(\Delta^{-5}).
\end{equation}

\section{Spinlocking the {\rm g-f} qubit}
\label{spinlock_g_f}

In this appendix we demonstrate how the interaction between the $\ket{\rm g}$ and $\ket{\rm f}$ levels of a transmon, Eq.~\eqref{eq:gfDrive}, can be engineered using a two-photon XY drive.
The Hamiltonian describing the driven transmon is $H=H_0+H_{\rm d}$, where
\begin{subequations}
\begin{align}
&H_0=\omega a^{\dagger}a+\frac{\eta}{2}a^{\dagger}a^{\dagger}aa,\label{eq:Htrans}\\
&H_{\rm d} =  2\epsilon_{\rm d}\cos\left(\omega_{\rm d}t-\varphi_{\rm d}\right)\left(a^{\dagger}+a\right), 
\end{align}
\end{subequations}
$\omega$ is the energy gap between $\left|{\rm g}\right\rangle$ and $\left|{\rm e}\right\rangle $, $\eta$ is the transmon nonlinearity, $\epsilon_{\rm d}$ is the drive amplitude, $\omega_{\rm d}$ and $\varphi_{\rm d}$ are the drive frequency and phase, respectively.
We choose the drive frequency to be half the transition frequency between $\ket{\rm g}$ and $\ket{\rm f}$, namely $\omega_{\rm d}=\omega-\eta/2$, anticipating a resonant transition at second order in $\epsilon_{\rm d}$.
Moving to the interaction picture with respect to $H_0$, and limiting our analysis to the four lowest states of the transmon, $\left|{\rm g}\right\rangle ,\left|{\rm e}\right\rangle $,
$\left|{\rm f}\right\rangle $ and $\left|{\rm h}\right\rangle$, we have (ignoring fast rotating terms)
\begin{equation}
\label{eq:rotating_frame_gf}
H_{\rm I} =\epsilon_{\rm d}\exp[i\left({\eta t/2}+\varphi_{\rm d}\right)]\ketbra{{\rm e}}{{\rm g}}
+\sqrt{2}\epsilon_{\rm d}\exp[-i\left({\eta t/2}-\varphi_{\rm d}\right)]\ketbra{{\rm f}}{{\rm e}}
+\sqrt{3}\epsilon_{\rm d}\exp[-i\left(3\eta t/2 -\varphi_{\rm d}\right)]\ketbra{{\rm h}}{{\rm f}} + {\rm h.c.}
\end{equation}
Adiabatically eliminating the states $\left|{\rm e}\right\rangle $ and
$\left|{\rm h}\right\rangle $ that are never resonantly excited, we obtain the effective Hamiltonian to leading order in $\epsilon_{\rm d}/\eta$, i.e.,
\begin{equation}
H_{\rm eff} =\frac{2\epsilon_{\rm d}^{2}}{\eta} \left(\ketbra{{\rm g}}{{\rm g}}
+\ketbra{{\rm f}}{{\rm f}}\right)
+\frac{\sqrt{8}\epsilon_{\rm d}^{2}}{\eta}\left[\exp(-2i\varphi_{\rm d})\ketbra{{\rm g}}{{\rm f}}
+{\rm h.c.}\right] + \mathcal{O}\left(\epsilon^3_{\rm d}/\eta^2\right).
\end{equation}

The first term is proportional to the identity operator when acting in the computational subspace and can be ignored. The second term is the desired interaction between the $\ket{\rm g}$ and $\ket{\rm f}$ levels, which motivates our choice of the computational basis for the g-f qubit from Eq.~\eqref{eq:gfCompBase}.
Finally, upon going back from the interaction picture this term yields Eq.~\eqref{eq:gfDrive} with $\Omega_0=4\sqrt{2}\epsilon_{\rm d}^2/\eta$ and $\varphi_{\rm R}=2\varphi_{\rm d}$.

As mentioned in Sec.~\ref{sub:gf}, it is important to keep the population of the $\ket{\rm e}$ level low in order to avoid undetected errors. To estimate the $\ket{\rm e}$ population due to the two-photon drive we evolve each of the computational basis states by integrating Eq.~\eqref{eq:rotating_frame_gf} to first order, projecting onto the $\ket{\rm e}$ levels, and averaging over time, i.e.,
\begin{equation}
P_{\rm e}^{(b)} = \overline{\left|\bra{{\rm e}}\left( 1 - i\int_0^t H_{\rm I}(t^\prime)\d t^\prime \right)\ket{b}\right|^{2}}
=\frac{2\left(1+(-1)^b\sqrt{2}\right)^{2}\epsilon_{\rm d}^{2}}{\eta^2}\overline{2\sin^{2}\left(\eta t/4\right)}=\frac{2\left(1+(-1)^b\sqrt{2}\right)^{2}\epsilon_{\rm d}^{2}}{\eta^2},\quad b=0,1.
\end{equation}
For the equally-weighted mixed state of $\ket{0}$ and $\ket{1}$, we get  $P_{\rm e}=(P^{(0)}_{\rm e}+P^{(1)}_{\rm e})/2=6{\epsilon_{\rm d}^{2}}/{\eta^2}$ as provided in Sec.~\ref{sub:gf}. 
Using the same procedure, we find the average population of $\ket{{\rm h}}$ to be $P_{\rm h} = 4\epsilon_d^2/3 \eta^2$.

\section{The effective noise on the {\rm g-f} qubit}
\label{t1_decay_g_f}

In this section we discuss the amplitude damping noise affecting a three-level transmon and show that, due to spin-locking, the effective noise on the g-f qubit is the heralded erasure noise, as captured by Eq.~\eqref{eq_gfnoise}.
We consider a transmon whose unitary dynamics is governed by the Hamiltonian $H=H_0+H_{\rm R}$, where $H_0$ is the bare transmon Hamiltonian given in Eq.~\eqref{eq:Htrans} and $H_{\rm R}$ describes the spin-locking drive given in Eq.~\eqref{eq:gfDrive}.
The master equation describing the dynamics of the g-f qubit in the interaction picture with respect to $H_0$ can be written as
\begin{equation}
\frac{{\rm d}}{{\rm d}t}\rho_{\rm I} = -i\frac{\Omega_0}{2}[{\ketbra{\rm g}{f}}+{\ketbra{\rm f}{g}},\rho_{\rm I}]  + \Gamma_1  \mathcal{D} [a] \rho_{\rm I},
\end{equation}
where $\Omega_0$ is the spin-locking Rabi frequency, $\Gamma_1$ is the amplitude-damping rate, $\mathcal{D}[a]\rho=a\rho a^\dagger - \tfrac{1}{2}(a^\dagger a\rho + \rho a^\dagger a)$, and $a$ is the annihilation operator for the transmon.

In typical settings the spin-locking Rabi frequency $\Omega_0$ is two or three orders of magnitude larger than $\Gamma_1$. We therefore move to the interaction picture with respect to $\Omega_0$, drop terms oscillating at frequency $\Omega_0$ and obtain
\begin{equation}\label{eq:DMgf}
\frac{{\rm d}}{{\rm d}t}\rho_{\rm II} = \Gamma_1  \mathcal{D} [{\ketbra{\rm e}{0}}] \rho_{\rm II} + \Gamma_1  \mathcal{D} [{\ketbra{\rm e}{1}}] \rho_{\rm II}\\
+ \frac{1}{2}\Gamma_1  \mathcal{D} [{\ketbra{0}{\rm e}}] \rho_{\rm II} + \frac{1}{2}\Gamma_1 \mathcal{D} [{\ketbra{1}{\rm e}}] \rho_{\rm II},
\end{equation}
where we write the jump operators in terms of the computational basis states, $\ket{0}$ and $\ket{1}$, defined in Eq.~\eqref{eq:gfCompBase}.
Integrating Eq.~\eqref{eq:DMgf} over a short period of time $\delta t$ and projecting $\rho_{\rm II}$ onto the computational subspace results in $\mathcal N(\rho) = (1-\gamma)\rho + \gamma \ketbra{{\rm e}}{{\rm e}}$ with $\gamma = \Gamma_1\delta t$, as given in Eq.~\eqref{eq_gfnoise}.

\section{Derivation of the effective Hamiltonian for two {\rm g-f} qubits}
\label{ZZ_interaction}

In this appendix we derive the effective Hamiltonian in Eq.~\eqref{eq:coupling_gf_qubits} that describes the interaction between two g-f qubits.
The Hamiltonian describing the coupling between two driven transmons can be written as $H=H_0+H_{\rm c}+H_{\rm R}^{(i)}$, where
\begin{subequations}\label{eq:H2Qgf}
\begin{align}
&H_0 = \sum_{i=1}^{2}\omega_{i}a_{i}^{\dagger}a_{i}+\frac{\eta}{2}a_{i}^{\dagger}a_{i}^{\dagger}a_{i}a_{i},
\\
&H_{\rm c} = g_{\rm c}\left(a_{1}^{\dagger}a_{2}+\text{h.c.}\right),
\end{align}
\end{subequations}
and $H_{\rm R}^{\left(i\right)}\left(t\right)$ is the spin-locking drive term for each of the g-f qubits given in Eq.~\eqref{eq:gfDrive}.
We take the transmon frequencies to be detuned with respect to each other, $\omega_2=\omega_1+\Delta$.
In the dispersive limit, where $|\Delta|\gg g_{\rm c}$, we can obtain an effective Hamiltonian by applying second-order Magnus expansion, namely
\begin{equation}
H_{\rm eff} = \frac{1}{2it}\int_0^t {\rm d}t_1 \int_0^{t_1} {\rm d}t_2 [H_{\rm I}(t_1),H_{\rm I}(t_2)],
\end{equation}
where $H_{\rm I}(t)$ is the Hamiltonian of Eq.~\eqref{eq:H2Qgf} written in the interaction picture with respect to $H_0$. Dropping fast-oscillating terms and projecting to the computational subspace, we have 
\begin{equation}
H_{\rm eff} = \frac{1}{2}\Omega_0(\sigma^{\rm Z}_1+\sigma^{\rm Z}_2) + h^{(1)}_{\rm X} \sigma^{\rm X}_1 + h^{(2)}_{\rm X} \sigma^{\rm X}_2 + g_{\rm XX} \sigma^{\rm X}_1\sigma^{\rm X}_2,
\end{equation}
where $h^{(1)}_{\rm X} = 2 g_{\rm c}^{2}(\Delta-3\eta)/(\Delta^{2}-\eta^{2})$, $h^{(2)}_{\rm X} = - 2 g_{\rm c}^{2}(\Delta+3\eta)/(\Delta^{2}-\eta^{2})$, and $g_{\rm XX} = 4g_{\rm c}^{2}\eta/(\Delta^{2}-\eta^{2})$.
As in the case of the dual-rail 2Q interaction, we can cancel the effect of the term $\sigma^{\rm X}_i$ by appropriately detuning the frequency of the spin-locking drive of qubit $i$, resulting in the Hamiltonian of Eq.~\eqref{eq:coupling_gf_qubits}.

\section{Dynamical decoupling and the ability to correct amplitude damping errors}
\label{sec_dynamical}

In this appendix we argue that dynamical decoupling can negatively affect the ability to correct amplitude damping errors.
To illustrate our discussion, we start with a simple example of a qubit engineered with four transmons that is based on the four-qubit code.
We choose the computational basis states for our qubit to be
\begin{equation}
\label{eq_fourqubit}
\ket{0} = (\ket{\rm gggg}+\ket{\rm eeee})/\sqrt{2},\quad
\ket{1} = (\ket{\rm ggee}+\ket{\rm eegg})/\sqrt{2}.
\end{equation}
Then, as explained in Ref.~\cite{Leung1997}, for an arbitrary state $\ket\psi = \alpha\ket{0}+\beta\ket{1}$, where $\alpha,\beta\in\mathbb C$ and $|\alpha|^2 + |\beta|^2 = 1$, any single amplitude damping error (or no error) can be detected and the state $\ket\psi$ approximately recovered.

Recall that in our scheme that we discuss in Sec.~\ref{sec:SurfCode} we crucially rely on the $T_\phi/T_1$ bias that needs to be engineered via, e.g., dynamical decoupling in the form of a series many of $\pi$ pulses.
To be concrete, consider implementing each $\pi$ pulse (which applies a Pauli $\sigma^{\rm X}$ operator in the computational subspace) as a unitary $U\otimes U\otimes I\otimes I$, where $U = \ketbra{{\rm g}}{{\rm e}} + \ketbra{{\rm g}}{{\rm e}}$ and $I$ is the identity operator.
Further, assume that a single amplitude damping error happens with equal probability between any two consecutive $\pi$ pulses but we do not know when~\footnote{One way to overcome this problem is to realize a measurement that detects a single amplitude damping error after every $\pi$ pulse, however, at least in the superconducting scenario this is not realistic.}.
One can verify that if the amplitude damping error happens on either the third or fourth transmon, then the qubit will be in an equal probabilistic mixture of two states
$(\alpha\ket{{\rm ee}}+\beta\ket{{\rm gg}})\otimes \ket{{\rm gg}}$
and $(\alpha\ket{{\rm gg}}+\beta\ket{{\rm ee}})\otimes \ket{{\rm gg}}$, and the state $\ket\psi$ cannot be reliably recovered.
We thus conclude that dynamical decoupling renders the four-qubit code incapable of correcting any single amplitude damping error, unless we know when the error happens.

We also point out that the computational basis states specified by Eq.~\eqref{eq_fourqubit} suffer from dephasing due to the inherent energy gap between the states $\ket{\rm g}$ and $\ket{\rm e}$.
We can, however, modify the four-qubit code and choose a different computational subspace, for instance
\begin{equation}
\ket{b} = 
\frac{1}{2} \left( \ket{{\rm ge}} - (-1)^b \ket{{\rm eg}} \right)^{\otimes 2}, \quad b=0,1,
\end{equation}
so that the resulting qubit is insensitive to dephasing.
Unfortunately, dynamical decoupling would still make correcting any single amplitude damping error impossible, unless we know when the error happens.

Lastly, we remark that an issue would arise if we engineered a qubit with two three-level transmons.
For instance, if we choose the computational basis states to be
\begin{equation}
\ket{b} = [\ket{{\rm g}} - (-1)^b\ket{{\rm f}}]/2, \quad b=0,1,
\end{equation}
then, in the absence of dynamical decoupling, any single amplitude damping error (or no error) can be detected and approximately corrected.
On the other hand, dynamical decoupling renders correction impossible.

\bibliographystyle{apsrev4-2}
\bibliography{Bib_Alex}

\begin{thebibliography}{59}%
\makeatletter
\providecommand \@ifxundefined [1]{%
 \@ifx{#1\undefined}
}%
\providecommand \@ifnum [1]{%
 \ifnum #1\expandafter \@firstoftwo
 \else \expandafter \@secondoftwo
 \fi
}%
\providecommand \@ifx [1]{%
 \ifx #1\expandafter \@firstoftwo
 \else \expandafter \@secondoftwo
 \fi
}%
\providecommand \natexlab [1]{#1}%
\providecommand \enquote  [1]{``#1''}%
\providecommand \bibnamefont  [1]{#1}%
\providecommand \bibfnamefont [1]{#1}%
\providecommand \citenamefont [1]{#1}%
\providecommand \href@noop [0]{\@secondoftwo}%
\providecommand \href [0]{\begingroup \@sanitize@url \@href}%
\providecommand \@href[1]{\@@startlink{#1}\@@href}%
\providecommand \@@href[1]{\endgroup#1\@@endlink}%
\providecommand \@sanitize@url [0]{\catcode `\\12\catcode `\$12\catcode
  `\&12\catcode `\#12\catcode `\^12\catcode `\_12\catcode `\%12\relax}%
\providecommand \@@startlink[1]{}%
\providecommand \@@endlink[0]{}%
\providecommand \url  [0]{\begingroup\@sanitize@url \@url }%
\providecommand \@url [1]{\endgroup\@href {#1}{\urlprefix }}%
\providecommand \urlprefix  [0]{URL }%
\providecommand \Eprint [0]{\href }%
\providecommand \doibase [0]{https://doi.org/}%
\providecommand \selectlanguage [0]{\@gobble}%
\providecommand \bibinfo  [0]{\@secondoftwo}%
\providecommand \bibfield  [0]{\@secondoftwo}%
\providecommand \translation [1]{[#1]}%
\providecommand \BibitemOpen [0]{}%
\providecommand \bibitemStop [0]{}%
\providecommand \bibitemNoStop [0]{.\EOS\space}%
\providecommand \EOS [0]{\spacefactor3000\relax}%
\providecommand \BibitemShut  [1]{\csname bibitem#1\endcsname}%
\let\auto@bib@innerbib\@empty
\bibitem [{\citenamefont {Aliferis}\ and\ \citenamefont
  {Preskill}(2008)}]{Aliferis2008}%
  \BibitemOpen
  \bibfield  {author} {\bibinfo {author} {\bibfnamefont {P.}~\bibnamefont
  {Aliferis}}\ and\ \bibinfo {author} {\bibfnamefont {J.}~\bibnamefont
  {Preskill}},\ }\href {https://doi.org/10.1103/PhysRevA.78.052331} {\bibfield
  {journal} {\bibinfo  {journal} {Physical Review A}\ }\textbf {\bibinfo
  {volume} {78}},\ \bibinfo {pages} {052331} (\bibinfo {year}
  {2008})}\BibitemShut {NoStop}%
\bibitem [{\citenamefont {Aliferis}\ \emph {et~al.}(2009)\citenamefont
  {Aliferis}, \citenamefont {Brito}, \citenamefont {DiVincenzo}, \citenamefont
  {Preskill}, \citenamefont {Steffen},\ and\ \citenamefont
  {Terhal}}]{Aliferis2009}%
  \BibitemOpen
  \bibfield  {author} {\bibinfo {author} {\bibfnamefont {P.}~\bibnamefont
  {Aliferis}}, \bibinfo {author} {\bibfnamefont {F.}~\bibnamefont {Brito}},
  \bibinfo {author} {\bibfnamefont {D.~P.}\ \bibnamefont {DiVincenzo}},
  \bibinfo {author} {\bibfnamefont {J.}~\bibnamefont {Preskill}}, \bibinfo
  {author} {\bibfnamefont {M.}~\bibnamefont {Steffen}},\ and\ \bibinfo {author}
  {\bibfnamefont {B.~M.}\ \bibnamefont {Terhal}},\ }\href
  {https://doi.org/10.1088/1367-2630/11/1/013061} {\bibfield  {journal}
  {\bibinfo  {journal} {New Journal of Physics}\ }\textbf {\bibinfo {volume}
  {11}},\ \bibinfo {pages} {013061} (\bibinfo {year} {2009})}\BibitemShut
  {NoStop}%
\bibitem [{\citenamefont {Sarvepalli}\ \emph {et~al.}(2009)\citenamefont
  {Sarvepalli}, \citenamefont {Klappenecker},\ and\ \citenamefont
  {Rötteler}}]{Sarvepalli2009}%
  \BibitemOpen
  \bibfield  {author} {\bibinfo {author} {\bibfnamefont {P.~K.}\ \bibnamefont
  {Sarvepalli}}, \bibinfo {author} {\bibfnamefont {A.}~\bibnamefont
  {Klappenecker}},\ and\ \bibinfo {author} {\bibfnamefont {M.}~\bibnamefont
  {Rötteler}},\ }\href {https://doi.org/10.1098/rspa.2008.0439} {\bibfield
  {journal} {\bibinfo  {journal} {Proceedings of the Royal Society A:
  Mathematical, Physical and Engineering Sciences}\ }\textbf {\bibinfo {volume}
  {465}},\ \bibinfo {pages} {1645} (\bibinfo {year} {2009})}\BibitemShut
  {NoStop}%
\bibitem [{\citenamefont {Brooks}\ and\ \citenamefont
  {Preskill}(2013)}]{Brooks2013}%
  \BibitemOpen
  \bibfield  {author} {\bibinfo {author} {\bibfnamefont {P.}~\bibnamefont
  {Brooks}}\ and\ \bibinfo {author} {\bibfnamefont {J.}~\bibnamefont
  {Preskill}},\ }\href {https://doi.org/10.1103/PhysRevA.87.032310} {\bibfield
  {journal} {\bibinfo  {journal} {Physical Review A}\ }\textbf {\bibinfo
  {volume} {87}},\ \bibinfo {pages} {032310} (\bibinfo {year}
  {2013})}\BibitemShut {NoStop}%
\bibitem [{\citenamefont {Stephens}\ \emph {et~al.}(2013)\citenamefont
  {Stephens}, \citenamefont {Munro},\ and\ \citenamefont
  {Nemoto}}]{Stephens2013}%
  \BibitemOpen
  \bibfield  {author} {\bibinfo {author} {\bibfnamefont {A.~M.}\ \bibnamefont
  {Stephens}}, \bibinfo {author} {\bibfnamefont {W.~J.}\ \bibnamefont
  {Munro}},\ and\ \bibinfo {author} {\bibfnamefont {K.}~\bibnamefont
  {Nemoto}},\ }\href {https://doi.org/10.1103/PhysRevA.88.060301} {\bibfield
  {journal} {\bibinfo  {journal} {Physical Review A}\ }\textbf {\bibinfo
  {volume} {88}},\ \bibinfo {pages} {060301} (\bibinfo {year}
  {2013})}\BibitemShut {NoStop}%
\bibitem [{\citenamefont {Webster}\ \emph {et~al.}(2015)\citenamefont
  {Webster}, \citenamefont {Bartlett},\ and\ \citenamefont
  {Poulin}}]{Webster2015}%
  \BibitemOpen
  \bibfield  {author} {\bibinfo {author} {\bibfnamefont {P.}~\bibnamefont
  {Webster}}, \bibinfo {author} {\bibfnamefont {S.~D.}\ \bibnamefont
  {Bartlett}},\ and\ \bibinfo {author} {\bibfnamefont {D.}~\bibnamefont
  {Poulin}},\ }\href {https://doi.org/10.1103/PhysRevA.92.062309} {\bibfield
  {journal} {\bibinfo  {journal} {Physical Review A}\ }\textbf {\bibinfo
  {volume} {92}},\ \bibinfo {pages} {062309} (\bibinfo {year}
  {2015})}\BibitemShut {NoStop}%
\bibitem [{\citenamefont {Gefen}\ \emph {et~al.}(2017)\citenamefont {Gefen},
  \citenamefont {Cohen}, \citenamefont {Cohen},\ and\ \citenamefont
  {Retzker}}]{gefen2017enhancing}%
  \BibitemOpen
  \bibfield  {author} {\bibinfo {author} {\bibfnamefont {T.}~\bibnamefont
  {Gefen}}, \bibinfo {author} {\bibfnamefont {D.}~\bibnamefont {Cohen}},
  \bibinfo {author} {\bibfnamefont {I.}~\bibnamefont {Cohen}},\ and\ \bibinfo
  {author} {\bibfnamefont {A.}~\bibnamefont {Retzker}},\ }\href
  {https://doi.org/10.1103/PhysRevA.95.032314} {\bibfield  {journal} {\bibinfo
  {journal} {Physical Review A}\ }\textbf {\bibinfo {volume} {95}},\ \bibinfo
  {pages} {032314} (\bibinfo {year} {2017})}\BibitemShut {NoStop}%
\bibitem [{\citenamefont {Tuckett}\ \emph {et~al.}(2019)\citenamefont
  {Tuckett}, \citenamefont {Darmawan}, \citenamefont {Chubb}, \citenamefont
  {Bravyi}, \citenamefont {Bartlett},\ and\ \citenamefont
  {Flammia}}]{tuckett2019tailoring}%
  \BibitemOpen
  \bibfield  {author} {\bibinfo {author} {\bibfnamefont {D.~K.}\ \bibnamefont
  {Tuckett}}, \bibinfo {author} {\bibfnamefont {A.~S.}\ \bibnamefont
  {Darmawan}}, \bibinfo {author} {\bibfnamefont {C.~T.}\ \bibnamefont {Chubb}},
  \bibinfo {author} {\bibfnamefont {S.}~\bibnamefont {Bravyi}}, \bibinfo
  {author} {\bibfnamefont {S.~D.}\ \bibnamefont {Bartlett}},\ and\ \bibinfo
  {author} {\bibfnamefont {S.~T.}\ \bibnamefont {Flammia}},\ }\href
  {https://doi.org/10.1103/PhysRevX.9.041031} {\bibfield  {journal} {\bibinfo
  {journal} {Physical Review X}\ }\textbf {\bibinfo {volume} {9}},\ \bibinfo
  {pages} {041031} (\bibinfo {year} {2019})}\BibitemShut {NoStop}%
\bibitem [{\citenamefont {Tuckett}\ \emph {et~al.}(2018)\citenamefont
  {Tuckett}, \citenamefont {Bartlett},\ and\ \citenamefont
  {Flammia}}]{tuckett2018ultrahigh}%
  \BibitemOpen
  \bibfield  {author} {\bibinfo {author} {\bibfnamefont {D.~K.}\ \bibnamefont
  {Tuckett}}, \bibinfo {author} {\bibfnamefont {S.~D.}\ \bibnamefont
  {Bartlett}},\ and\ \bibinfo {author} {\bibfnamefont {S.~T.}\ \bibnamefont
  {Flammia}},\ }\href {https://doi.org/10.1103/PhysRevLett.120.050505}
  {\bibfield  {journal} {\bibinfo  {journal} {Physical Review Letters}\
  }\textbf {\bibinfo {volume} {120}},\ \bibinfo {pages} {050505} (\bibinfo
  {year} {2018})}\BibitemShut {NoStop}%
\bibitem [{\citenamefont {Tuckett}\ \emph {et~al.}(2020)\citenamefont
  {Tuckett}, \citenamefont {Bartlett}, \citenamefont {Flammia},\ and\
  \citenamefont {Brown}}]{tuckett2020fault}%
  \BibitemOpen
  \bibfield  {author} {\bibinfo {author} {\bibfnamefont {D.~K.}\ \bibnamefont
  {Tuckett}}, \bibinfo {author} {\bibfnamefont {S.~D.}\ \bibnamefont
  {Bartlett}}, \bibinfo {author} {\bibfnamefont {S.~T.}\ \bibnamefont
  {Flammia}},\ and\ \bibinfo {author} {\bibfnamefont {B.~J.}\ \bibnamefont
  {Brown}},\ }\href {https://doi.org/10.1103/PhysRevLett.124.130501} {\bibfield
   {journal} {\bibinfo  {journal} {Physical Review Letters}\ }\textbf {\bibinfo
  {volume} {124}},\ \bibinfo {pages} {130501} (\bibinfo {year}
  {2020})}\BibitemShut {NoStop}%
\bibitem [{\citenamefont {Bonilla~Ataides}\ \emph {et~al.}(2021)\citenamefont
  {Bonilla~Ataides}, \citenamefont {Tuckett}, \citenamefont {Bartlett},
  \citenamefont {Flammia},\ and\ \citenamefont {Brown}}]{bonilla2021xzzx}%
  \BibitemOpen
  \bibfield  {author} {\bibinfo {author} {\bibfnamefont {J.~P.}\ \bibnamefont
  {Bonilla~Ataides}}, \bibinfo {author} {\bibfnamefont {D.~K.}\ \bibnamefont
  {Tuckett}}, \bibinfo {author} {\bibfnamefont {S.~D.}\ \bibnamefont
  {Bartlett}}, \bibinfo {author} {\bibfnamefont {S.~T.}\ \bibnamefont
  {Flammia}},\ and\ \bibinfo {author} {\bibfnamefont {B.~J.}\ \bibnamefont
  {Brown}},\ }\href {https://doi.org/10.1038/s41467-021-22274-1} {\bibfield
  {journal} {\bibinfo  {journal} {Nature Communications}\ }\textbf {\bibinfo
  {volume} {12}},\ \bibinfo {pages} {1} (\bibinfo {year} {2021})}\BibitemShut
  {NoStop}%
\bibitem [{\citenamefont {Dua}\ \emph {et~al.}(2022)\citenamefont {Dua},
  \citenamefont {Kubica}, \citenamefont {Jiang}, \citenamefont {Flammia},\ and\
  \citenamefont {Gullans}}]{Dua2022}%
  \BibitemOpen
  \bibfield  {author} {\bibinfo {author} {\bibfnamefont {A.}~\bibnamefont
  {Dua}}, \bibinfo {author} {\bibfnamefont {A.}~\bibnamefont {Kubica}},
  \bibinfo {author} {\bibfnamefont {L.}~\bibnamefont {Jiang}}, \bibinfo
  {author} {\bibfnamefont {S.~T.}\ \bibnamefont {Flammia}},\ and\ \bibinfo
  {author} {\bibfnamefont {M.~J.}\ \bibnamefont {Gullans}},\ }\href
  {https://doi.org/10.48550/ARXIV.2201.07802} {\bibinfo {title}
  {Clifford-deformed surface codes}},\ \bibinfo {howpublished}
  {arXiv:2201.07802} (\bibinfo {year} {2022})\BibitemShut {NoStop}%
\bibitem [{\citenamefont {Higgott}\ \emph {et~al.}(2022)\citenamefont
  {Higgott}, \citenamefont {Bohdanowicz}, \citenamefont {Kubica}, \citenamefont
  {Flammia},\ and\ \citenamefont {Campbell}}]{Higgott2022}%
  \BibitemOpen
  \bibfield  {author} {\bibinfo {author} {\bibfnamefont {O.}~\bibnamefont
  {Higgott}}, \bibinfo {author} {\bibfnamefont {T.~C.}\ \bibnamefont
  {Bohdanowicz}}, \bibinfo {author} {\bibfnamefont {A.}~\bibnamefont {Kubica}},
  \bibinfo {author} {\bibfnamefont {S.~T.}\ \bibnamefont {Flammia}},\ and\
  \bibinfo {author} {\bibfnamefont {E.~T.}\ \bibnamefont {Campbell}},\ }\href
  {https://doi.org/10.48550/ARXIV.2203.04948} {\bibinfo {title} {Fragile
  boundaries of tailored surface codes}},\ \bibinfo {howpublished}
  {arXiv:2203.04948} (\bibinfo {year} {2022})\BibitemShut {NoStop}%
\bibitem [{\citenamefont {Xu}\ \emph {et~al.}(2022)\citenamefont {Xu},
  \citenamefont {Mannucci}, \citenamefont {Seif}, \citenamefont {Kubica},
  \citenamefont {Flammia},\ and\ \citenamefont {Jiang}}]{Xu2022}%
  \BibitemOpen
  \bibfield  {author} {\bibinfo {author} {\bibfnamefont {Q.}~\bibnamefont
  {Xu}}, \bibinfo {author} {\bibfnamefont {N.}~\bibnamefont {Mannucci}},
  \bibinfo {author} {\bibfnamefont {A.}~\bibnamefont {Seif}}, \bibinfo {author}
  {\bibfnamefont {A.}~\bibnamefont {Kubica}}, \bibinfo {author} {\bibfnamefont
  {S.~T.}\ \bibnamefont {Flammia}},\ and\ \bibinfo {author} {\bibfnamefont
  {L.}~\bibnamefont {Jiang}},\ }\href
  {https://doi.org/10.48550/ARXIV.2203.16486} {\bibinfo {title} {Tailored xzzx
  codes for biased noise}},\ \bibinfo {howpublished} {arXiv:2203.16486}
  (\bibinfo {year} {2022})\BibitemShut {NoStop}%
\bibitem [{\citenamefont {Cochrane}\ \emph {et~al.}(1999)\citenamefont
  {Cochrane}, \citenamefont {Milburn},\ and\ \citenamefont
  {Munro}}]{cochrane1999macroscopically}%
  \BibitemOpen
  \bibfield  {author} {\bibinfo {author} {\bibfnamefont {P.~T.}\ \bibnamefont
  {Cochrane}}, \bibinfo {author} {\bibfnamefont {G.~J.}\ \bibnamefont
  {Milburn}},\ and\ \bibinfo {author} {\bibfnamefont {W.~J.}\ \bibnamefont
  {Munro}},\ }\href {https://doi.org/10.1103/PhysRevA.59.2631} {\bibfield
  {journal} {\bibinfo  {journal} {Physical Review A}\ }\textbf {\bibinfo
  {volume} {59}},\ \bibinfo {pages} {2631} (\bibinfo {year}
  {1999})}\BibitemShut {NoStop}%
\bibitem [{\citenamefont {Mirrahimi}\ \emph {et~al.}(2014)\citenamefont
  {Mirrahimi}, \citenamefont {Leghtas}, \citenamefont {Albert}, \citenamefont
  {Touzard}, \citenamefont {Schoelkopf}, \citenamefont {Jiang},\ and\
  \citenamefont {Devoret}}]{mirrahimi2014dynamically}%
  \BibitemOpen
  \bibfield  {author} {\bibinfo {author} {\bibfnamefont {M.}~\bibnamefont
  {Mirrahimi}}, \bibinfo {author} {\bibfnamefont {Z.}~\bibnamefont {Leghtas}},
  \bibinfo {author} {\bibfnamefont {V.~V.}\ \bibnamefont {Albert}}, \bibinfo
  {author} {\bibfnamefont {S.}~\bibnamefont {Touzard}}, \bibinfo {author}
  {\bibfnamefont {R.~J.}\ \bibnamefont {Schoelkopf}}, \bibinfo {author}
  {\bibfnamefont {L.}~\bibnamefont {Jiang}},\ and\ \bibinfo {author}
  {\bibfnamefont {M.~H.}\ \bibnamefont {Devoret}},\ }\href
  {https://doi.org/10.1088/1367-2630/16/4/045014} {\bibfield  {journal}
  {\bibinfo  {journal} {New Journal of Physics}\ }\textbf {\bibinfo {volume}
  {16}},\ \bibinfo {pages} {045014} (\bibinfo {year} {2014})}\BibitemShut
  {NoStop}%
\bibitem [{\citenamefont {Ofek}\ \emph {et~al.}(2016)\citenamefont {Ofek},
  \citenamefont {Petrenko}, \citenamefont {Heeres}, \citenamefont {Reinhold},
  \citenamefont {Leghtas}, \citenamefont {Vlastakis}, \citenamefont {Liu},
  \citenamefont {Frunzio}, \citenamefont {Girvin}, \citenamefont {Jiang} \emph
  {et~al.}}]{ofek2016extending}%
  \BibitemOpen
  \bibfield  {author} {\bibinfo {author} {\bibfnamefont {N.}~\bibnamefont
  {Ofek}}, \bibinfo {author} {\bibfnamefont {A.}~\bibnamefont {Petrenko}},
  \bibinfo {author} {\bibfnamefont {R.}~\bibnamefont {Heeres}}, \bibinfo
  {author} {\bibfnamefont {P.}~\bibnamefont {Reinhold}}, \bibinfo {author}
  {\bibfnamefont {Z.}~\bibnamefont {Leghtas}}, \bibinfo {author} {\bibfnamefont
  {B.}~\bibnamefont {Vlastakis}}, \bibinfo {author} {\bibfnamefont
  {Y.}~\bibnamefont {Liu}}, \bibinfo {author} {\bibfnamefont {L.}~\bibnamefont
  {Frunzio}}, \bibinfo {author} {\bibfnamefont {S.}~\bibnamefont {Girvin}},
  \bibinfo {author} {\bibfnamefont {L.}~\bibnamefont {Jiang}}, \emph {et~al.},\
  }\href {https://doi.org/10.1038/nature18949} {\bibfield  {journal} {\bibinfo
  {journal} {Nature}\ }\textbf {\bibinfo {volume} {536}},\ \bibinfo {pages}
  {441} (\bibinfo {year} {2016})}\BibitemShut {NoStop}%
\bibitem [{\citenamefont {Puri}\ \emph {et~al.}(2019)\citenamefont {Puri},
  \citenamefont {Grimm}, \citenamefont {Campagne-Ibarcq}, \citenamefont
  {Eickbusch}, \citenamefont {Noh}, \citenamefont {Roberts}, \citenamefont
  {Jiang}, \citenamefont {Mirrahimi}, \citenamefont {Devoret},\ and\
  \citenamefont {Girvin}}]{puri2019stabilized}%
  \BibitemOpen
  \bibfield  {author} {\bibinfo {author} {\bibfnamefont {S.}~\bibnamefont
  {Puri}}, \bibinfo {author} {\bibfnamefont {A.}~\bibnamefont {Grimm}},
  \bibinfo {author} {\bibfnamefont {P.}~\bibnamefont {Campagne-Ibarcq}},
  \bibinfo {author} {\bibfnamefont {A.}~\bibnamefont {Eickbusch}}, \bibinfo
  {author} {\bibfnamefont {K.}~\bibnamefont {Noh}}, \bibinfo {author}
  {\bibfnamefont {G.}~\bibnamefont {Roberts}}, \bibinfo {author} {\bibfnamefont
  {L.}~\bibnamefont {Jiang}}, \bibinfo {author} {\bibfnamefont
  {M.}~\bibnamefont {Mirrahimi}}, \bibinfo {author} {\bibfnamefont {M.~H.}\
  \bibnamefont {Devoret}},\ and\ \bibinfo {author} {\bibfnamefont {S.~M.}\
  \bibnamefont {Girvin}},\ }\href {https://doi.org/10.1103/PhysRevX.9.041009}
  {\bibfield  {journal} {\bibinfo  {journal} {Physical Review X}\ }\textbf
  {\bibinfo {volume} {9}},\ \bibinfo {pages} {041009} (\bibinfo {year}
  {2019})}\BibitemShut {NoStop}%
\bibitem [{\citenamefont {Guillaud}\ and\ \citenamefont
  {Mirrahimi}(2019)}]{guillaud2019repetition}%
  \BibitemOpen
  \bibfield  {author} {\bibinfo {author} {\bibfnamefont {J.}~\bibnamefont
  {Guillaud}}\ and\ \bibinfo {author} {\bibfnamefont {M.}~\bibnamefont
  {Mirrahimi}},\ }\href {https://doi.org/10.1103/PhysRevX.9.041053} {\bibfield
  {journal} {\bibinfo  {journal} {Phys. Rev. X}\ }\textbf {\bibinfo {volume}
  {9}},\ \bibinfo {pages} {041053} (\bibinfo {year} {2019})}\BibitemShut
  {NoStop}%
\bibitem [{\citenamefont {Yan}\ \emph {et~al.}(2018)\citenamefont {Yan},
  \citenamefont {Campbell}, \citenamefont {Krantz}, \citenamefont {Kjaergaard},
  \citenamefont {Kim}, \citenamefont {Yoder}, \citenamefont {Hover},
  \citenamefont {Sears}, \citenamefont {Kerman}, \citenamefont {Orlando},
  \citenamefont {Gustavsson},\ and\ \citenamefont
  {Oliver}}]{PhysRevLett.120.260504}%
  \BibitemOpen
  \bibfield  {author} {\bibinfo {author} {\bibfnamefont {F.}~\bibnamefont
  {Yan}}, \bibinfo {author} {\bibfnamefont {D.}~\bibnamefont {Campbell}},
  \bibinfo {author} {\bibfnamefont {P.}~\bibnamefont {Krantz}}, \bibinfo
  {author} {\bibfnamefont {M.}~\bibnamefont {Kjaergaard}}, \bibinfo {author}
  {\bibfnamefont {D.}~\bibnamefont {Kim}}, \bibinfo {author} {\bibfnamefont
  {J.~L.}\ \bibnamefont {Yoder}}, \bibinfo {author} {\bibfnamefont
  {D.}~\bibnamefont {Hover}}, \bibinfo {author} {\bibfnamefont
  {A.}~\bibnamefont {Sears}}, \bibinfo {author} {\bibfnamefont {A.~J.}\
  \bibnamefont {Kerman}}, \bibinfo {author} {\bibfnamefont {T.~P.}\
  \bibnamefont {Orlando}}, \bibinfo {author} {\bibfnamefont {S.}~\bibnamefont
  {Gustavsson}},\ and\ \bibinfo {author} {\bibfnamefont {W.~D.}\ \bibnamefont
  {Oliver}},\ }\href {https://doi.org/10.1103/PhysRevLett.120.260504}
  {\bibfield  {journal} {\bibinfo  {journal} {Physical Review Letters}\
  }\textbf {\bibinfo {volume} {120}},\ \bibinfo {pages} {260504} (\bibinfo
  {year} {2018})}\BibitemShut {NoStop}%
\bibitem [{\citenamefont {Burnett}\ \emph {et~al.}(2019)\citenamefont
  {Burnett}, \citenamefont {Bengtsson}, \citenamefont {Scigliuzzo},
  \citenamefont {Niepce}, \citenamefont {Kudra}, \citenamefont {Delsing},\ and\
  \citenamefont {Bylander}}]{burnett2019decoherence}%
  \BibitemOpen
  \bibfield  {author} {\bibinfo {author} {\bibfnamefont {J.~J.}\ \bibnamefont
  {Burnett}}, \bibinfo {author} {\bibfnamefont {A.}~\bibnamefont {Bengtsson}},
  \bibinfo {author} {\bibfnamefont {M.}~\bibnamefont {Scigliuzzo}}, \bibinfo
  {author} {\bibfnamefont {D.}~\bibnamefont {Niepce}}, \bibinfo {author}
  {\bibfnamefont {M.}~\bibnamefont {Kudra}}, \bibinfo {author} {\bibfnamefont
  {P.}~\bibnamefont {Delsing}},\ and\ \bibinfo {author} {\bibfnamefont
  {J.}~\bibnamefont {Bylander}},\ }\href
  {https://doi.org/10.1038/s41534-019-0168-5} {\bibfield  {journal} {\bibinfo
  {journal} {npj Quantum Information}\ }\textbf {\bibinfo {volume} {5}},\
  \bibinfo {pages} {1} (\bibinfo {year} {2019})}\BibitemShut {NoStop}%
\bibitem [{\citenamefont {Cao}\ \emph {et~al.}(2020)\citenamefont {Cao},
  \citenamefont {Yang}, \citenamefont {Gong}, \citenamefont {Yu}, \citenamefont
  {Retzker}, \citenamefont {Plenio}, \citenamefont {M{\"u}ller}, \citenamefont
  {Tomek}, \citenamefont {Naydenov}, \citenamefont {McGuinness} \emph
  {et~al.}}]{cao2020protecting}%
  \BibitemOpen
  \bibfield  {author} {\bibinfo {author} {\bibfnamefont {Q.-Y.}\ \bibnamefont
  {Cao}}, \bibinfo {author} {\bibfnamefont {P.-C.}\ \bibnamefont {Yang}},
  \bibinfo {author} {\bibfnamefont {M.-S.}\ \bibnamefont {Gong}}, \bibinfo
  {author} {\bibfnamefont {M.}~\bibnamefont {Yu}}, \bibinfo {author}
  {\bibfnamefont {A.}~\bibnamefont {Retzker}}, \bibinfo {author} {\bibfnamefont
  {M.~B.}\ \bibnamefont {Plenio}}, \bibinfo {author} {\bibfnamefont
  {C.}~\bibnamefont {M{\"u}ller}}, \bibinfo {author} {\bibfnamefont
  {N.}~\bibnamefont {Tomek}}, \bibinfo {author} {\bibfnamefont
  {B.}~\bibnamefont {Naydenov}}, \bibinfo {author} {\bibfnamefont
  {L.}~\bibnamefont {McGuinness}}, \emph {et~al.},\ }\href
  {https://doi.org/10.1103/PhysRevApplied.13.024021} {\bibfield  {journal}
  {\bibinfo  {journal} {Physical Review Applied}\ }\textbf {\bibinfo {volume}
  {13}},\ \bibinfo {pages} {024021} (\bibinfo {year} {2020})}\BibitemShut
  {NoStop}%
\bibitem [{\citenamefont {Balasubramanian}\ \emph {et~al.}(2009)\citenamefont
  {Balasubramanian}, \citenamefont {Neumann}, \citenamefont {Twitchen},
  \citenamefont {Markham}, \citenamefont {Kolesov}, \citenamefont {Mizuochi},
  \citenamefont {Isoya}, \citenamefont {Achard}, \citenamefont {Beck},
  \citenamefont {Tissler}, \citenamefont {Jacques}, \citenamefont {Hemmer},
  \citenamefont {Jelezko},\ and\ \citenamefont
  {Wrachtrup}}]{balasubramanian2009ultralong}%
  \BibitemOpen
  \bibfield  {author} {\bibinfo {author} {\bibfnamefont {G.}~\bibnamefont
  {Balasubramanian}}, \bibinfo {author} {\bibfnamefont {P.}~\bibnamefont
  {Neumann}}, \bibinfo {author} {\bibfnamefont {D.}~\bibnamefont {Twitchen}},
  \bibinfo {author} {\bibfnamefont {M.}~\bibnamefont {Markham}}, \bibinfo
  {author} {\bibfnamefont {R.}~\bibnamefont {Kolesov}}, \bibinfo {author}
  {\bibfnamefont {N.}~\bibnamefont {Mizuochi}}, \bibinfo {author}
  {\bibfnamefont {J.}~\bibnamefont {Isoya}}, \bibinfo {author} {\bibfnamefont
  {J.}~\bibnamefont {Achard}}, \bibinfo {author} {\bibfnamefont
  {J.}~\bibnamefont {Beck}}, \bibinfo {author} {\bibfnamefont {J.}~\bibnamefont
  {Tissler}}, \bibinfo {author} {\bibfnamefont {V.}~\bibnamefont {Jacques}},
  \bibinfo {author} {\bibfnamefont {P.}~\bibnamefont {Hemmer}}, \bibinfo
  {author} {\bibfnamefont {F.}~\bibnamefont {Jelezko}},\ and\ \bibinfo {author}
  {\bibfnamefont {J.}~\bibnamefont {Wrachtrup}},\ }\href
  {https://doi.org/10.1038/nmat2420} {\bibfield  {journal} {\bibinfo  {journal}
  {Nature Materials}\ }\textbf {\bibinfo {volume} {8}},\ \bibinfo {pages} {383}
  (\bibinfo {year} {2009})}\BibitemShut {NoStop}%
\bibitem [{\citenamefont {Herbschleb}\ \emph {et~al.}(2019)\citenamefont
  {Herbschleb}, \citenamefont {Kato}, \citenamefont {Maruyama}, \citenamefont
  {Danjo}, \citenamefont {Makino}, \citenamefont {Yamasaki}, \citenamefont
  {Ohki}, \citenamefont {Hayashi}, \citenamefont {Morishita}, \citenamefont
  {Fujiwara} \emph {et~al.}}]{herbschleb2019ultra}%
  \BibitemOpen
  \bibfield  {author} {\bibinfo {author} {\bibfnamefont {E.}~\bibnamefont
  {Herbschleb}}, \bibinfo {author} {\bibfnamefont {H.}~\bibnamefont {Kato}},
  \bibinfo {author} {\bibfnamefont {Y.}~\bibnamefont {Maruyama}}, \bibinfo
  {author} {\bibfnamefont {T.}~\bibnamefont {Danjo}}, \bibinfo {author}
  {\bibfnamefont {T.}~\bibnamefont {Makino}}, \bibinfo {author} {\bibfnamefont
  {S.}~\bibnamefont {Yamasaki}}, \bibinfo {author} {\bibfnamefont
  {I.}~\bibnamefont {Ohki}}, \bibinfo {author} {\bibfnamefont {K.}~\bibnamefont
  {Hayashi}}, \bibinfo {author} {\bibfnamefont {H.}~\bibnamefont {Morishita}},
  \bibinfo {author} {\bibfnamefont {M.}~\bibnamefont {Fujiwara}}, \emph
  {et~al.},\ }\href {https://doi.org/10.1038/s41467-019-11776-8} {\bibfield
  {journal} {\bibinfo  {journal} {Nature Communications}\ }\textbf {\bibinfo
  {volume} {10}},\ \bibinfo {pages} {1} (\bibinfo {year} {2019})}\BibitemShut
  {NoStop}%
\bibitem [{\citenamefont {Sosnova}(2020)}]{Sosnova_Thesis}%
  \BibitemOpen
  \bibfield  {author} {\bibinfo {author} {\bibfnamefont {K.}~\bibnamefont
  {Sosnova}},\ }\emph {\bibinfo {title} {Mixed-species ion chains for quantum
  networks}},\ \href {https://doi.org/10.13016/jjsc-jlae} {Ph.D. thesis},\
  \bibinfo  {school} {University of Maryland, College Park} (\bibinfo {year}
  {2020})\BibitemShut {NoStop}%
\bibitem [{\citenamefont {Leung}\ \emph {et~al.}(1997)\citenamefont {Leung},
  \citenamefont {Nielsen}, \citenamefont {Chuang},\ and\ \citenamefont
  {Yamamoto}}]{Leung1997}%
  \BibitemOpen
  \bibfield  {author} {\bibinfo {author} {\bibfnamefont {D.~W.}\ \bibnamefont
  {Leung}}, \bibinfo {author} {\bibfnamefont {M.~A.}\ \bibnamefont {Nielsen}},
  \bibinfo {author} {\bibfnamefont {I.~L.}\ \bibnamefont {Chuang}},\ and\
  \bibinfo {author} {\bibfnamefont {Y.}~\bibnamefont {Yamamoto}},\ }\href
  {https://doi.org/10.1103/PhysRevA.56.2567} {\bibfield  {journal} {\bibinfo
  {journal} {Physical Review A}\ }\textbf {\bibinfo {volume} {56}},\ \bibinfo
  {pages} {2567} (\bibinfo {year} {1997})}\BibitemShut {NoStop}%
\bibitem [{\citenamefont {Laflamme}\ \emph {et~al.}(1996)\citenamefont
  {Laflamme}, \citenamefont {Miquel}, \citenamefont {Paz},\ and\ \citenamefont
  {Zurek}}]{Laflamme1996}%
  \BibitemOpen
  \bibfield  {author} {\bibinfo {author} {\bibfnamefont {R.}~\bibnamefont
  {Laflamme}}, \bibinfo {author} {\bibfnamefont {C.}~\bibnamefont {Miquel}},
  \bibinfo {author} {\bibfnamefont {J.~P.}\ \bibnamefont {Paz}},\ and\ \bibinfo
  {author} {\bibfnamefont {W.~H.}\ \bibnamefont {Zurek}},\ }\href
  {https://doi.org/10.48550/ARXIV.QUANT-PH/9602019} {\bibinfo {title} {Perfect
  quantum error correction code}},\ \bibinfo {howpublished}
  {arXiv:quant-ph/9602019} (\bibinfo {year} {1996})\BibitemShut {NoStop}%
\bibitem [{\citenamefont {Kitaev}(2003)}]{kitaev2003fault}%
  \BibitemOpen
  \bibfield  {author} {\bibinfo {author} {\bibfnamefont {A.~Y.}\ \bibnamefont
  {Kitaev}},\ }\href {https://doi.org/10.1016/S0003-4916(02)00018-0} {\bibfield
   {journal} {\bibinfo  {journal} {Annals of Physics}\ }\textbf {\bibinfo
  {volume} {303}},\ \bibinfo {pages} {2} (\bibinfo {year} {2003})}\BibitemShut
  {NoStop}%
\bibitem [{\citenamefont {Bravyi}\ and\ \citenamefont
  {Kitaev}(1998)}]{bravyi1998}%
  \BibitemOpen
  \bibfield  {author} {\bibinfo {author} {\bibfnamefont {S.~B.}\ \bibnamefont
  {Bravyi}}\ and\ \bibinfo {author} {\bibfnamefont {A.~Y.}\ \bibnamefont
  {Kitaev}},\ }\href {https://doi.org/10.48550/ARXIV.QUANT-PH/9811052}
  {\bibinfo {title} {Quantum codes on a lattice with boundary}},\ \bibinfo
  {howpublished} {arXiv:quant-ph/9811052} (\bibinfo {year} {1998})\BibitemShut
  {NoStop}%
\bibitem [{\citenamefont {Darmawan}\ and\ \citenamefont
  {Poulin}(2017)}]{Darmawan2017}%
  \BibitemOpen
  \bibfield  {author} {\bibinfo {author} {\bibfnamefont {A.~S.}\ \bibnamefont
  {Darmawan}}\ and\ \bibinfo {author} {\bibfnamefont {D.}~\bibnamefont
  {Poulin}},\ }\href {https://doi.org/10.1103/PhysRevLett.119.040502}
  {\bibfield  {journal} {\bibinfo  {journal} {Physical Review Letters}\
  }\textbf {\bibinfo {volume} {119}},\ \bibinfo {pages} {040502} (\bibinfo
  {year} {2017})}\BibitemShut {NoStop}%
\bibitem [{\citenamefont {Bombin}\ \emph {et~al.}(2012)\citenamefont {Bombin},
  \citenamefont {Andrist}, \citenamefont {Ohzeki}, \citenamefont {Katzgraber},\
  and\ \citenamefont {Martin-Delgado}}]{Bombin2012}%
  \BibitemOpen
  \bibfield  {author} {\bibinfo {author} {\bibfnamefont {H.}~\bibnamefont
  {Bombin}}, \bibinfo {author} {\bibfnamefont {R.~S.}\ \bibnamefont {Andrist}},
  \bibinfo {author} {\bibfnamefont {M.}~\bibnamefont {Ohzeki}}, \bibinfo
  {author} {\bibfnamefont {H.~G.}\ \bibnamefont {Katzgraber}},\ and\ \bibinfo
  {author} {\bibfnamefont {M.~A.}\ \bibnamefont {Martin-Delgado}},\ }\href
  {https://doi.org/10.1103/PhysRevX.2.021004} {\bibfield  {journal} {\bibinfo
  {journal} {Physical Review X}\ }\textbf {\bibinfo {volume} {2}},\ \bibinfo
  {pages} {021004} (\bibinfo {year} {2012})}\BibitemShut {NoStop}%
\bibitem [{\citenamefont {Grassl}\ \emph {et~al.}(1997)\citenamefont {Grassl},
  \citenamefont {Beth},\ and\ \citenamefont {Pellizzari}}]{Grassl1997}%
  \BibitemOpen
  \bibfield  {author} {\bibinfo {author} {\bibfnamefont {M.}~\bibnamefont
  {Grassl}}, \bibinfo {author} {\bibfnamefont {T.}~\bibnamefont {Beth}},\ and\
  \bibinfo {author} {\bibfnamefont {T.}~\bibnamefont {Pellizzari}},\ }\href
  {https://doi.org/10.1103/PhysRevA.56.33} {\bibfield  {journal} {\bibinfo
  {journal} {Physical Review A}\ }\textbf {\bibinfo {volume} {56}},\ \bibinfo
  {pages} {33} (\bibinfo {year} {1997})}\BibitemShut {NoStop}%
\bibitem [{\citenamefont {Stace}\ \emph {et~al.}(2009)\citenamefont {Stace},
  \citenamefont {Barrett},\ and\ \citenamefont {Doherty}}]{Stace2009}%
  \BibitemOpen
  \bibfield  {author} {\bibinfo {author} {\bibfnamefont {T.~M.}\ \bibnamefont
  {Stace}}, \bibinfo {author} {\bibfnamefont {S.~D.}\ \bibnamefont {Barrett}},\
  and\ \bibinfo {author} {\bibfnamefont {A.~C.}\ \bibnamefont {Doherty}},\
  }\href {https://doi.org/10.1103/PhysRevLett.102.200501} {\bibfield  {journal}
  {\bibinfo  {journal} {Physical Review Letters}\ }\textbf {\bibinfo {volume}
  {102}},\ \bibinfo {pages} {200501} (\bibinfo {year} {2009})}\BibitemShut
  {NoStop}%
\bibitem [{\citenamefont {Dennis}\ \emph {et~al.}(2002)\citenamefont {Dennis},
  \citenamefont {Kitaev}, \citenamefont {Landahl},\ and\ \citenamefont
  {Preskill}}]{Dennis2002}%
  \BibitemOpen
  \bibfield  {author} {\bibinfo {author} {\bibfnamefont {E.}~\bibnamefont
  {Dennis}}, \bibinfo {author} {\bibfnamefont {A.}~\bibnamefont {Kitaev}},
  \bibinfo {author} {\bibfnamefont {A.}~\bibnamefont {Landahl}},\ and\ \bibinfo
  {author} {\bibfnamefont {J.}~\bibnamefont {Preskill}},\ }\href
  {https://doi.org/10.1063/1.1499754} {\bibfield  {journal} {\bibinfo
  {journal} {Journal of Mathematical Physics}\ }\textbf {\bibinfo {volume}
  {43}},\ \bibinfo {pages} {4452} (\bibinfo {year} {2002})}\BibitemShut
  {NoStop}%
\bibitem [{\citenamefont {Fowler}\ \emph {et~al.}(2012)\citenamefont {Fowler},
  \citenamefont {Mariantoni}, \citenamefont {Martinis},\ and\ \citenamefont
  {Cleland}}]{Fowler2012}%
  \BibitemOpen
  \bibfield  {author} {\bibinfo {author} {\bibfnamefont {A.~G.}\ \bibnamefont
  {Fowler}}, \bibinfo {author} {\bibfnamefont {M.}~\bibnamefont {Mariantoni}},
  \bibinfo {author} {\bibfnamefont {J.~M.}\ \bibnamefont {Martinis}},\ and\
  \bibinfo {author} {\bibfnamefont {A.~N.}\ \bibnamefont {Cleland}},\ }\href
  {https://doi.org/10.1103/PhysRevA.86.032324} {\bibfield  {journal} {\bibinfo
  {journal} {Physical Review A}\ }\textbf {\bibinfo {volume} {86}},\ \bibinfo
  {pages} {032324} (\bibinfo {year} {2012})}\BibitemShut {NoStop}%
\bibitem [{Note1()}]{Note1}%
  \BibitemOpen
  \bibinfo {note} {Note that a fully-depolarizing two-qubit Pauli channel has
  error rate of $15/16$.}\BibitemShut {Stop}%
\bibitem [{\citenamefont {Duan}\ \emph {et~al.}(2010)\citenamefont {Duan},
  \citenamefont {Grassl}, \citenamefont {Ji},\ and\ \citenamefont
  {Zeng}}]{Duan2010}%
  \BibitemOpen
  \bibfield  {author} {\bibinfo {author} {\bibfnamefont {R.}~\bibnamefont
  {Duan}}, \bibinfo {author} {\bibfnamefont {M.}~\bibnamefont {Grassl}},
  \bibinfo {author} {\bibfnamefont {Z.}~\bibnamefont {Ji}},\ and\ \bibinfo
  {author} {\bibfnamefont {B.}~\bibnamefont {Zeng}},\ }in\ \href
  {https://doi.org/10.1109/ISIT.2010.5513648} {\emph {\bibinfo {booktitle}
  {2010 IEEE International Symposium on Information Theory}}}\ (\bibinfo {year}
  {2010})\ pp.\ \bibinfo {pages} {2672--2676}\BibitemShut {NoStop}%
\bibitem [{\citenamefont {Shim}\ and\ \citenamefont
  {Tahan}(2016)}]{Shim2016semiconductor}%
  \BibitemOpen
  \bibfield  {author} {\bibinfo {author} {\bibfnamefont {Y.-P.}\ \bibnamefont
  {Shim}}\ and\ \bibinfo {author} {\bibfnamefont {C.}~\bibnamefont {Tahan}},\
  }\href {https://doi.org/10.1038/ncomms11059} {\bibfield  {journal} {\bibinfo
  {journal} {Nature Communications}\ }\textbf {\bibinfo {volume} {7}},\
  \bibinfo {pages} {1} (\bibinfo {year} {2016})}\BibitemShut {NoStop}%
\bibitem [{\citenamefont {Campbell}\ \emph {et~al.}(2020)\citenamefont
  {Campbell}, \citenamefont {Shim}, \citenamefont {Kannan}, \citenamefont
  {Winik}, \citenamefont {Kim}, \citenamefont {Melville}, \citenamefont
  {Niedzielski}, \citenamefont {Yoder}, \citenamefont {Tahan}, \citenamefont
  {Gustavsson} \emph {et~al.}}]{campbell2020universal}%
  \BibitemOpen
  \bibfield  {author} {\bibinfo {author} {\bibfnamefont {D.~L.}\ \bibnamefont
  {Campbell}}, \bibinfo {author} {\bibfnamefont {Y.-P.}\ \bibnamefont {Shim}},
  \bibinfo {author} {\bibfnamefont {B.}~\bibnamefont {Kannan}}, \bibinfo
  {author} {\bibfnamefont {R.}~\bibnamefont {Winik}}, \bibinfo {author}
  {\bibfnamefont {D.~K.}\ \bibnamefont {Kim}}, \bibinfo {author} {\bibfnamefont
  {A.}~\bibnamefont {Melville}}, \bibinfo {author} {\bibfnamefont {B.~M.}\
  \bibnamefont {Niedzielski}}, \bibinfo {author} {\bibfnamefont {J.~L.}\
  \bibnamefont {Yoder}}, \bibinfo {author} {\bibfnamefont {C.}~\bibnamefont
  {Tahan}}, \bibinfo {author} {\bibfnamefont {S.}~\bibnamefont {Gustavsson}},
  \emph {et~al.},\ }\href {https://doi.org/10.1103/PhysRevX.10.041051}
  {\bibfield  {journal} {\bibinfo  {journal} {Physical Review X}\ }\textbf
  {\bibinfo {volume} {10}},\ \bibinfo {pages} {041051} (\bibinfo {year}
  {2020})}\BibitemShut {NoStop}%
\bibitem [{Note2()}]{Note2}%
  \BibitemOpen
  \bibinfo {note} {$T_1$ may fluctuate and differ considerably for different
  transmons, leading to an extra term in Eq.~\protect \textup {\hbox
  {\mathsurround \z@ \protect \normalfont (\ignorespaces \ref
  {eq_dualrail_amp}\unskip \@@italiccorr )}} capturing dephasing at the rate
  $\Delta \Gamma = |\Gamma _1 - \Gamma _2|$, where $\Gamma _i$ is the amplitude
  damping rate for transmon $i = 1, 2$. However, the strong coupling between
  the two transmons comprising the dual-rail qubit substantially reduces
  dephasing as long as $g_{\protect \rm c} \gg \Gamma _1, \Gamma
  _2$.}\BibitemShut {Stop}%
\bibitem [{Note3()}]{Note3}%
  \BibitemOpen
  \bibinfo {note} {The numerical prefactor in the definition of $T_\phi $ was
  chosen such that the short-time expansion of the decoherence function is
  consistent with that of a single transmon, $\protect \qopname \relax
  o{exp}[-(t/T^{\protect \rm trans}_\phi )^2] \simeq 1-(t/T^{\protect \rm
  trans}_\phi )^2$.}\BibitemShut {Stop}%
\bibitem [{\citenamefont {Blais}\ \emph {et~al.}(2004)\citenamefont {Blais},
  \citenamefont {Huang}, \citenamefont {Wallraff}, \citenamefont {Girvin},\
  and\ \citenamefont {Schoelkopf}}]{Blais2004Cavity}%
  \BibitemOpen
  \bibfield  {author} {\bibinfo {author} {\bibfnamefont {A.}~\bibnamefont
  {Blais}}, \bibinfo {author} {\bibfnamefont {R.-S.}\ \bibnamefont {Huang}},
  \bibinfo {author} {\bibfnamefont {A.}~\bibnamefont {Wallraff}}, \bibinfo
  {author} {\bibfnamefont {S.~M.}\ \bibnamefont {Girvin}},\ and\ \bibinfo
  {author} {\bibfnamefont {R.~J.}\ \bibnamefont {Schoelkopf}},\ }\href
  {https://doi.org/10.1103/PhysRevA.69.062320} {\bibfield  {journal} {\bibinfo
  {journal} {Physical Review A}\ }\textbf {\bibinfo {volume} {69}},\ \bibinfo
  {pages} {062320} (\bibinfo {year} {2004})}\BibitemShut {NoStop}%
\bibitem [{\citenamefont {Wallraff}\ \emph {et~al.}(2004)\citenamefont
  {Wallraff}, \citenamefont {Schuster}, \citenamefont {Blais}, \citenamefont
  {Frunzio}, \citenamefont {Huang}, \citenamefont {Majer}, \citenamefont
  {Kumar}, \citenamefont {Girvin},\ and\ \citenamefont
  {Schoelkopf}}]{Wallraff2004strong}%
  \BibitemOpen
  \bibfield  {author} {\bibinfo {author} {\bibfnamefont {A.}~\bibnamefont
  {Wallraff}}, \bibinfo {author} {\bibfnamefont {D.~I.}\ \bibnamefont
  {Schuster}}, \bibinfo {author} {\bibfnamefont {A.}~\bibnamefont {Blais}},
  \bibinfo {author} {\bibfnamefont {L.}~\bibnamefont {Frunzio}}, \bibinfo
  {author} {\bibfnamefont {R.-S.}\ \bibnamefont {Huang}}, \bibinfo {author}
  {\bibfnamefont {J.}~\bibnamefont {Majer}}, \bibinfo {author} {\bibfnamefont
  {S.}~\bibnamefont {Kumar}}, \bibinfo {author} {\bibfnamefont {S.~M.}\
  \bibnamefont {Girvin}},\ and\ \bibinfo {author} {\bibfnamefont {R.~J.}\
  \bibnamefont {Schoelkopf}},\ }\href {https://doi.org/10.1038/nature02851}
  {\bibfield  {journal} {\bibinfo  {journal} {Nature}\ }\textbf {\bibinfo
  {volume} {431}},\ \bibinfo {pages} {162} (\bibinfo {year}
  {2004})}\BibitemShut {NoStop}%
\bibitem [{\citenamefont {Sete}\ \emph {et~al.}(2015)\citenamefont {Sete},
  \citenamefont {Martinis},\ and\ \citenamefont {Korotkov}}]{Sete2015quantum}%
  \BibitemOpen
  \bibfield  {author} {\bibinfo {author} {\bibfnamefont {E.~A.}\ \bibnamefont
  {Sete}}, \bibinfo {author} {\bibfnamefont {J.~M.}\ \bibnamefont {Martinis}},\
  and\ \bibinfo {author} {\bibfnamefont {A.~N.}\ \bibnamefont {Korotkov}},\
  }\href {https://doi.org/10.1103/PhysRevA.92.012325} {\bibfield  {journal}
  {\bibinfo  {journal} {Physical Review A}\ }\textbf {\bibinfo {volume} {92}},\
  \bibinfo {pages} {012325} (\bibinfo {year} {2015})}\BibitemShut {NoStop}%
\bibitem [{Note4()}]{Note4}%
  \BibitemOpen
  \bibinfo {note} {We assumed here that the fluctuations in $n_{\protect \rm
  c}=\protect \tilde {c}^\dagger \protect \tilde {c}$ happen on a long enough
  time scale such that we can diagonalize the qubit Hamiltonian in the
  adiabatic basis.}\BibitemShut {Stop}%
\bibitem [{\citenamefont {Duan}\ and\ \citenamefont
  {Guo}(1997)}]{PhysRevLett.79.1953}%
  \BibitemOpen
  \bibfield  {author} {\bibinfo {author} {\bibfnamefont {L.-M.}\ \bibnamefont
  {Duan}}\ and\ \bibinfo {author} {\bibfnamefont {G.-C.}\ \bibnamefont {Guo}},\
  }\href {https://doi.org/10.1103/PhysRevLett.79.1953} {\bibfield  {journal}
  {\bibinfo  {journal} {Phys. Rev. Lett.}\ }\textbf {\bibinfo {volume} {79}},\
  \bibinfo {pages} {1953} (\bibinfo {year} {1997})}\BibitemShut {NoStop}%
\bibitem [{\citenamefont {Lidar}\ \emph {et~al.}(1998)\citenamefont {Lidar},
  \citenamefont {Chuang},\ and\ \citenamefont {Whaley}}]{PhysRevLett.81.2594}%
  \BibitemOpen
  \bibfield  {author} {\bibinfo {author} {\bibfnamefont {D.~A.}\ \bibnamefont
  {Lidar}}, \bibinfo {author} {\bibfnamefont {I.~L.}\ \bibnamefont {Chuang}},\
  and\ \bibinfo {author} {\bibfnamefont {K.~B.}\ \bibnamefont {Whaley}},\
  }\href {https://doi.org/10.1103/PhysRevLett.81.2594} {\bibfield  {journal}
  {\bibinfo  {journal} {Phys. Rev. Lett.}\ }\textbf {\bibinfo {volume} {81}},\
  \bibinfo {pages} {2594} (\bibinfo {year} {1998})}\BibitemShut {NoStop}%
\bibitem [{Note5()}]{Note5}%
  \BibitemOpen
  \bibinfo {note} {We assume here that false-positive errors are dominated by
  amplitude decay events during the measurement, as opposed to errors in
  resolving the cavity quantum state.}\BibitemShut {Stop}%
\bibitem [{\citenamefont {Yan}\ \emph {et~al.}(2013)\citenamefont {Yan},
  \citenamefont {Gustavsson}, \citenamefont {Bylander}, \citenamefont {Jin},
  \citenamefont {Yoshihara}, \citenamefont {Cory}, \citenamefont {Nakamura},
  \citenamefont {Orlando},\ and\ \citenamefont {Oliver}}]{Yan2013rotating}%
  \BibitemOpen
  \bibfield  {author} {\bibinfo {author} {\bibfnamefont {F.}~\bibnamefont
  {Yan}}, \bibinfo {author} {\bibfnamefont {S.}~\bibnamefont {Gustavsson}},
  \bibinfo {author} {\bibfnamefont {J.}~\bibnamefont {Bylander}}, \bibinfo
  {author} {\bibfnamefont {X.}~\bibnamefont {Jin}}, \bibinfo {author}
  {\bibfnamefont {F.}~\bibnamefont {Yoshihara}}, \bibinfo {author}
  {\bibfnamefont {D.~G.}\ \bibnamefont {Cory}}, \bibinfo {author}
  {\bibfnamefont {Y.}~\bibnamefont {Nakamura}}, \bibinfo {author}
  {\bibfnamefont {T.~P.}\ \bibnamefont {Orlando}},\ and\ \bibinfo {author}
  {\bibfnamefont {W.~D.}\ \bibnamefont {Oliver}},\ }\href
  {https://doi.org/10.1038/ncomms3337} {\bibfield  {journal} {\bibinfo
  {journal} {Nature Communications}\ }\textbf {\bibinfo {volume} {4}},\
  \bibinfo {pages} {1} (\bibinfo {year} {2013})}\BibitemShut {NoStop}%
\bibitem [{Note6()}]{Note6}%
  \BibitemOpen
  \bibinfo {note} {While the g-f qubit frequency is insensitive to noise in the
  transmon frequency, it is sensitive to noise in the spin-locking drive
  amplitude, Eq.~\protect \textup {\hbox {\mathsurround \z@ \protect
  \normalfont (\ignorespaces \ref {eq:gfDrive}\unskip \@@italiccorr )}}. The
  latter can nevertheless be dealt with by applying concatenated decoupling
  schemes~\cite {cai2012robust,Genov2019mixed}, which involves pulsing the
  phase of the drive.}\BibitemShut {Stop}%
\bibitem [{\citenamefont {Wu}\ \emph {et~al.}(2022)\citenamefont {Wu},
  \citenamefont {Kolkowitz}, \citenamefont {Puri},\ and\ \citenamefont
  {Thompson}}]{Wu2022erasure}%
  \BibitemOpen
  \bibfield  {author} {\bibinfo {author} {\bibfnamefont {Y.}~\bibnamefont
  {Wu}}, \bibinfo {author} {\bibfnamefont {S.}~\bibnamefont {Kolkowitz}},
  \bibinfo {author} {\bibfnamefont {S.}~\bibnamefont {Puri}},\ and\ \bibinfo
  {author} {\bibfnamefont {J.~D.}\ \bibnamefont {Thompson}},\ }\href
  {https://doi.org/10.48550/ARXIV.2201.03540} {\bibinfo {title} {Erasure
  conversion for fault-tolerant quantum computing in alkaline earth rydberg
  atom arrays}},\ \bibinfo {howpublished} {arXiv:2201.03540} (\bibinfo {year}
  {2022})\BibitemShut {NoStop}%
\bibitem [{\citenamefont {Chamberland}\ \emph {et~al.}(2022)\citenamefont
  {Chamberland}, \citenamefont {Noh}, \citenamefont {Arrangoiz-Arriola},
  \citenamefont {Campbell}, \citenamefont {Hann}, \citenamefont {Iverson},
  \citenamefont {Putterman}, \citenamefont {Bohdanowicz}, \citenamefont
  {Flammia}, \citenamefont {Keller}, \citenamefont {Refael}, \citenamefont
  {Preskill}, \citenamefont {Jiang}, \citenamefont {Safavi-Naeini},
  \citenamefont {Painter},\ and\ \citenamefont
  {Brand\~ao}}]{Chamberland2022building}%
  \BibitemOpen
  \bibfield  {author} {\bibinfo {author} {\bibfnamefont {C.}~\bibnamefont
  {Chamberland}}, \bibinfo {author} {\bibfnamefont {K.}~\bibnamefont {Noh}},
  \bibinfo {author} {\bibfnamefont {P.}~\bibnamefont {Arrangoiz-Arriola}},
  \bibinfo {author} {\bibfnamefont {E.~T.}\ \bibnamefont {Campbell}}, \bibinfo
  {author} {\bibfnamefont {C.~T.}\ \bibnamefont {Hann}}, \bibinfo {author}
  {\bibfnamefont {J.}~\bibnamefont {Iverson}}, \bibinfo {author} {\bibfnamefont
  {H.}~\bibnamefont {Putterman}}, \bibinfo {author} {\bibfnamefont {T.~C.}\
  \bibnamefont {Bohdanowicz}}, \bibinfo {author} {\bibfnamefont {S.~T.}\
  \bibnamefont {Flammia}}, \bibinfo {author} {\bibfnamefont {A.}~\bibnamefont
  {Keller}}, \bibinfo {author} {\bibfnamefont {G.}~\bibnamefont {Refael}},
  \bibinfo {author} {\bibfnamefont {J.}~\bibnamefont {Preskill}}, \bibinfo
  {author} {\bibfnamefont {L.}~\bibnamefont {Jiang}}, \bibinfo {author}
  {\bibfnamefont {A.~H.}\ \bibnamefont {Safavi-Naeini}}, \bibinfo {author}
  {\bibfnamefont {O.}~\bibnamefont {Painter}},\ and\ \bibinfo {author}
  {\bibfnamefont {F.~G.}\ \bibnamefont {Brand\~ao}},\ }\href
  {https://doi.org/10.1103/PRXQuantum.3.010329} {\bibfield  {journal} {\bibinfo
   {journal} {PRX Quantum}\ }\textbf {\bibinfo {volume} {3}},\ \bibinfo {pages}
  {010329} (\bibinfo {year} {2022})}\BibitemShut {NoStop}%
\bibitem [{Note7()}]{Note7}%
  \BibitemOpen
  \bibinfo {note} {Concatenation of the four-qubit code and the surface code
  has already been considered in the context of the standard circuit noise with
  no amplitude damping errors in Ref.~\cite {Criger2016}.}\BibitemShut {Stop}%
\bibitem [{\citenamefont {Bergli}\ \emph {et~al.}(2009)\citenamefont {Bergli},
  \citenamefont {Galperin},\ and\ \citenamefont
  {Altshuler}}]{Bergli2009decoherence}%
  \BibitemOpen
  \bibfield  {author} {\bibinfo {author} {\bibfnamefont {J.}~\bibnamefont
  {Bergli}}, \bibinfo {author} {\bibfnamefont {Y.~M.}\ \bibnamefont
  {Galperin}},\ and\ \bibinfo {author} {\bibfnamefont {B.~L.}\ \bibnamefont
  {Altshuler}},\ }\href {https://doi.org/10.1088/1367-2630/11/2/025002}
  {\bibfield  {journal} {\bibinfo  {journal} {New Journal of Physics}\ }\textbf
  {\bibinfo {volume} {11}},\ \bibinfo {pages} {025002} (\bibinfo {year}
  {2009})}\BibitemShut {NoStop}%
\bibitem [{\citenamefont {O'Malley}(2016)}]{Omalley2016superconducting}%
  \BibitemOpen
  \bibfield  {author} {\bibinfo {author} {\bibfnamefont {P.~J.~J.}\
  \bibnamefont {O'Malley}},\ }\emph {\bibinfo {title} {Superconducting qubits:
  dephasing and quantum chemistry}},\ \href
  {https://web.physics.ucsb.edu/~martinisgroup/theses/OMalley2016.pdf} {Ph.D.
  thesis},\ \bibinfo  {school} {University of California, Santa Barbara}
  (\bibinfo {year} {2016})\BibitemShut {NoStop}%
\bibitem [{Note8()}]{Note8}%
  \BibitemOpen
  \bibinfo {note} {One way to overcome this problem is to realize a measurement
  that detects a single amplitude damping error after every $\pi $ pulse,
  however, at least in the superconducting scenario this is not
  realistic.}\BibitemShut {Stop}%
\bibitem [{\citenamefont {Cai}\ \emph {et~al.}(2012)\citenamefont {Cai},
  \citenamefont {Naydenov}, \citenamefont {Pfeiffer}, \citenamefont
  {McGuinness}, \citenamefont {Jahnke}, \citenamefont {Jelezko}, \citenamefont
  {Plenio},\ and\ \citenamefont {Retzker}}]{cai2012robust}%
  \BibitemOpen
  \bibfield  {author} {\bibinfo {author} {\bibfnamefont {J.}~\bibnamefont
  {Cai}}, \bibinfo {author} {\bibfnamefont {B.}~\bibnamefont {Naydenov}},
  \bibinfo {author} {\bibfnamefont {R.}~\bibnamefont {Pfeiffer}}, \bibinfo
  {author} {\bibfnamefont {L.~P.}\ \bibnamefont {McGuinness}}, \bibinfo
  {author} {\bibfnamefont {K.~D.}\ \bibnamefont {Jahnke}}, \bibinfo {author}
  {\bibfnamefont {F.}~\bibnamefont {Jelezko}}, \bibinfo {author} {\bibfnamefont
  {M.~B.}\ \bibnamefont {Plenio}},\ and\ \bibinfo {author} {\bibfnamefont
  {A.}~\bibnamefont {Retzker}},\ }\href
  {https://doi.org/10.1088/1367-2630/14/11/113023} {\bibfield  {journal}
  {\bibinfo  {journal} {New Journal of Physics}\ }\textbf {\bibinfo {volume}
  {14}},\ \bibinfo {pages} {113023} (\bibinfo {year} {2012})}\BibitemShut
  {NoStop}%
\bibitem [{\citenamefont {Genov}\ \emph {et~al.}(2019)\citenamefont {Genov},
  \citenamefont {Aharon}, \citenamefont {Jelezko},\ and\ \citenamefont
  {Retzker}}]{Genov2019mixed}%
  \BibitemOpen
  \bibfield  {author} {\bibinfo {author} {\bibfnamefont {G.~T.}\ \bibnamefont
  {Genov}}, \bibinfo {author} {\bibfnamefont {N.}~\bibnamefont {Aharon}},
  \bibinfo {author} {\bibfnamefont {F.}~\bibnamefont {Jelezko}},\ and\ \bibinfo
  {author} {\bibfnamefont {A.}~\bibnamefont {Retzker}},\ }\href
  {https://doi.org/10.1088/2058-9565/ab2afd} {\bibfield  {journal} {\bibinfo
  {journal} {Quantum Science and Technology}\ }\textbf {\bibinfo {volume}
  {4}},\ \bibinfo {pages} {035010} (\bibinfo {year} {2019})}\BibitemShut
  {NoStop}%
\bibitem [{\citenamefont {Criger}\ and\ \citenamefont
  {Terhal}(2016)}]{Criger2016}%
  \BibitemOpen
  \bibfield  {author} {\bibinfo {author} {\bibfnamefont {B.}~\bibnamefont
  {Criger}}\ and\ \bibinfo {author} {\bibfnamefont {B.}~\bibnamefont
  {Terhal}},\ }\href {https://doi.org/10.26421/qic16.15-16} {\bibfield
  {journal} {\bibinfo  {journal} {Quantum Information \& Computation}\ }\textbf
  {\bibinfo {volume} {16}},\ \bibinfo {pages} {1261} (\bibinfo {year}
  {2016})}\BibitemShut {NoStop}%
\end{thebibliography}%

\end{document}